\documentclass[final,3p,times]{elsarticle}
\usepackage{amssymb}
\usepackage{multicol}
\usepackage{graphicx}
\usepackage{epsfig}
\usepackage{fancybox}
\usepackage{pstricks,pst-node,pst-text,pst-3d}
\usepackage{subfigure}
\journal{Physics Letters A}
\begin{document}
\begin{frontmatter}
\title{Monte Carlo simulations of dynamic phase transitions in ultrathin Blume-Capel films }
\author{Yusuf Y\"{u}ksel}
\cortext[cor1]{Corresponding author. Tel.: +90 2324119518; fax: +90 2324534188.}
\ead{yusuf.yuksel@deu.edu.tr}
\address{Department of Physics, Dokuz Eyl\"{u}l University, TR-35160 Izmir, Turkey}
\begin{abstract}
Dynamic phase transition phenomena in ultrathin films described by Blume-Capel model have been investigated using Monte Carlo simulations. Hysteresis loops, micromagnetic structures, and hysteresis loop area curves, as well as dynamic correlation between the magnetization and the external field have been studied as functions of the field, as well as the film parameters. The variation of critical coupling of the modified film surface at which the transition temperature becomes independent of film thickness have been clarified for varying system parameters. Frequency dispersion of hysteresis loop area have been found to obey a power law for low and moderate frequencies for both ordinary and enhanced surfaces.
\end{abstract}
\end{frontmatter}
\section{Introduction}\label{intro}
Recently, magnetic materials with finite sizes such as thin films have attracted considerable amount of interest both in theoretical and experimental manner and the research of thin film magnetism is
a current topic of critical phenomena \cite{falicov,kaneyoshi1,pleimling,Kaneyoshi2,hasenbusch}.  In magnetic systems with thin film geometry, due to their reduced coordination number, the surface atoms may have a lower symmetry in comparison with that of the inner atoms, meanwhile, the exchange interactions between the surface atoms may be different from those between the corresponding bulk counterparts, leading to a phenomenon known as surface enhancement in which the surface may exhibit an ordered phase even if the bulk itself is disordered. This phenomenon has already been experimentally observed \cite{ran,polak,tang}. In addition to the fundamental purpose, thin film materials have great importance in technological applications. For instance, ultrathin Au/Co/Au magnetic films, CoPt-alloy films, CoNi/Pt and Tb/Fe multilayers are considered as potential candidates of raw materials in ultrahigh-density magneto-optic recording devices \cite{lyberatos}.

Due to the presence of modified exchange couplings at the surface, magnetic thin film systems may exhibit an extraordinary phase transition at which the surface transition temperature is higher than that of the bulk whereas in the ordinary case, the transition temperature of the film is mainly determined by the bulk region. In order to investigate the thermal and magnetic properties of thin films, the models based on an Ising type spin Hamiltonian are well suited since many thin films such as the Fe/Ag(100) system \cite{qiu} exhibit a strong uniaxial anisotropy. A great many theoretical efforts have shown that there exists a critical value of surface to bulk ratio of exchange interactions $R_{c}$ above which the surface effects are dominant and the transition temperature of the entire film is determined by the surface magnetization whereas below $R_{c}$, the transition characteristics of the film are governed by the bulk magnetization. The critical value $R_{c}$ itself is called as the special point at which the transition temperature of the film becomes independent of thickness, and the numerical value of this point has been examined within various theoretical techniques for spin-1/2 case \cite{aguilera,binder1,kaneyoshi3,burkhart,landau,sarmento1,neto,akinci1,hasenbusch1}. On the other hand, the problem has also been extended to higher spins using a number of techniques \cite{zhong,jia,tucker1,bahmad,zaim,saber1,kaneyoshi4}. Among these works, using extensive Monte Carlo (MC) simulations, the effect of surface exchange enhancement on ultrathin spin-1 films has been studied by Tucker \cite{tucker1}, and it was concluded that the $R_{c}$ value is spin dependent. However, in a recent work \cite{zaim}, using MC simulations, the influence of crystal field
interaction (or single ion anisotropy) on the critical behavior of a magnetic spin-1 film has been studied, and it has been argued that $R_{c}$ is independent of the crystal field interaction. This latter outcome clearly contradicts with that reported in a very recent study based on the effective-field theory (EFT) \cite{yuksel1}. Moreover, Wang et al. \cite{wang} have studied the magnetic properties and critical behavior of a molecular-based magnetic film within the framework of MC simulations. They have acquired two special points at which the compensation and critical temperatures are independent of the layer thickness and the surface single-ion anisotropy.

Apart from these, in contrast to the large number of works concerning the equilibrium behavior of thin films, there are very few studies in the literature devoted to the investigation of nonequilibrium phase transition characteristics of thin film systems \cite{aktas}, and there exist several uncertain points remaining to be clarified in the latter case. For instance, the effect of a time dependent perturbation such as the presence of an oscillating external field on the magnetic behavior of film should be clarified. When a ferromagnetic material is subjected to a periodically oscillating magnetic field, the response of the ferromagnet may be delayed depending on the competition between the period of the external field and the relaxation time of the system. This competition shows itself as a phase lag between the external perturbation and response (i.e. magnetization) of the system. It has been previously shown by Acharyya \cite{acharyya1} that for bulk systems near the transition point, the phase lag gives a peak by achieving its maximum value. Critical phenomena in nonequilibrium bulk systems have been clarified using several techniques such as MC simulations \cite{acharyya1,lo,chakrabarti}, mean-field theory (MFT) \cite{tome}, and recently EFT \cite{deviren1}. In addition, related to this nonequilibrium phenomena, a few theoretical studies have been devoted to the investigation of dynamical aspects of phase transition properties of finite systems such as cylindrical Ising nanowire and nanotube systems \cite{deviren2,deviren3}, as well as ferrimagnetic nanoparticles \cite{yuksel2}. Moreover, using MC simulations, Laosiritaworn \cite{lao} has investigated the thickness dependence of hysteresis properties in Ising thin films, and it has been found that  the calculated hysteresis properties significantly change due to the stronger ferromagnetic coupling in thicker films. It has also been shown that  hysteresis properties obey the universal power law relations for varying thickness and field parameters. However, effects of the presence of modified surface exchange interactions were ignored in the aforementioned work. Recently, Park and Pleimling \cite{park} have shown that nonequilibrium surface exponents do not coincide with those of the equilibrium critical surface.

On the experimental side, a dynamic phase transition (DPT) occurring for the high-frequency magnetic fields was studied by Jiang et al. \cite{jiang} using the surface magneto-optical Kerr-effect technique for epitaxially grown ultrathin Co films on a Cu (001) surface. For a [Co(4{\AA})/Pt(7{\AA})] multilayer system with strong perpendicular anisotropy, an example of DPT has been observed by Robb et al. \cite{robb}. They found that the experimental nonequilibrium phase diagrams strongly resemble the dynamic behavior predicted from theoretical calculations of a kinetic Ising model. It is clear from these works that there exists strong evidence of qualitative consistency between theoretical and experimental studies.

Another important feature in nonequilibrium phase transitions is the hysteretic response of the system to the periodically oscillating magnetic fields. In dynamic systems (i.e. the systems in the presence of ac fields), the hysteresis phenomenon is related to a dynamic phase lag between instantaneous magnetization and periodic external magnetic field \cite{yuksel2}. In contrast to the behavior observed in static models (i.e. the systems with dc magnetic fields), where the strength of the external field does not change with time explicitly, dynamic hysteresis in nonequilibrium phase transitions is observed in dynamically paramagnetic phase. In other words,
in the presence of dc magnetic fields, a ferromagnetic material exhibits a nonzero coercivity. By the agency of a tunable parameter, such as increasing the temperature, this coercive field reduces to zero in the vicinity of ferromagnetic-paramagnetic phase transition region. However, the situation is clearly different in the presence of an ac field. Namely, at high oscillation frequencies of the external field, the relaxation time of the system is relatively much larger than the oscillation period of the external perturbation. Therefore, the system is not able to follow the driving field which results in a dynamic symmetry loss leading to asymmetric hysteresis curves. In this case, coercivity is not applicable since the magnetization never reaches to zero. As the field frequency decreases or the amplitude of the external field increases, the system approaches to dynamic paramagnetic phase, and we observe symmetric hysteresis loops with nonzero coercivity \cite{acharyya1,lo,tome}.

Explanation of static hysteresis behavior observed in the bulk magnetic systems in the presence of dc fields has been well established \cite{bertotti}. Jiles and Atherton \cite{jiles1} proposed a theory for explaining the magnetization process in ferromagnetic materials based on the two fundamental mechanisms. Namely, the propagation of domain walls under the influence of an applied field, and rotation of aligned moments within a domain towards the  field direction. They have also compared their theory \cite{jiles2} with experimental results for ferromagnetic steel by considering the various types of magnetization curves such as the anhysteretic, initial magnetization curve, and families of major  hysteresis loops. In all cases they found an excellent agreement. Zapperi et al. \cite{stanley} investigated the motion of a ferromagnetic domain wall driven by an external magnetic field through a disordered medium and observed remarkable agreement with experiments on $\mathrm{Fe_{21}Co_{64}B_{15}}$ amorphous alloy. Similar analysis have also been performed for hysteresis in ferroelectric materials based on the domain-wall theory \cite{smith}. On the other hand, for the systems in thin film geometry, Haas et al. \cite{haas} performed a comprehensive experimental study on the late stages of phase separation in thin polymer solution films with thicknesses $1.5$ and $10\mu m$, and estimated the average domain size as a function of the film thickness. Weir et al. \cite{weir}  determined a value for the domain-wall energy and investigated the magnetization reversal of CoNi/Pt multilayers supporting perpendicular magnetization. They have also shown that CoNi/Pt multilayers support irregular fine-scale domain structures. Nowak \cite{nowak} described the magnetic hysteresis behavior in thin ferromagnetic films with in plane anisotropy within the framework of Stoner-Wohlfarth model and obtained very realistic shapes of hysteresis loops. Moreover, a detailed review of comprehensive works on field-induced magnetization reversal in ultrathin ferromagnetic films can be found in \cite{ferre}.

It is clear from above discussions that the equilibrium phase transition properties, as well as the dc hysteresis process in thin films are well known. However, in nonequilibrium case, the problem deserves particular attention. Therefore, in the present paper, we have extended the model of Tucker \cite{tucker1} by taking into account a time dependent perturbation in the thin film system described by spin-1 Blume-Capel Hamiltonian. For this purpose, we have calculated the phase diagrams and dynamic specific heat curves, as well as  several characteristic quantities peculiar to kinetic models such as hysteresis loop area and dynamic correlation between magnetization and ac fields, and we consider the effect of the system parameters on these quantities. We have also discussed the micromagnetic structure of the system for selected system parameters at certain stages of the oscillating field.

The paper is organized as follows: In Section \ref{formulation}, we describe the model and its related dynamic quantities which have been calculated throughout the analysis. Section \ref{results} contains the numerical results and discussions. Finally, Section \ref{conclude} is devoted to our conclusions.

\section{Formulation}\label{formulation}
We consider a ferromagnetic thin film with thickness $L$ described by conventional spin-1 Blume-Capel Hamiltonian \cite{blume,capel}
\begin{equation}\label{eq1}
\mathcal{H}=-\sum_{<ij>}J_{ij}S_{i}S_{j}-D\sum_{i}(S_{i})^{2}-h(t)\sum_{i}S_{i},
\end{equation}
where $S_{i}=\pm1,0$ is a three-state spin variable, and $J_{ij}$ is the nearest neighbor interaction energy. The summation in the  first term is taken over only the nearest neighbor interactions whereas the sums in the second and third terms are carried out over the all lattice sites. The second term in Eq. (\ref{eq1}) represents the crystal field energy. In the third term, $h(t)=h_{0}sin(\omega t)$ represents the oscillating magnetic field, where $h_{0}$ and $\omega$ are the amplitude and the angular frequency of the applied field, respectively. The period of the oscillating magnetic field is given by $\tau=2\pi/\omega$. If the lattice sites $i$ and $j$ belong to one of the two surfaces of the film we have $J_{ij}=J_{s}$, otherwise $J_{ij}=J_{b}$, where $J_{s}$ and $J_{b}$ denote the ferromagnetic surface and bulk exchange interactions, respectively.

In order to simulate the system, we employ the Metropolis MC simulation algorithm \cite{binder2,newman} to Eq. (\ref{eq1})  on an $N\times N\times L$ simple cubic lattice and we apply periodic (free) boundary conditions in direction(s) parallel (perpendicular) to film plane. We have studied films with thickness $L=3,4,5$ with $N=70$ to simulate an ultrathin film \cite{tucker1}. For simplicity, the exchange couplings are restricted to the ferromagnetic case. Due to the existence of large number of adjustable parameters, namely the thickness $L$, exchange couplings $J_{s},J_{b}$, external field parameters $h_{0},\tau$, and temperature $T$, we are constrained to restrict the simulations for $D=0$ case. In a very recent work \cite{yuksel1} where we have studied the equilibrium properties (i.e. $h_{0}=0$) of thin Blume-Capel thin films, we have found that in terms of the shift exponent $\lambda$, a ferromagnetic spin-1/2 thin film is in the same universality class with its spin-1 counterpart.  However, single-ion anisotropy effects in nonequilibrium case (i.e. $h_{0}\neq0$) can be studied in a separate work.

Configurations were generated by selecting the sites in sequence through the lattice and making single-spin-flip
attempts, which were accepted or rejected according to the Metropolis algorithm, and $N\times N\times L$ sites are visited at each time step (a time step is defined as a MC step per site or simply MCS). The frequency $f$ of the oscillating magnetic field is defined in terms of MCSs in such a way that
$
f=1/\kappa \theta_{s},
$
where $\kappa$ is the number of MCSs necessary for one complete cycle of the oscillating field and $\theta_{s}$ is the time interval. In our simulations, we choose $\theta_{s}=1$, hence we obtain $\tau=\kappa$. Oscillation period of the external field is kept fixed as $\tau=100$ for the most of the simulations. However, in evaluating the frequency dispersion of hysteresis loop area, we consider period values within the range $10\leq \tau \leq 10^{3}$. Data were generated over $50-100$ independent sample realizations by running most of the realizations for 200 cycles of external field (i.e. $200\tau$ Monte Carlo steps per site) after discarding the first few cycles.

Our program calculates the instantaneous values of the bulk and surface magnetizations $M_{b}$ and $M_{s}$, at the time $t$. These quantities are defined as
$
M_{s}(t)=\frac{1}{N_{s}}\sum_{i=1}^{N_{s}}S_{i}, \quad M_{b}(t)=\frac{1}{N_{b}}\sum_{j=1}^{N_{b}}S_{j},
$
where $N_{s}$ and $N_{b}$ denote the number of spins in the surface and bulk layers, respectively. Using Eq. (\ref{eq1}), we calculate the total energy per spin $E_{tot}$. Consequently, the specific heat is defined as $C=dE_{tot}/dT$. The hysteresis loop area which measures the energy loss in a complete cycle of the external field is given by $A=-\oint m dh=-h_{0}\omega\oint m(t)\cos(\omega t)dt$ \cite{acharyya1}. Finally, the dynamic correlation between total magnetization $m(t)$ of the film and external field $h(t)$ is defined as $c=\frac{\omega}{2\pi}\oint m(t) h(t)dt=\frac{h_{0}\omega}{2\pi}\oint m(t)\sin(\omega t)dt$ \cite{acharyya1}.

We also note that the value of the bulk exchange interaction $J_{b}$ is fixed to unity, and we also use the normalized surface to bulk ratio of exchange interactions $R=J_{s}/J_{b}$, as well as the reduced field amplitude $H_{0}=h_{0}/J_{b}$, and reduced temperature $\Theta=k_{B}T/J_{b}$. We also set $k_{B}=1$.

\section{Results and Discussion}\label{results}
In Fig. \ref{fig1}, in order to obtain a general overview of the dynamic phase transition properties of spin-1 thin film model, we plot the phase diagrams depicted in a $(\Theta-R)$ plane with selected film thickness values $L=3,4,5$ and for three selected values of field amplitude $H_{0}$. Oscillation period is fixed as $\tau=100$. Fig. \ref{fig1} represents a characteristic phenomena peculiar to thin film systems. Namely, due to the existence of modified surfaces, there exists a special value of surface to bulk ratio of exchange interactions $R_{c}$ at which the transition temperature of the film becomes independent of thickness $L$. For $R<R_{c}$, we have ordinary transition behavior where the bulk magnetism is dominant against the surface magnetism whereas for $R>R_{c}$, the surface may exhibit enhanced magnetic behavior in comparison with bulk. This is called extraordinary transition. Moreover, as shown in Fig. \ref{fig1}, for $R<R_{c}$, thicker films have greater transition temperatures while for $R>R_{c}$, the transition temperature of the film decreases with increasing thickness. In order to provide a testing ground for our calculations, we have primarily studied the case $H_{0}=0$. This model defines equilibrium properties of spin-1 thin film system and has been examined previously using MC simulations \cite{tucker1,zaim}. Our simulated data fits well with those obtained in these works. By comparing Figs. \ref{fig1}a-\ref{fig1}c, we see that in the presence of oscillating magnetic fields, effect of field amplitude $H_{0}$ on the transition characteristics of the film is also straightforward: As $H_{0}$ increases then the magnetic energy supplied by the external field dominates against the ferromagnetic exchange couplings, and consequently, the system can relax within the oscillation period $\tau$ of the external field which causes a reduction in the transition temperature. However, variation of $R_{c}$ as a function of $H_{0}$ is very slow according to Fig. \ref{fig1}, and we see that the location of $R_{c}$ barely deviates from its equilibrium value \cite{tucker1} with increasing $H_{0}$.

In the left and middle-left panels of Fig. \ref{fig2}, we represent the thermal variation of specific heat and loop area curves corresponding to the phase diagrams depicted in Fig. \ref{fig1}b. Hysteresis loop area is a measure of energy dissipation (i.e. loss) due to the hysteresis. The temperature values at which the specific heat curves exhibit a sharp maximum correspond to the transition temperature of thin film. On the other hand, loop area exhibits a smooth and rounded cusp at a certain temperature above the transition temperature of the film both in the ordinary $(R=0.5)$ and extraordinary $(R=2.5)$ transition regions. This behavior of thermal variation of loop area curves has also been reported for bulk systems represented by kinetic Ising model \cite{acharyya1}. By comparing the loop area curves depicted in the left and middle-left panels of Fig. \ref{fig2}, we see that hysteretic loss becomes fairly minimized (i.e. the maximum lossy point reduces) as the surface effects become prominent. In the middle-right and right panels, bulk and surface hysteresis behaviors are shown in the presence of ordinary $(R=0.5)$ and enhanced $(R=2.5)$ surfaces, respectively. Since the temperature is below the transition point $(\Theta<\Theta_{c})$, the observed non-symmetric loops correspond to dynamic ferromagnetic phase in bulk and surface layers. It is clear in these figures that ferromagnetism in bulk (surface) layers is dominant against surface (bulk) layers for $R=0.5$ $(R=2.5)$.

Effects of the film thickness $L$, and the field amplitude $H_{0}$ on the thermal variation of loop areas are shown in Fig. \ref{fig3}. By comparing Figs.\ref{fig3}a-\ref{fig3}b and Figs.\ref{fig3}c-\ref{fig3}d with each other, we see that the maximum lossy point \cite{acharyya1} observed in the curves originates at higher temperatures for thicker (thinner) films in the presence of ordinary (enhanced) surfaces. Moreover, the maximum hysteretic loss is always observed for thinner films. A similar outcome has been recently reported for magnetic nanoparticle systems \cite{pankhurst}: large particles exhibit narrow hysteresis loops. It is also clear from Fig. \ref{fig3} that the temperature value corresponding to maximum lossy point in loop area curves slides to lower temperatures with increasing field amplitude values.

Fig. \ref{fig4} represents the film thickness $L$, and the field amplitude $H_{0}$ dependence of dynamic correlation versus temperature curves. At sufficiently low temperatures, thermal energy is almost negligible, and the ferromagnetic exchange interactions are dominant against the field energy, hence the dynamic correlation between magnetization of the film and oscillating external field is close to zero. In this case, the system exhibits a dynamic ferromagnetic phase. In the vicinity of the dynamic ferromagnetic-paramagnetic phase transition temperature, dynamic correlation curves exhibit a dip which becomes negative for sufficiently high $H_{0}$ values. This behavior of dynamic correlation curves is identical to those observed in bulk models simulated by the kinetic Ising model \cite{acharyya1}. The thinner films in the presence of enhanced surfaces with high field amplitudes exhibit the deepest negative dip point as a result of maximized phase lag between the magnetization and magnetic field in the vicinity of the dynamic phase transition temperature.

The responses of the bulk and surface magnetization curves to the oscillating external magnetic field corresponding to various stages of a typical dynamic correlation versus temperature curve are depicted in Fig. \ref{fig5}. The system parameters have been selected as $L=3$, $R=0.5$, $H_{0}=0.5$, and $\tau=100$. At the stage-I, the temperature is sufficiently low so that both the bulk and surface layers of the film exhibit a dynamic ferromagnetism. In this case, the dynamic correlation between oscillating field and magnetization of the system is close to zero, since the bulk and surface magnetizations can not follow the external field. At the stage-II, the temperature is slightly increased, ferromagnetism is reduced, and the loop areas of both bulk and surface layers become wider. This results in an increment in the dynamic correlation. At the stages I and II, bulk ferromagnetism is dominant against the surface. As the system attains the stage-III, phase lag between the external field and bulk (as well as surface) magnetization becomes maximum, indicating a dynamic phase transition. Observed hysteresis loops are square shaped, and the dynamic correlation exhibits a dip. As the temperature is increased above the transition temperature, we reach the stage-IV at which the dynamic correlation is maximized. The stage-IV hysteresis loops exhibit sigmoidal shape with a wide loop area. Further increment in temperature (the stage-V) causes narrower sigmoidal loops. At the stages III, IV and V, coercivity of bulk and surface layers are almost identical, however, bulk hysteresis exhibits a relatively large remanent magnetization in comparison with that of the surface. This is due to the existence of ordinary surfaces $(R=0.5<R_{c})$. However, in the presence of enhanced surfaces (i.e. $R>R_{c}$), we expect a similar scenario, except that the remanence of surface hysteresis will be larger than that of the bulk one.

A detailed investigation of effect of the presence of modified surfaces on the thermal variation of hysteresis loop area and dynamic correlation curves is represented in Fig. \ref{fig6}. In this figure, we consider the parameters $L=3$, $\tau=100$ and two different field amplitude values $H_{0}=0.1,0.5$. As shown in Fig. \ref{fig6}, the locations of the maximum lossy point observed in loop area curves and the dip point of dynamic correlation curves slide to higher temperatures with increasing $R$ or decreasing $H_{0}$. This is an expected result. As the surface effects become prominent, presence of enhanced ferromagnetic exchange interactions compensates the reduced coordination number of the surface layer. Therefore, under these circumstances, in order to observe a dynamic phase transition in the system, the thermal energy supplied by heat bath or the magnetic energy provided by the external field should be increased. Another interesting result depicted in Fig. \ref{fig6} can be discerned by investigating the variation of the maximum energy loss due to the hysteresis as a function of surface to bulk ratio of exchange interactions $R$. In Figs. \ref{fig6}a and \ref{fig6}c, we  observe that the maximum magnetic energy loss due to the hysteresis can be achieved at $R=1.0$. According to our simulation data, this loss purely originates from the fact that hysteresis loops exhibit enhanced remanence at $R=1.0$.

In order to clarify the microscopic origin of the maximum lossy point observed in the thermal variation of hysteresis loop area curves, stochastic domain patterns of the bulk and surface magnetizations within a complete cycle of the external field are depicted in Figs. \ref{fig7} and \ref{fig8}, in the presence of ordinary and enhanced surfaces, respectively. The histogram data presented in Figs. \ref{fig7} and \ref{fig8} have been collected over 50 independent sample realizations, each of which has been produced over 100 complete cycles of the external field. The illustrated magnetic patterns however, represent typical configurations as $S=1$ (red), $S=-1$ (blue), and $S=0$ (green) states.  Due to the symmetric structure of the film, only the top surface is considered. In Fig. \ref{fig7}, we consider a thin film with an ordinary surface. The selected temperature and the other system parameters are those corresponding to the maximum lossy point observed in the curve labeled $R=0.0$ in Fig. \ref{fig6}c. From Fig. \ref{fig7}, we see that at time $t=\tau/4$, external field reaches to its maximum value, and a positive magnetic energy (which tends to align the spins in the field direction) is transferred to the system. On the other hand, due to the large thermal energy $(\Theta>\Theta_{c})$, thermal fluctuations are also present (this energy drives the spins to a random alignment configuration). Consequently, due to the presence of dominant ferromagnetic bulk exchange interactions, bulk magnetization exhibits large domains of magnetic $S=1$ and $S=-1$ states which are separated by large domain walls whereas at the  surface, since the formation of domains is not energetically preferable, we observe nucleated droplets of magnetic $(S=\pm1)$ and nonmagnetic $S=0$ states. At the end of the half cycle (i.e. $t=\tau/2$), the energy supplied by the external field and ferromagnetic bulk coupling completely dominates against the thermal energy, and the bulk layer magnetization becomes magnetically saturated, and exhibit almost a single domain configuration of $S=1$ state whereas due to the weak ferromagnetic exchange and reduced coordination number, we still observe nucleated droplets at the surface layer. As a result of this mechanism, bulk remanence is fairly enhanced against the surface, and the remanent magnetization of the entire film is determined by the bulk layer. Moreover, at $t=3\tau/4$, magnetic field reversal occurs, and we observe a phase separation \cite{conti,roy,das} of magnetic $S=\pm1$ states in the bulk layer. Magnetic and nonmagnetic droplets coexist in the surface layer. Finally, at the end of the complete cycle at $t=\tau$, bulk layer reaches to negative remanence (which is relatively much greater than the surface remanence) by forming a single domain of $S=-1$ state. On the basis of the curve labeled $R=2.5$ in Fig. \ref{fig6}c,  we have performed similar analysis for the films in the presence of enhanced surfaces in Fig. \ref{fig8}. Under the guidance of the arguments mentioned above, we see that the maximum energy loss observed in the loop area curve in Fig. \ref{fig6}c mainly originates from the response of the surface magnetization of the film to the oscillating magnetic field. In this case, enhanced remanence of the surface layer is much larger than that of the bulk layer.

In Fig. \ref{fig9}, we evaluate the frequency dispersion of hysteresis loop area curves which have been classified as type-I and type-II. A similar classification scheme was also defined for bulk models \cite{punya} previously. In this regard, type-I curves correspond to dynamically paramagnetic phase at low oscillation frequencies, and exhibit a dynamic phase transition between paramagnetic and ferromagnetic states with increasing frequency. This phase transition shows itself as a rounded peak in the dispersion curves. On the other hand, type-II curves exhibit only paramagnetism within the whole frequency range. Effect of the field amplitude $H_{0}$ on the curves are clear. Namely, the loop area increases with increasing field amplitude. The system does not exhibit sigmoidal loops unless $H_{0}$ is greater than the coercivity of the film. Frequency induced phase transitions in type-I curves are observed to shift towards the high frequency regime with increasing $H_{0}$ values. The physical explanation of this behavior is straightforward. As  $H_{0}$ increases then the magnetization can follow the external oscillating field with some delay. In this case paramagnetic behavior is favored. Accordingly, in order to observe ferromagnetism, the relaxation time of the film should be larger than the oscillation period of the external field which can be achieved in the high frequency regime. Finally, whether $H_{0}$ is low or high, the low frequency curves exhibit a power law behavior both in the presence of ordinary and extraordinary surfaces. We note that the thicker films also exhibit qualitatively the same phenomena. Some examples of bulk and surface hysteresis loops regarding the dispersion curves depicted in Fig. \ref{fig9} have been presented in Fig. \ref{fig10} for ordinary $(R=0.5)$, as well as enhanced $(R=2.5)$ surfaces where the low frequency paramagnetic loops may turn into high frequency ferromagnetic loops due to a frequency induced dynamic symmetry loss in the system. Coercivity of both the bulk and surface magnetizations reduces with increasing field frequency. Effect of the modified surfaces on the hysteresis profiles are also noticeable. Most of the observations discussed above have also been reported for the core-shell magnetic nanoparticles \cite{wu}.

\section{Conclusions}\label{conclude}
In conclusion, we have applied Monte Carlo simulations to study the dynamic phase transition phenomena in ultrathin ferromagnetic Blume-Capel films in the presence of oscillating magnetic fields. The foremost results obtained from simulation data can be summarized as follows: Critical value of surface to bulk ratio of exchange interactions $R_{c}$ at which the transition temperature is independent of film thickness is not apparently responsive to varying field amplitude values, but exhibits slow variation as a function of $H_{0}$. As reported for bulk systems, loop area exhibits a smooth and rounded cusp at a certain temperature above the transition temperature of the film both in the ordinary and extraordinary transition regions. Moreover, hysteretic loss becomes fairly minimized (i.e. the maximum of the lossy point reduces) as the surface effects become prominent. The maximum lossy point \cite{acharyya1} observed in the curves originates at higher temperatures for thicker (thinner) films in the presence of ordinary (enhanced) surfaces. Besides, the maximum hysteretic loss is always observed for thinner films. The temperature value corresponding to maximum lossy point in loop area curves slides to lower temperatures with increasing field amplitude values.

In the vicinity of the dynamic ferromagnetic-paramagnetic phase transition temperature, dynamic correlation curves exhibit a dip which becomes negative for sufficiently high $H_{0}$ values. This behavior of dynamic correlation curves is identical to those observed in bulk models simulated by the kinetic Ising model \cite{acharyya1}. The thinner films in the presence of enhanced surfaces with high field amplitudes exhibit the deepest negative dip point as a result of maximized phase lag between the magnetization and magnetic field in the vicinity of the dynamic phase transition temperature. We have observed square shaped hysteresis loops in the vicinity of the transition temperature. These square shaped loops evolve into sigmoidal curves above the transition temperature. Area of the loops becomes wider when the dynamic correlation curve as a function of the temperature is maximum.

We have observed that the maximum magnetic energy loss due to the hysteresis can be achieved at $R=1.0$. According to our simulation data, this loss purely originates from the fact that hysteresis loops exhibit enhanced remanence at $R=1.0$. Furthermore, based on the information provided by the micromagnetic structure of the layer magnetizations, the maximum lossy point of loop area curves originates from enhanced remanence of bulk (surface) in the presence of ordinary (extraordinary) surfaces.

The frequency dispersion of hysteresis loop area curves exhibit two distinct behaviors which have been classified as type-I and type-II. In this regard, type-I curves correspond to dynamically paramagnetic phase at low oscillation frequencies, and exhibit a dynamic phase transition between paramagnetic and ferromagnetic states with increasing frequency. On the other hand, type-II curves exhibit only paramagnetism within the whole frequency range. Whether $H_{0}$ is low or high, the low frequency curves exhibit a power law behavior both in the presence of ordinary and extraordinary surfaces, and the low frequency paramagnetic loops may turn into high frequency ferromagnetic loops due to a frequency induced dynamic symmetry loss in the system.

Thin magnetic films are not only of great technological importance, but  these magnetic structures also present rich physics of fundamental interest. For instance, in Ref. \cite{wang}, it was reported that a molecular-based magnetic film may exhibit two special points at which the compensation and critical temperatures are independent of the film thickness, respectively. Dynamical aspects of such kinds of physical phenomena are worth to examine in the presence of oscillating magnetic fields.

Finally, we note that the use of natural Monte Carlo modeling for dynamic studies limits the upper bound of oscillation frequency as approximately $10^{-1}$ $\mathrm{(mcs^{-1})}$. It could also be
interesting to treat the problem presented in this study within the framework of a time-quantified Monte Carlo technique \cite{nowak2}.

\section*{Acknowledgements}
A critical reading of the manuscript by E. Vatansever from Dokuz Eyl\"{u}l University, and his many helpful suggestions are gratefully acknowledged.

\section*{References}

\newpage

\begin{figure}
\center
\includegraphics[width=8cm]{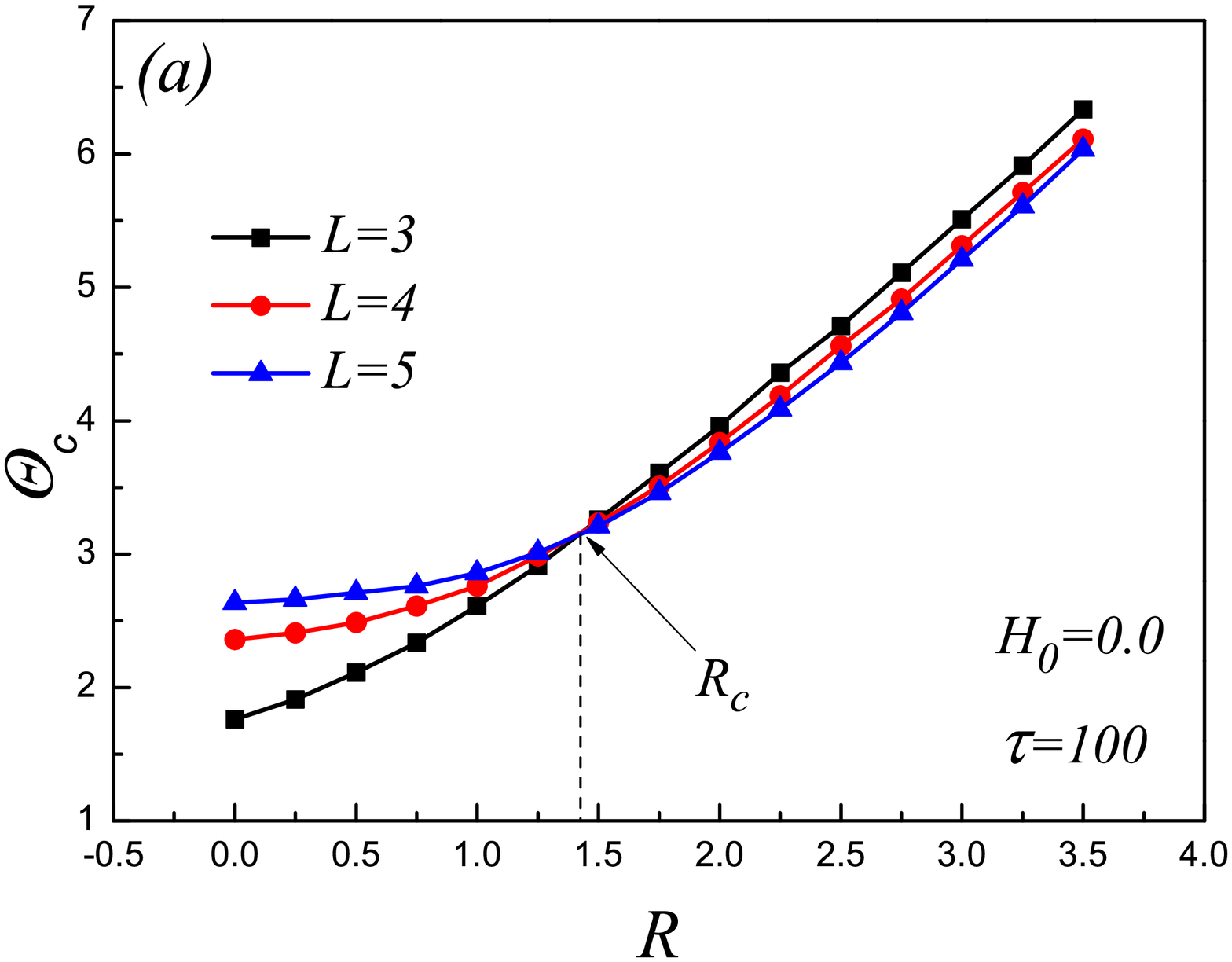}\\
\includegraphics[width=8cm]{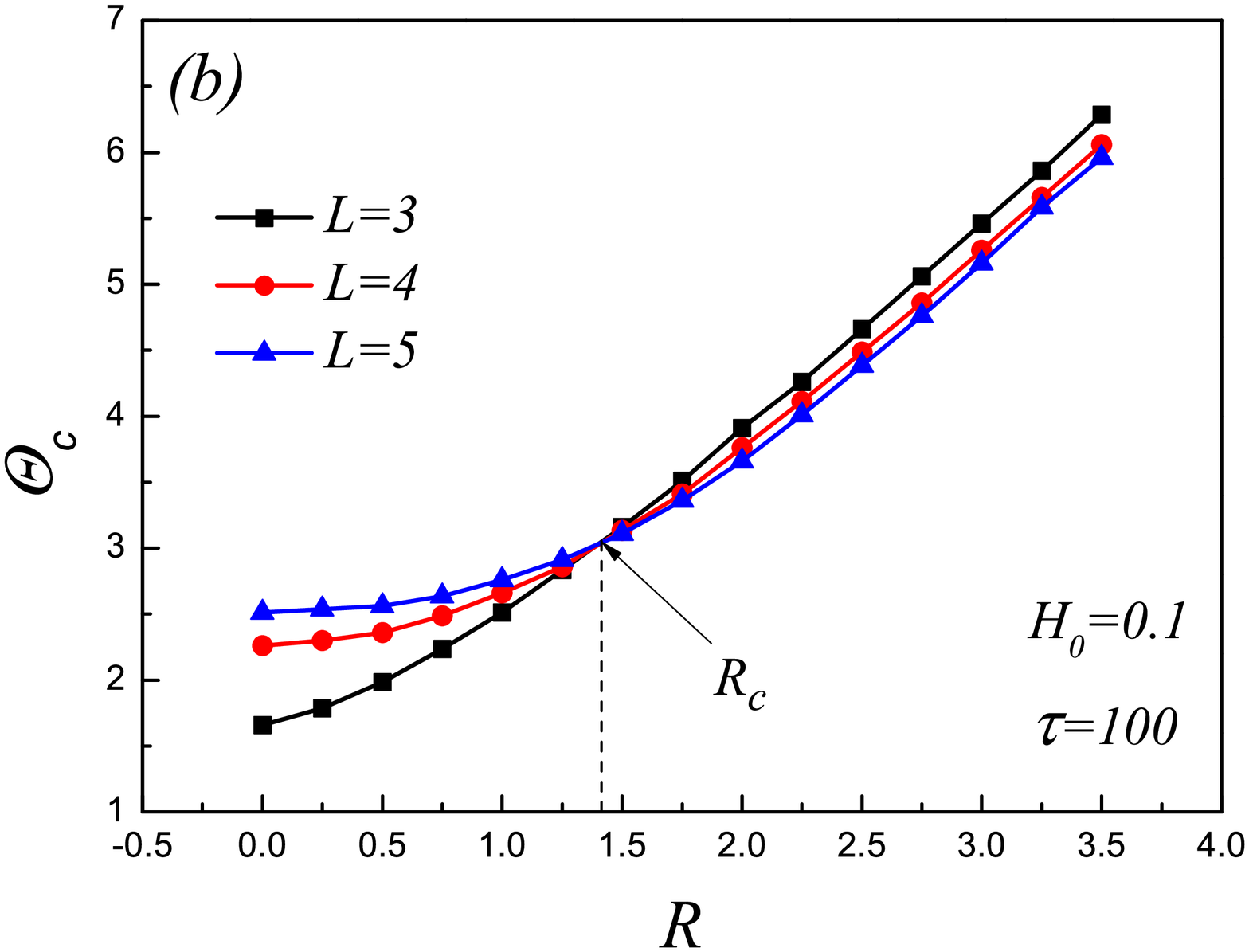}\\
\center
\includegraphics[width=8cm]{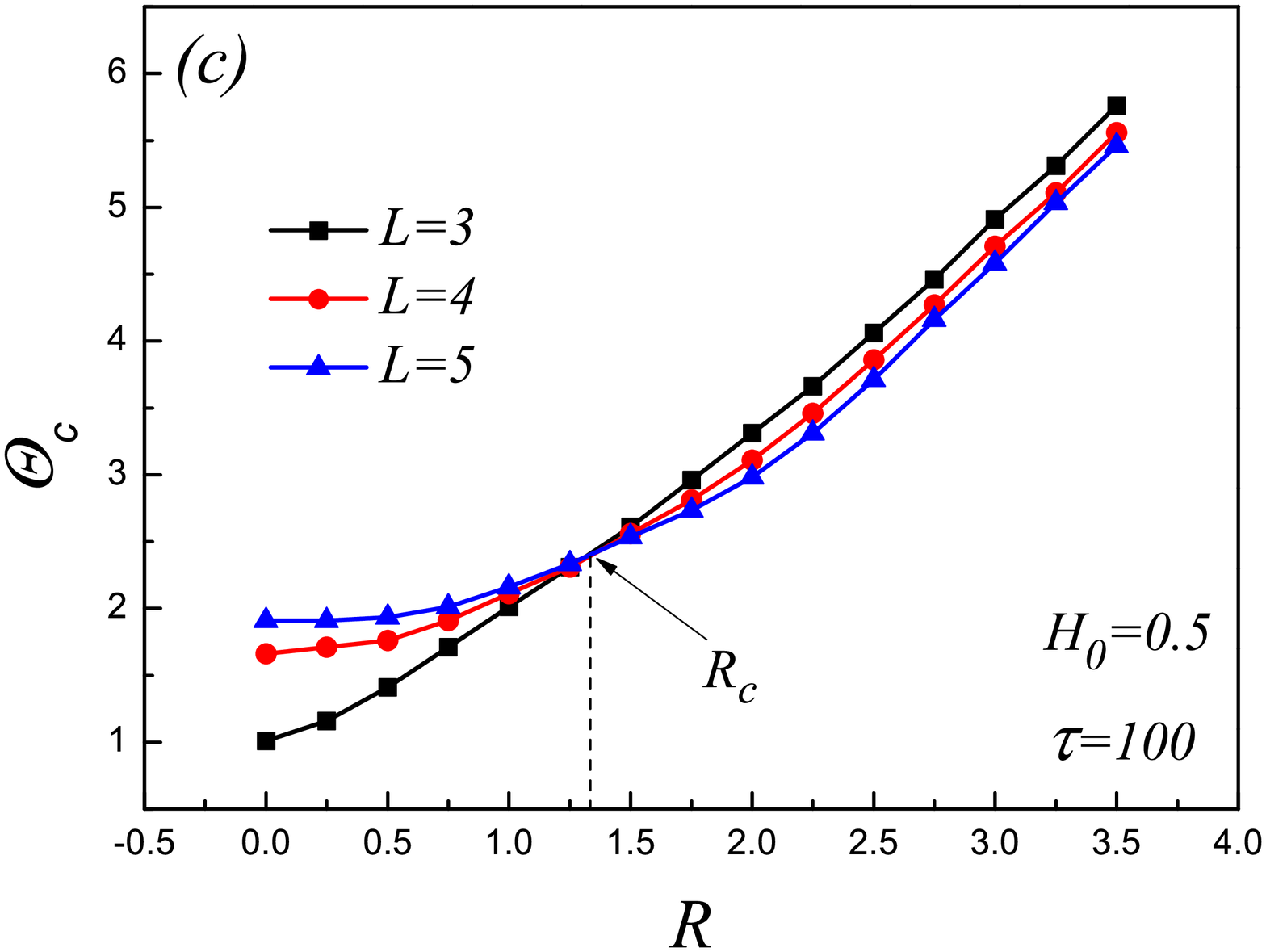}\\
\caption{}\label{fig1}
\end{figure}

\begin{figure}
\center
\includegraphics[width=8.5cm]{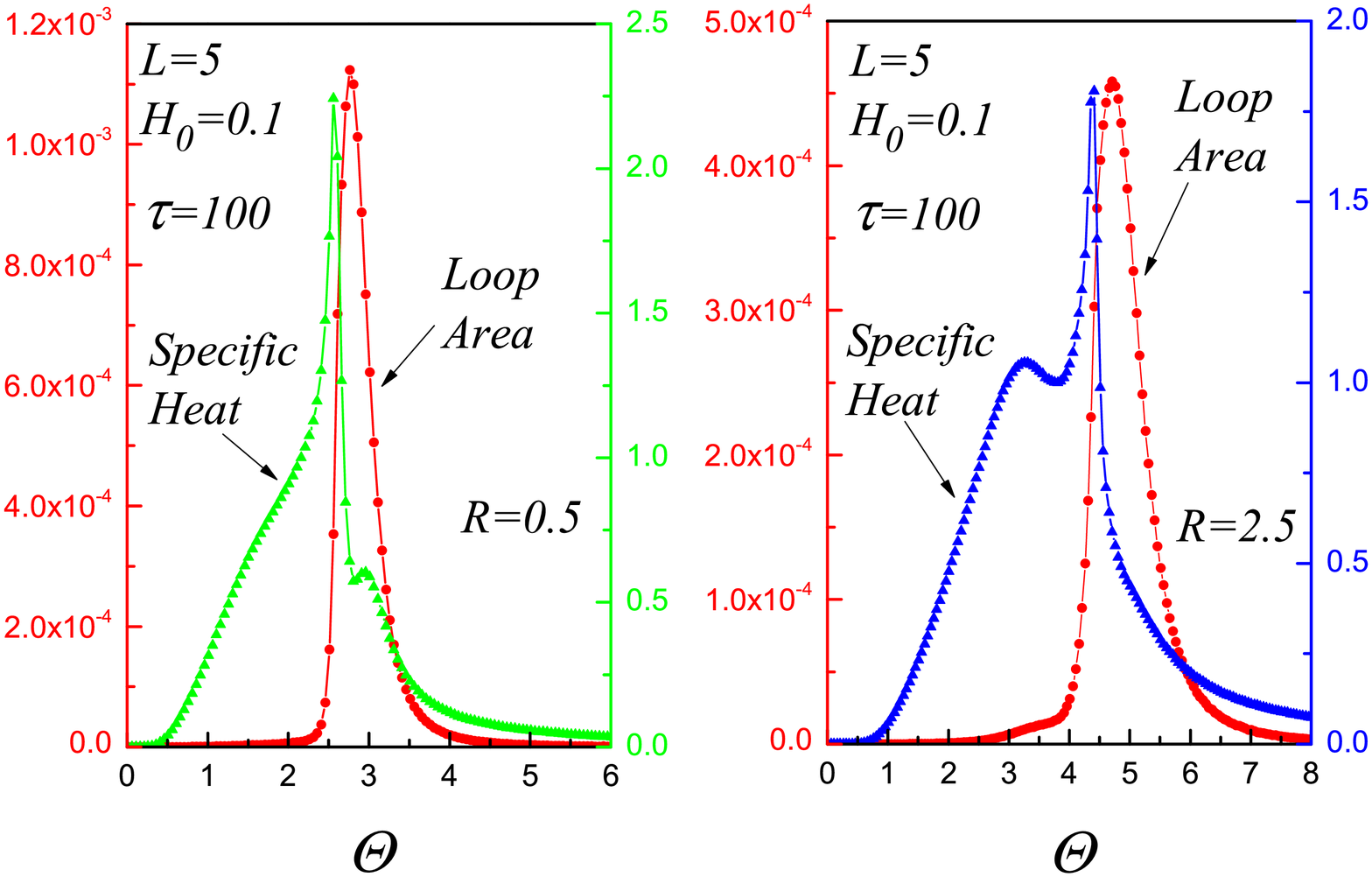}
\includegraphics[width=7cm]{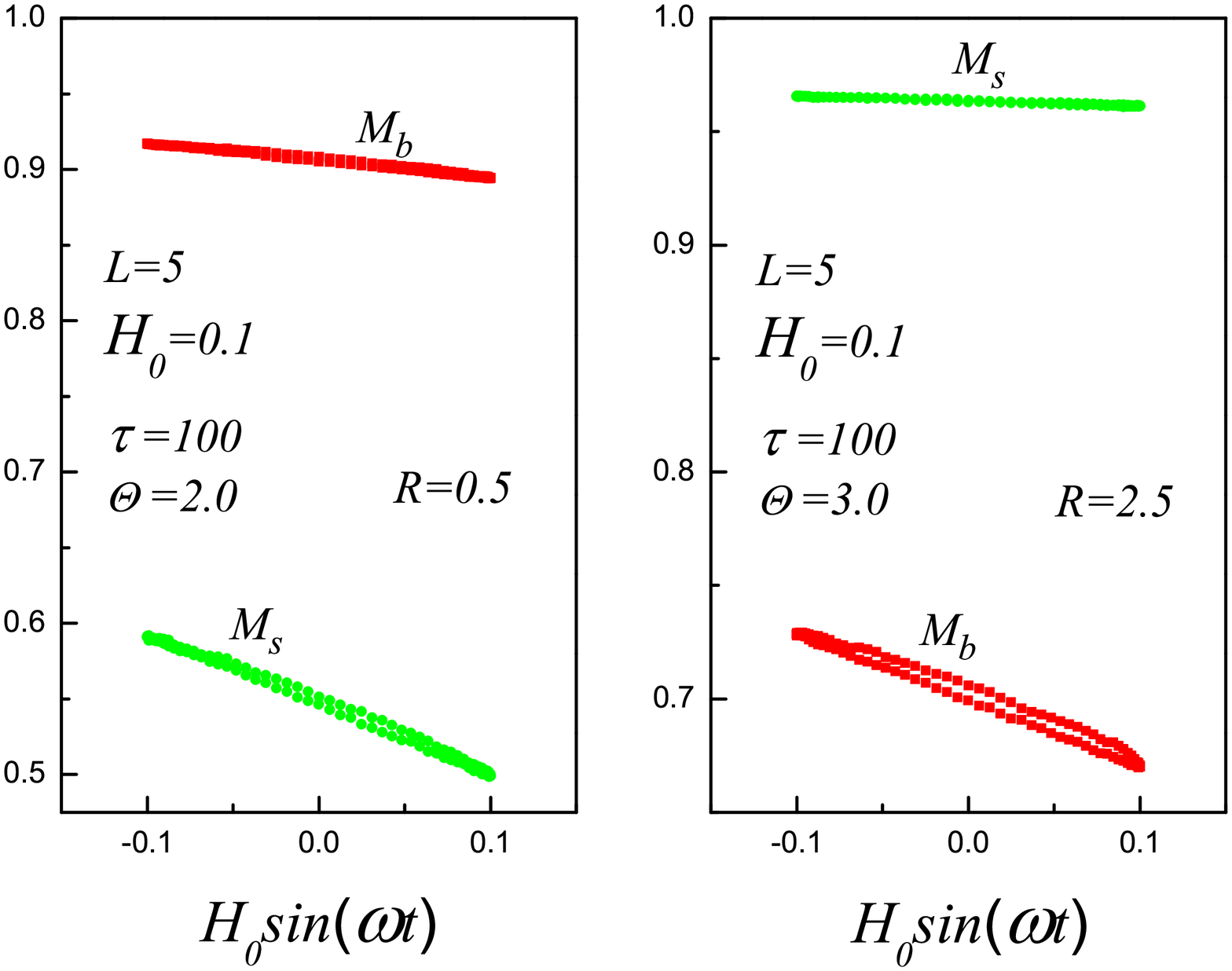}\\
\caption{}\label{fig2}
\end{figure}

\begin{figure}
\center
\includegraphics[width=8cm]{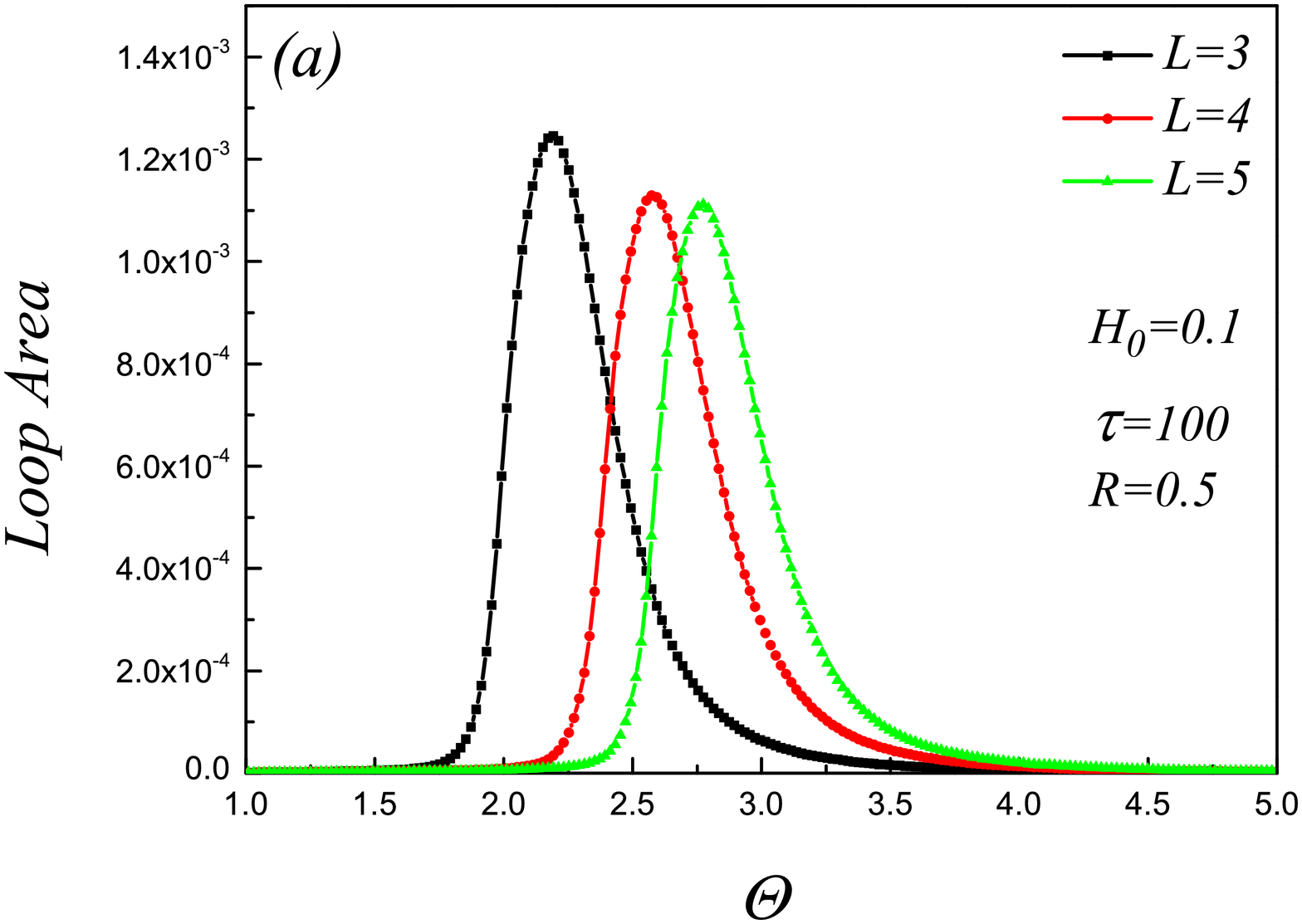}
\includegraphics[width=8cm]{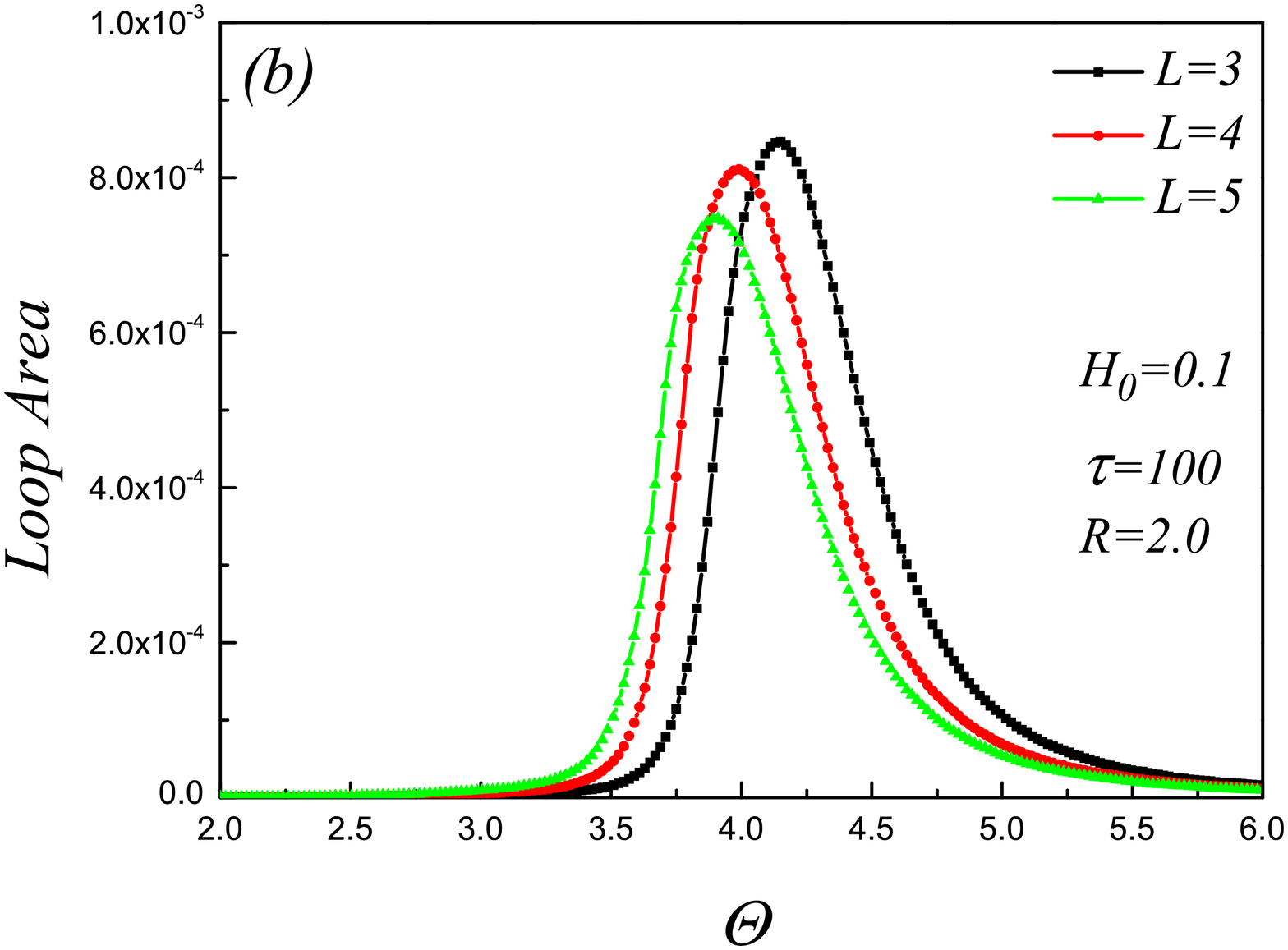}\\
\includegraphics[width=8cm]{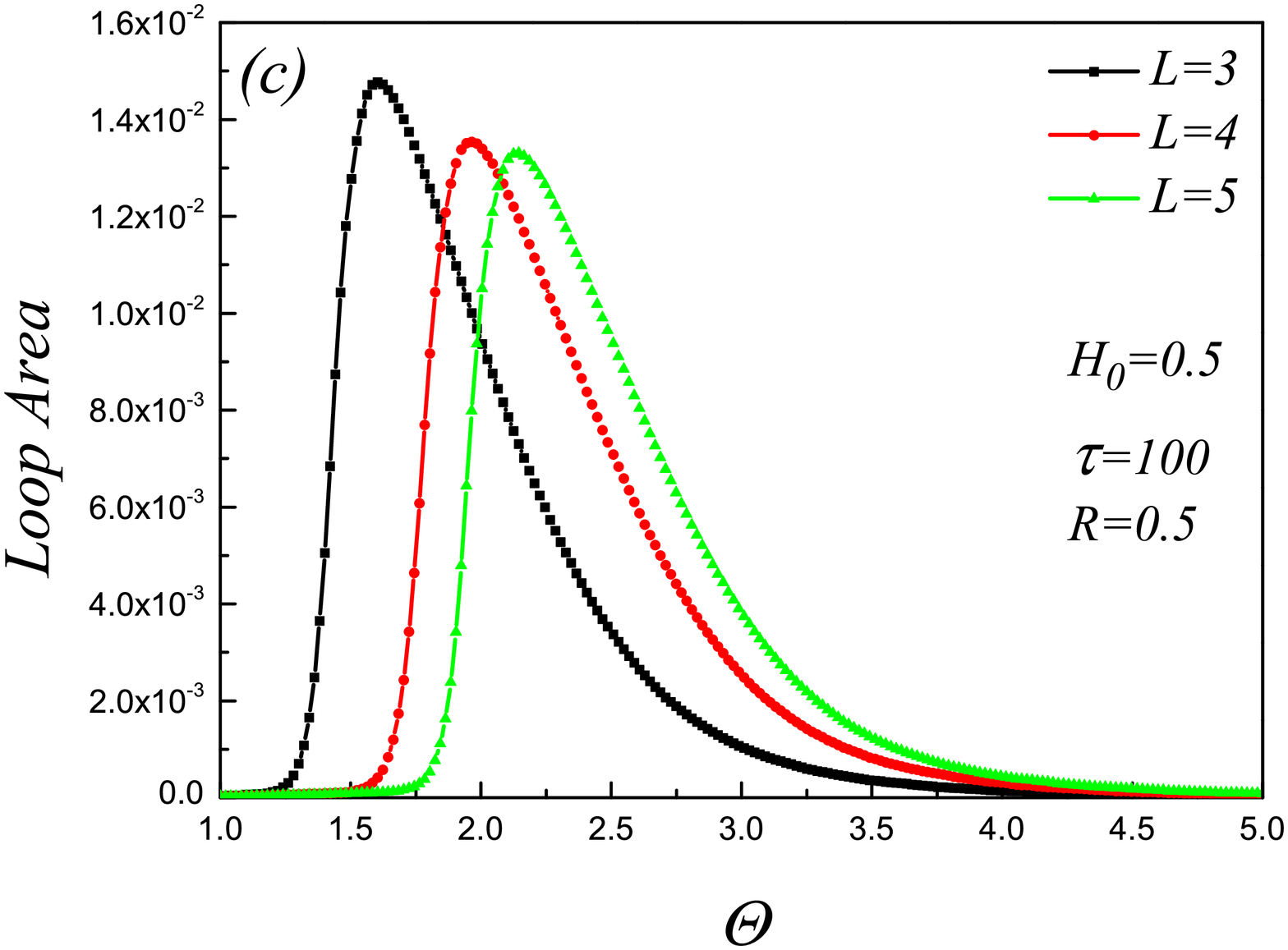}
\includegraphics[width=8cm]{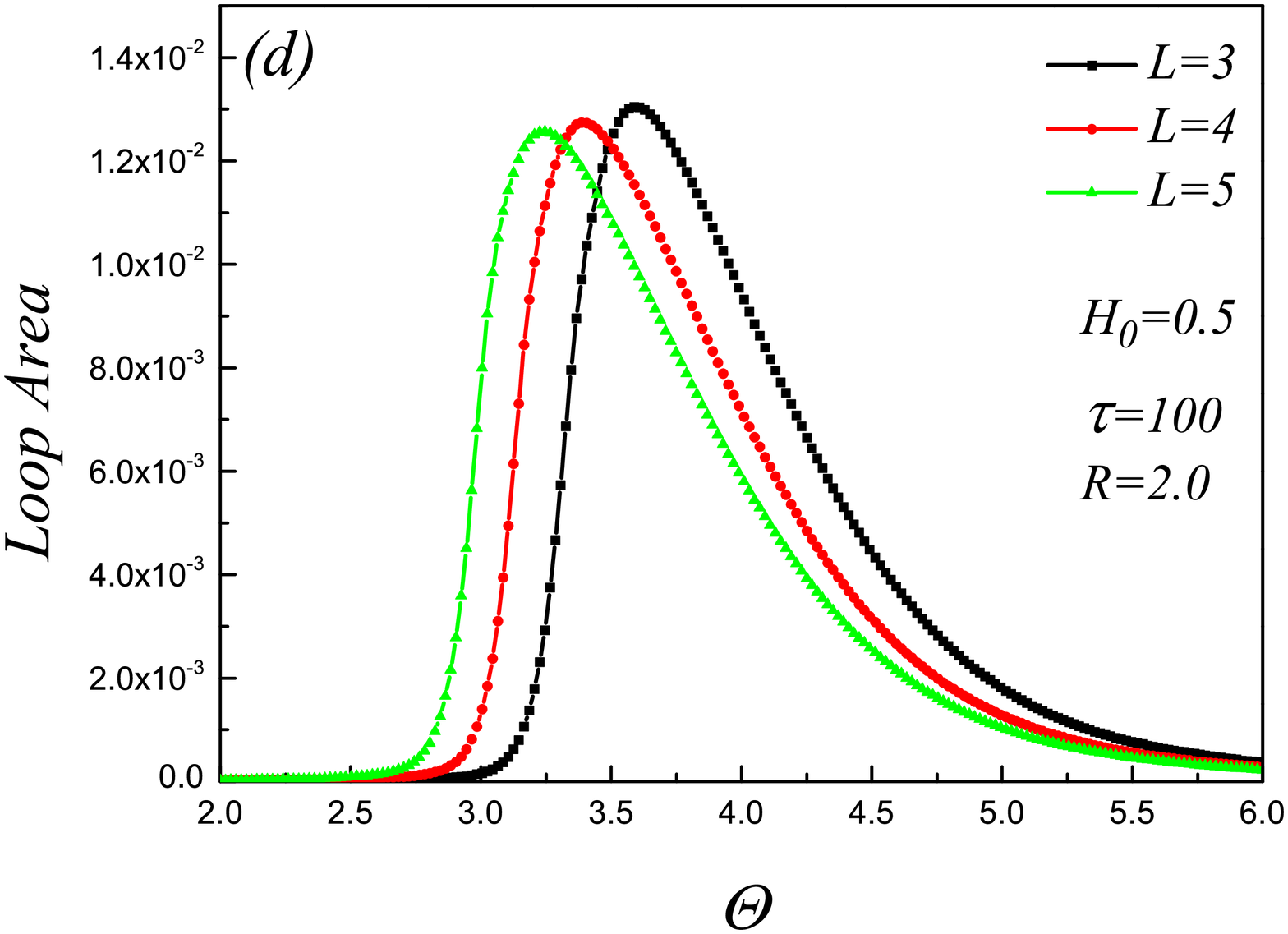}\\
\caption{ }\label{fig3}
\end{figure}

\begin{figure}
\center
\includegraphics[width=8cm]{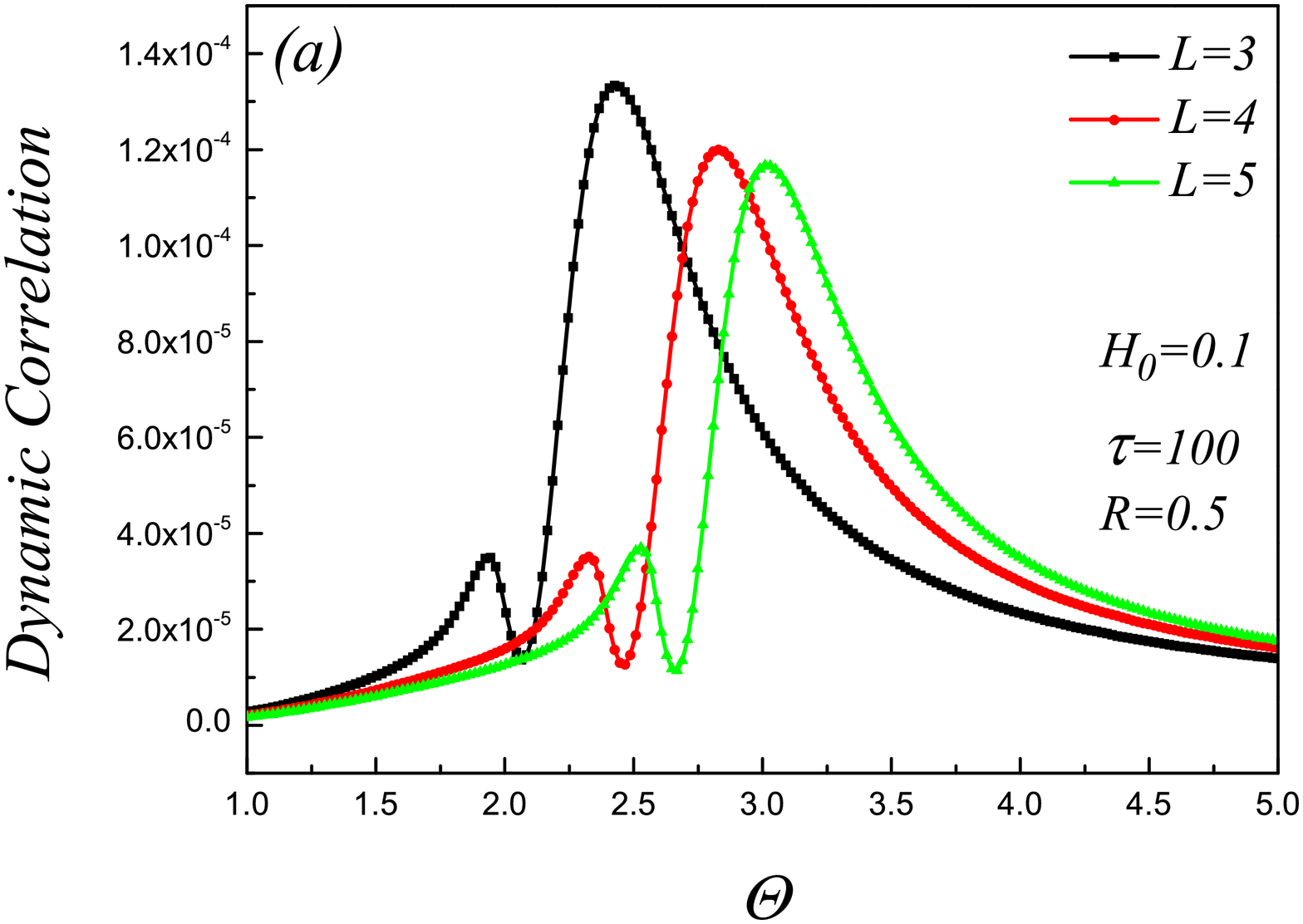}
\includegraphics[width=8cm]{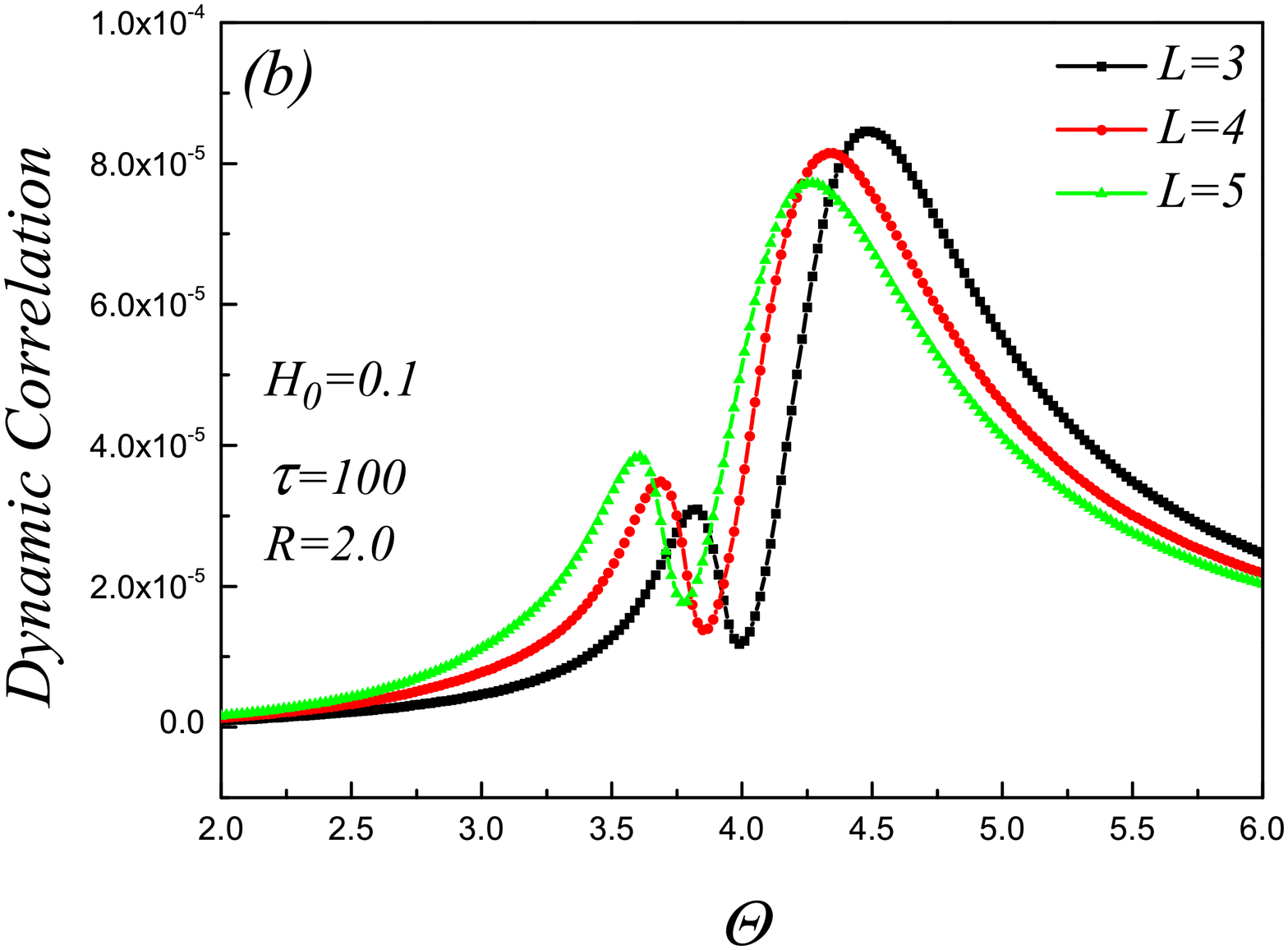}\\
\includegraphics[width=8cm]{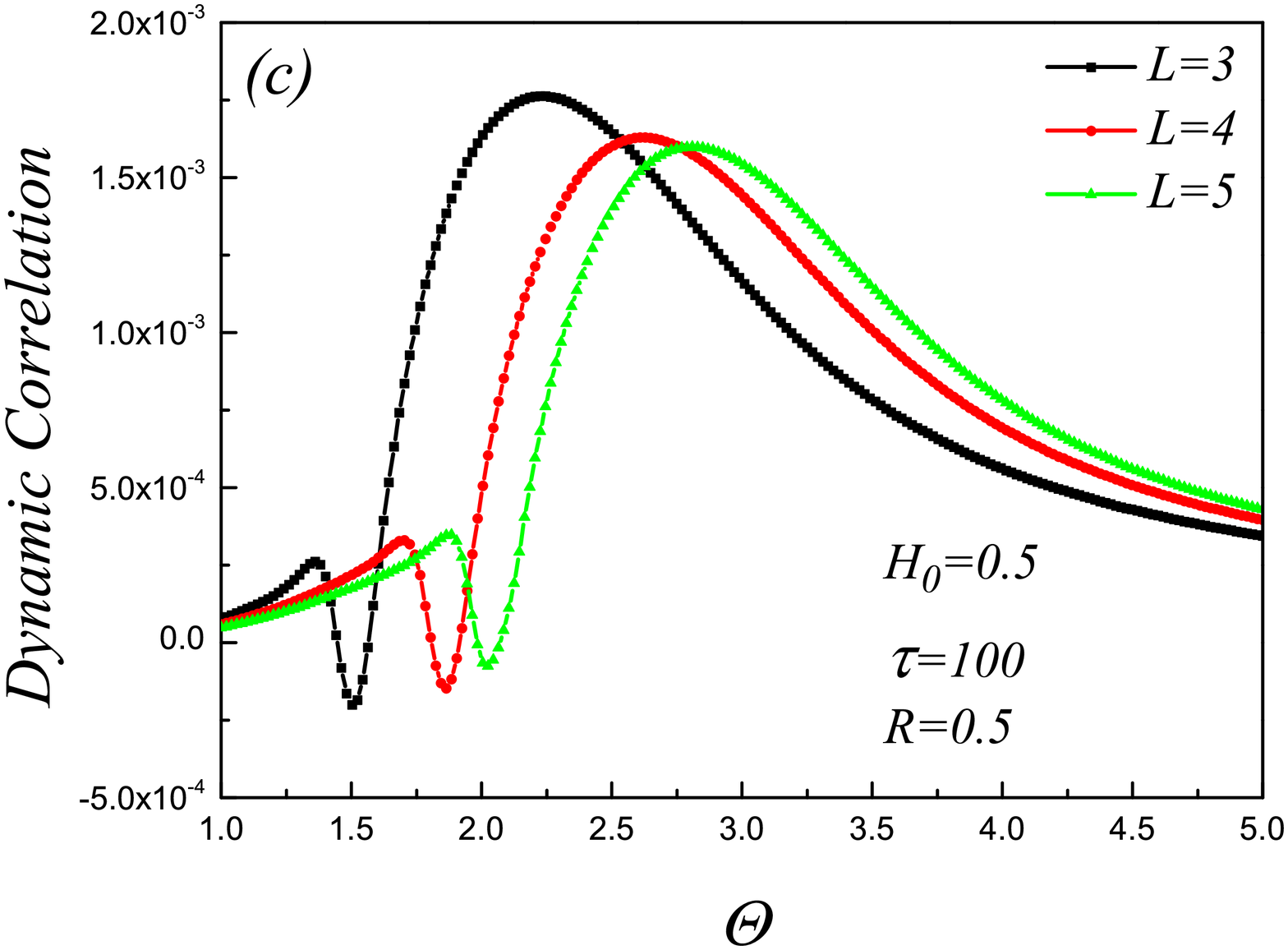}
\includegraphics[width=8cm]{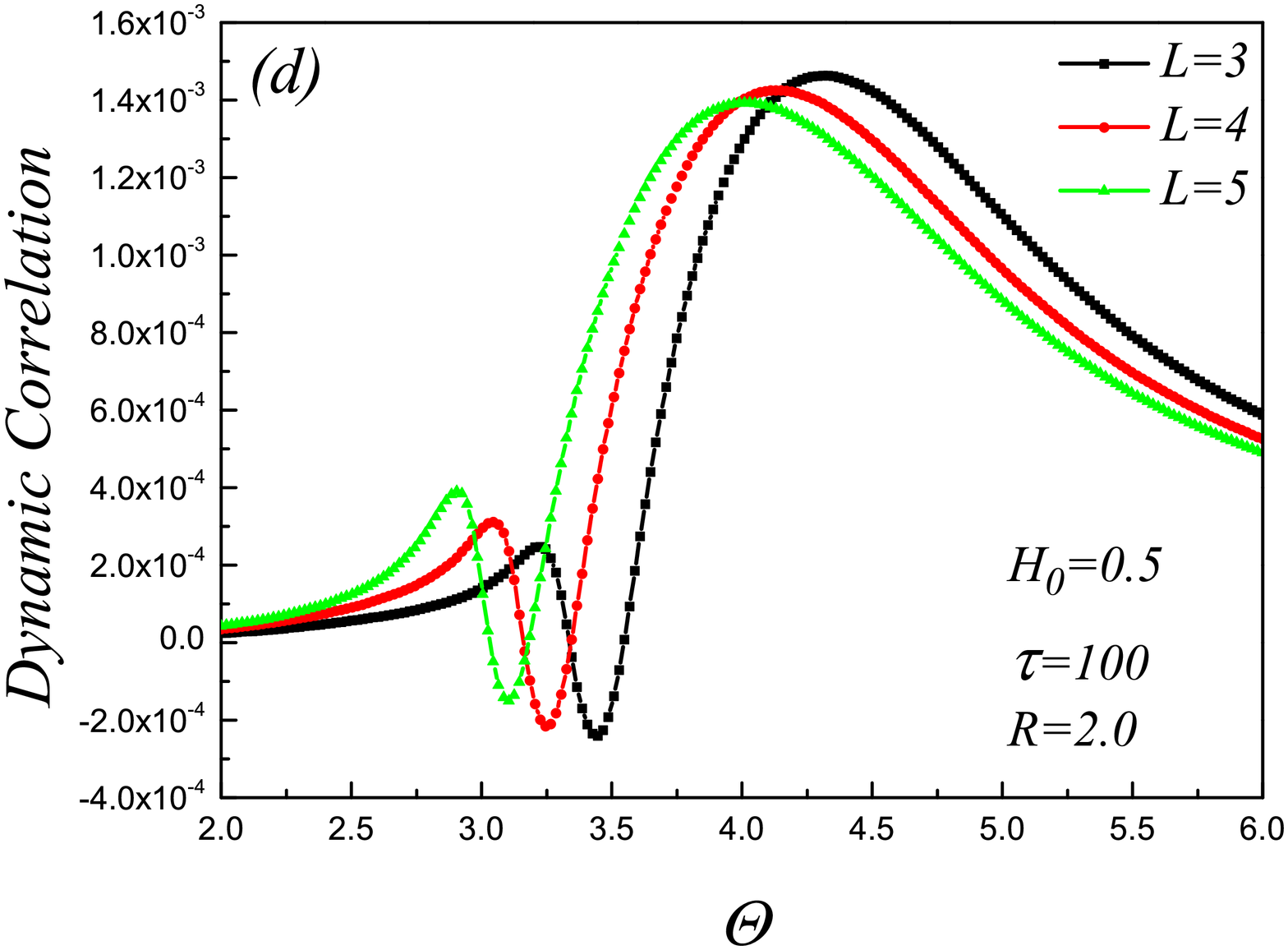}\\
\caption{ }\label{fig4}
\end{figure}

\begin{figure}
\center
\includegraphics[width=4.5cm]{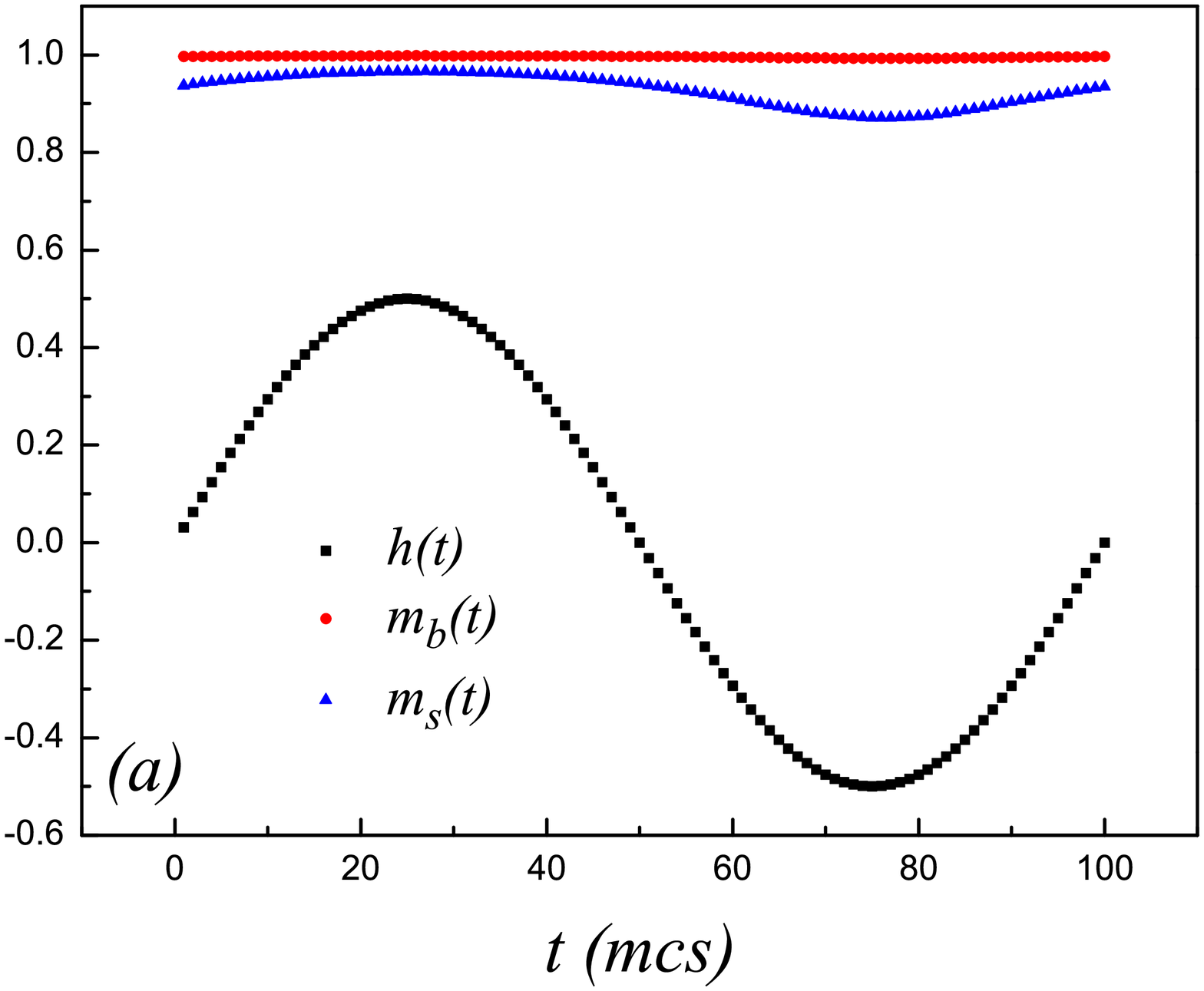}
\includegraphics[width=4.5cm]{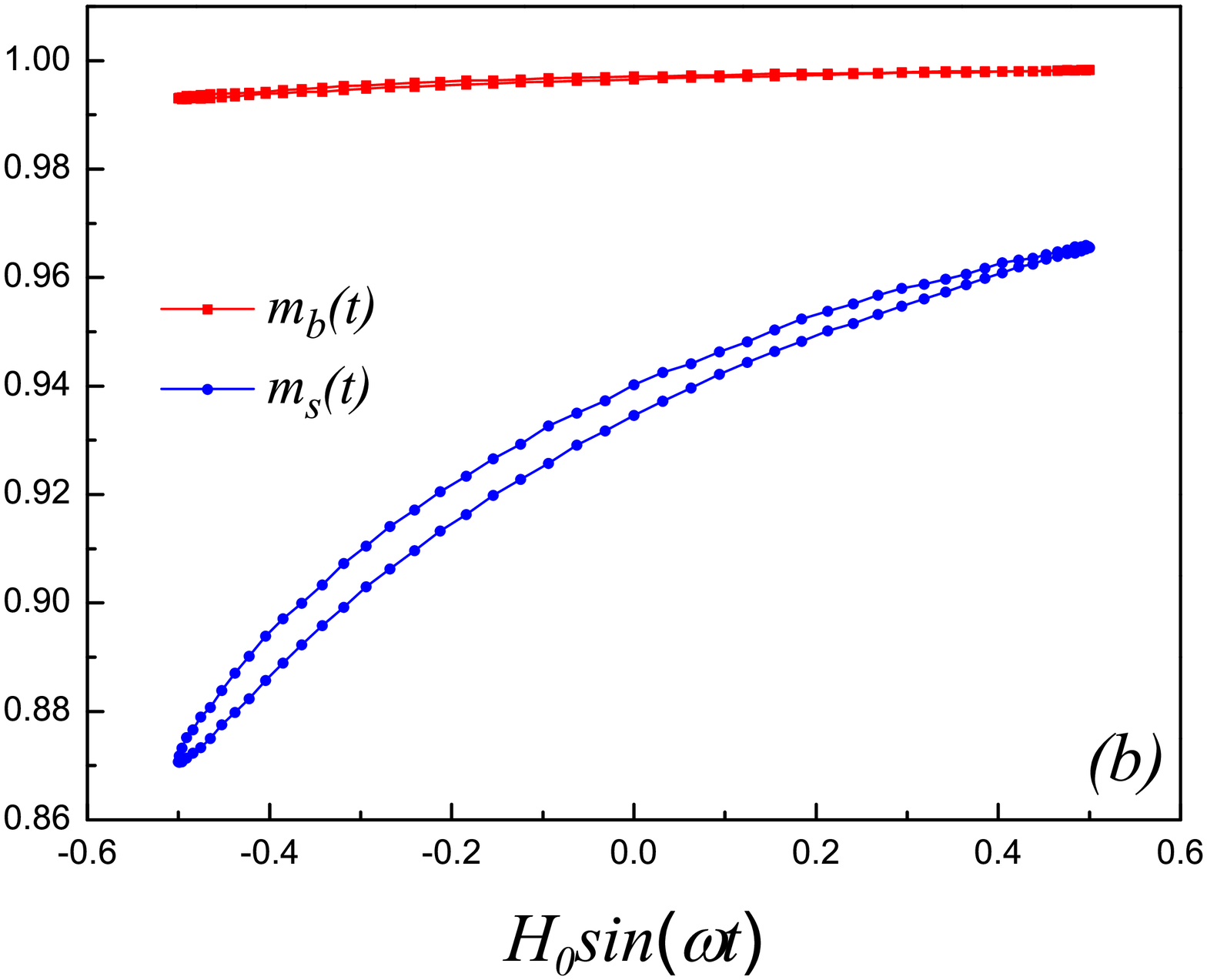}
\includegraphics[width=5.2cm]{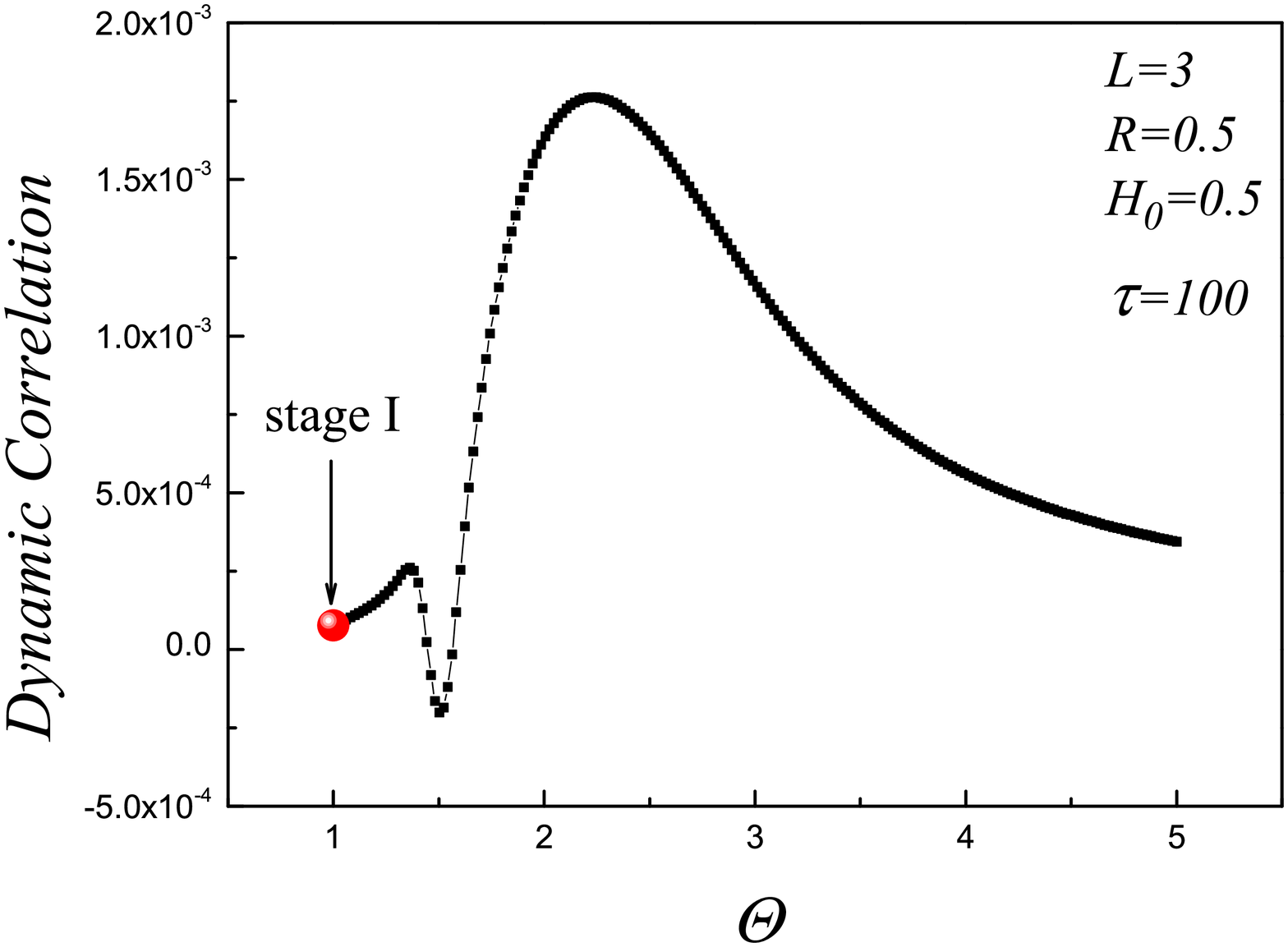}\\
\includegraphics[width=4.5cm]{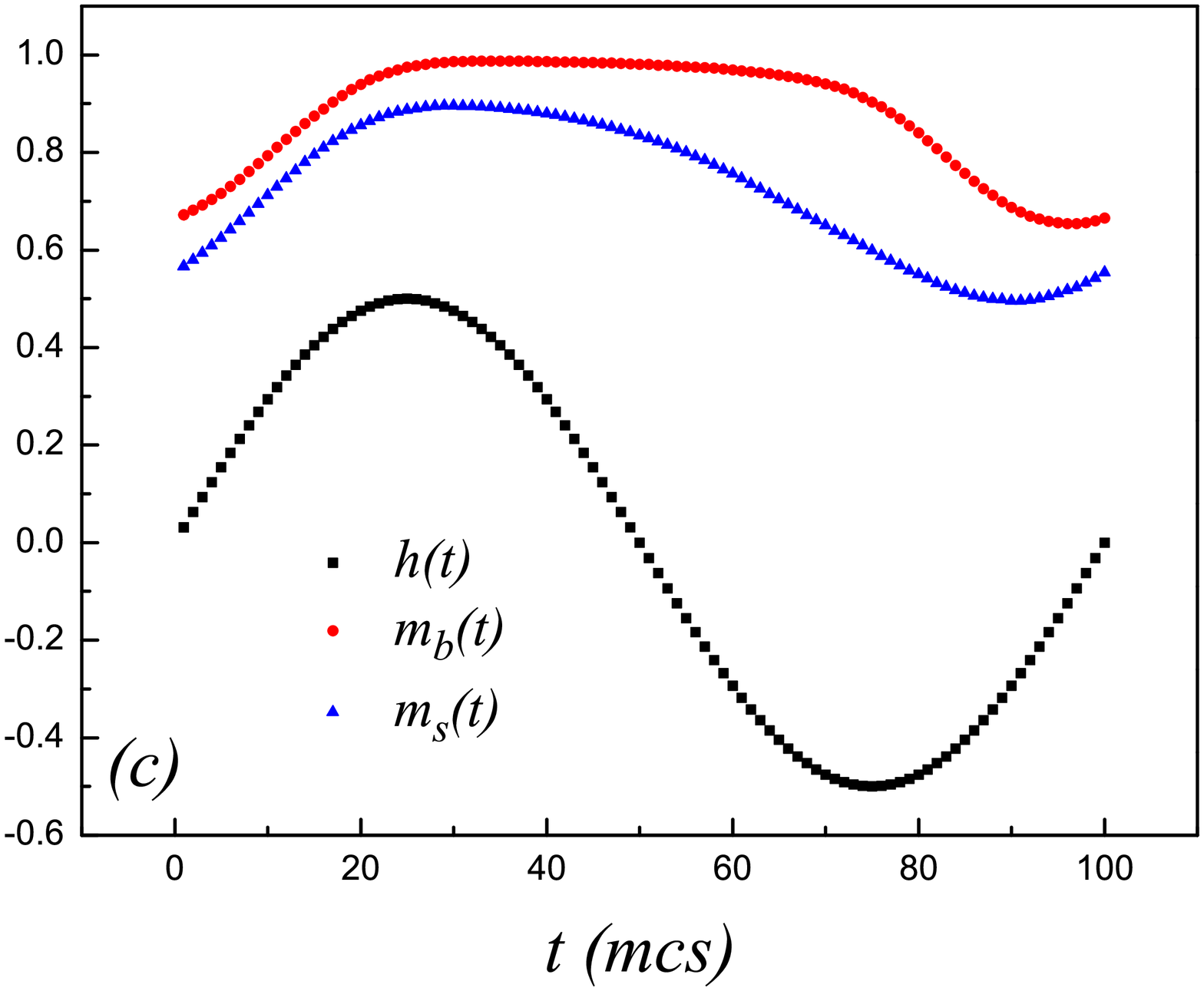}
\includegraphics[width=4.5cm]{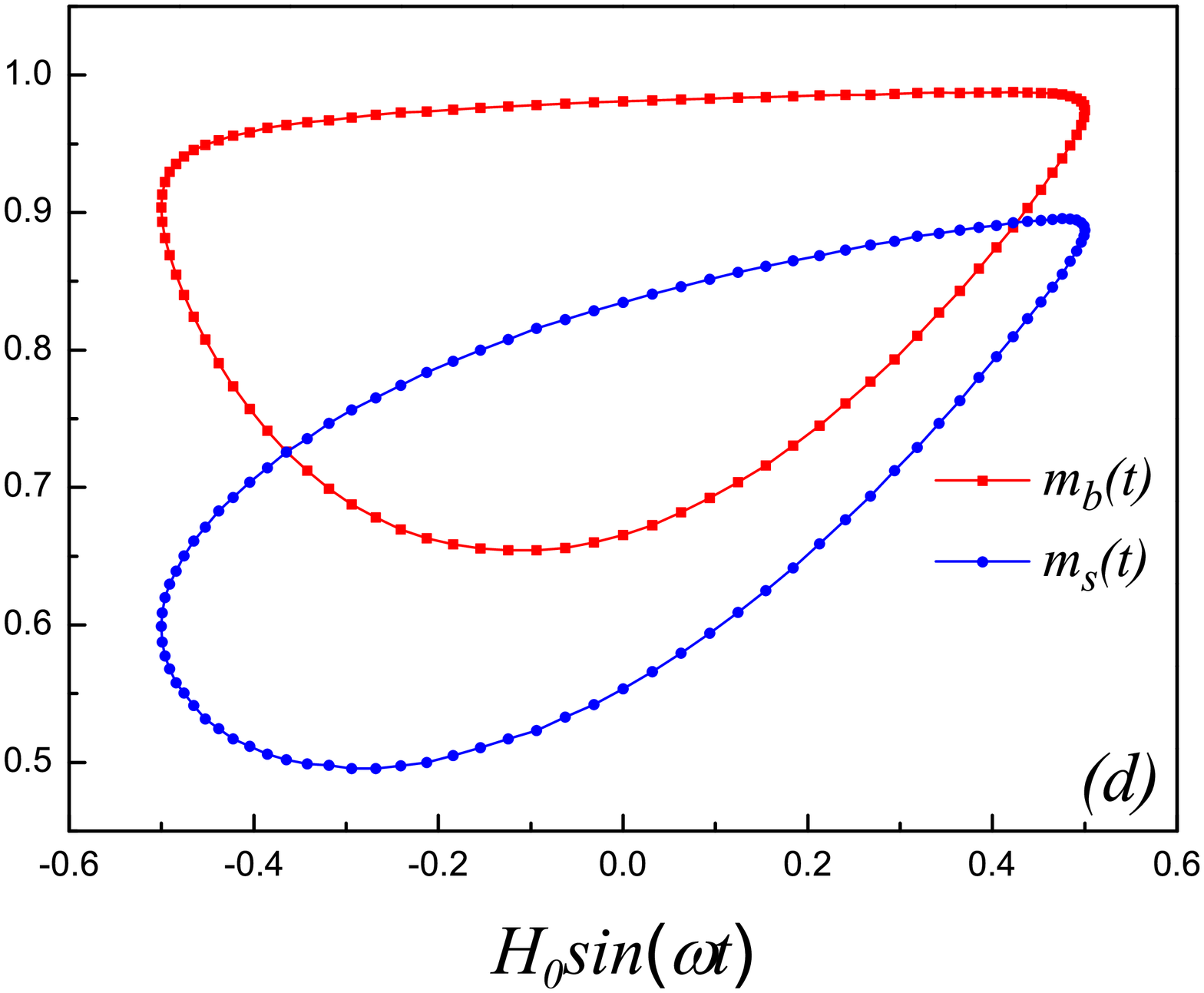}
\includegraphics[width=5.2cm]{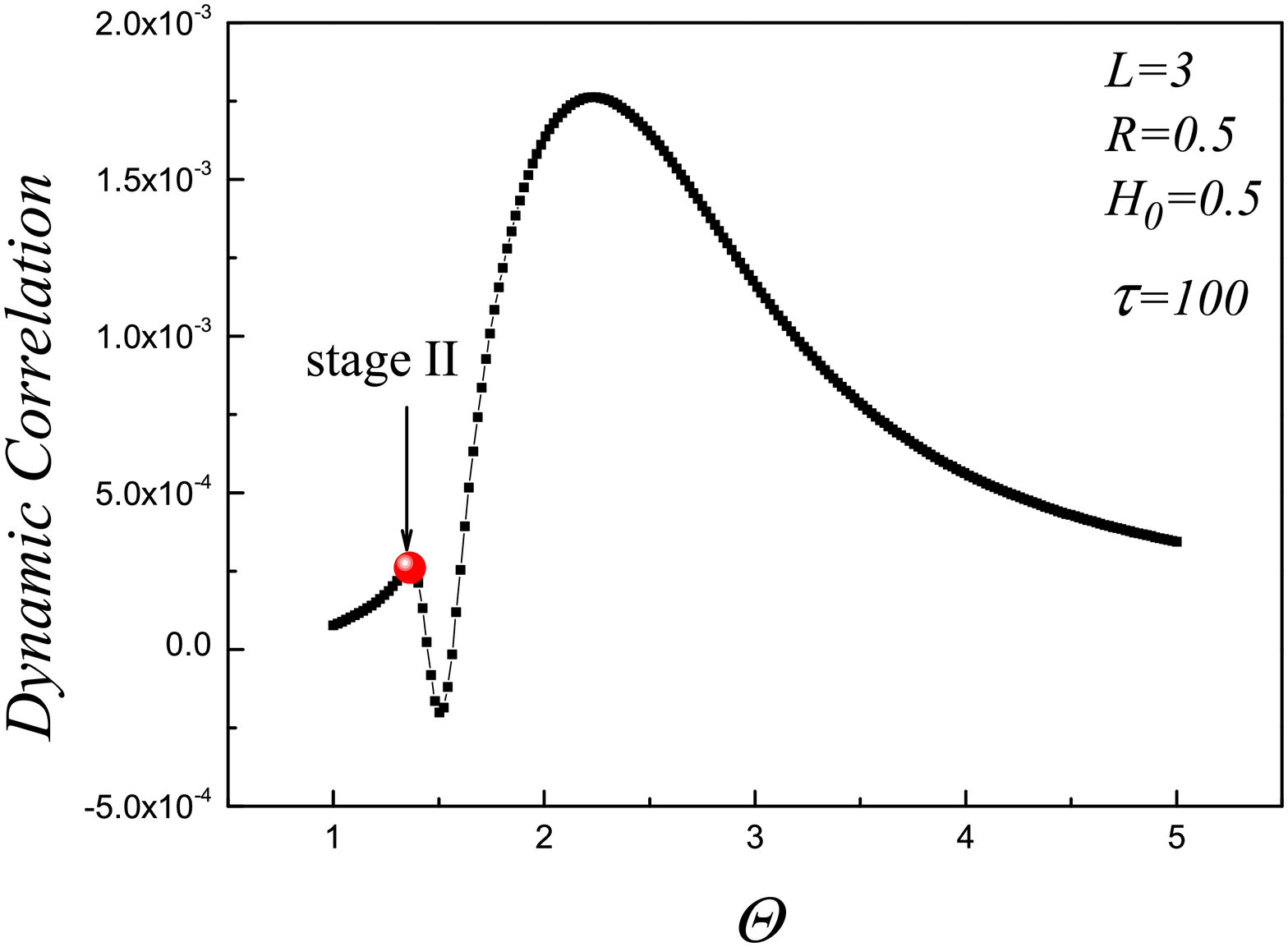}\\
\includegraphics[width=4.5cm]{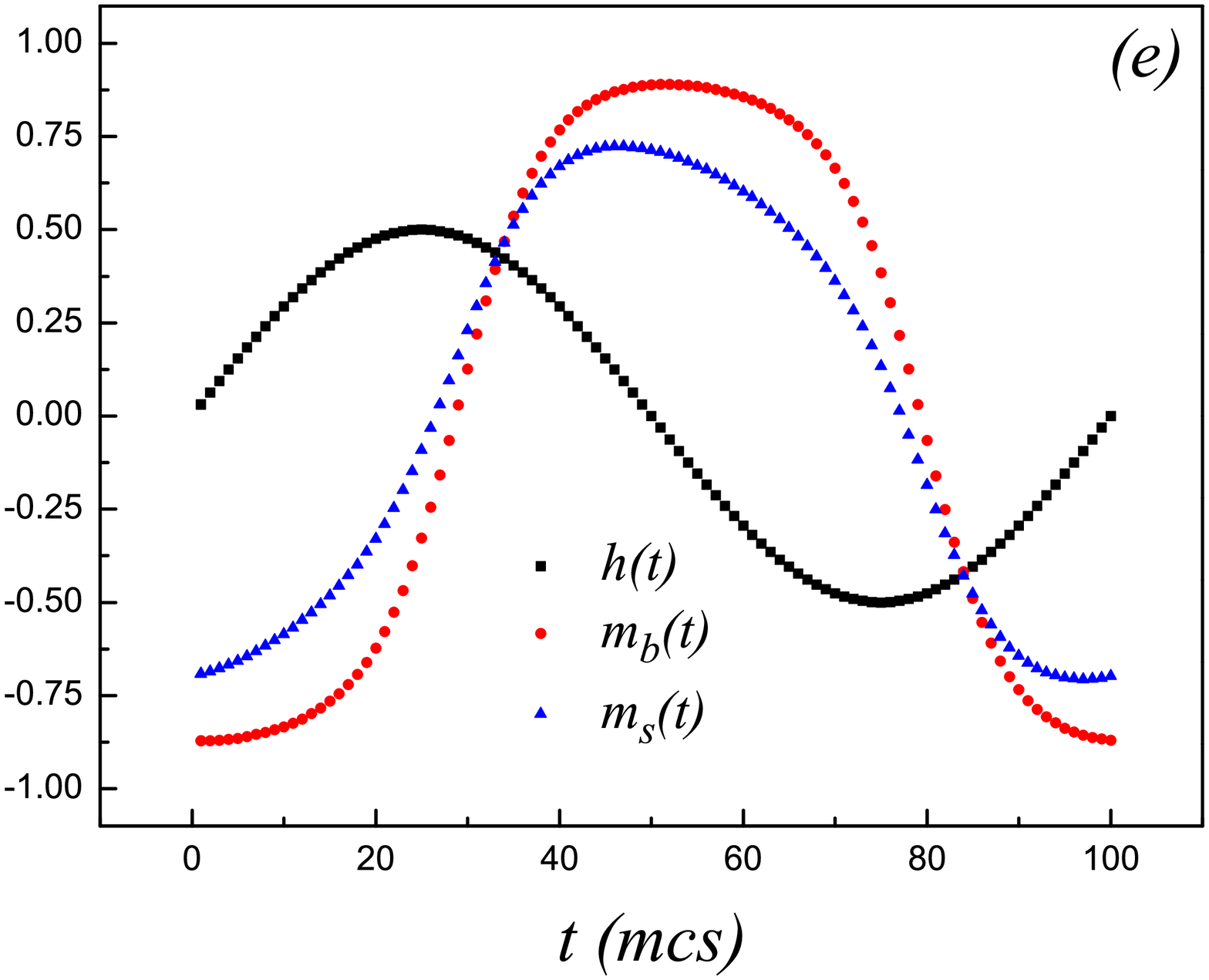}
\includegraphics[width=4.5cm]{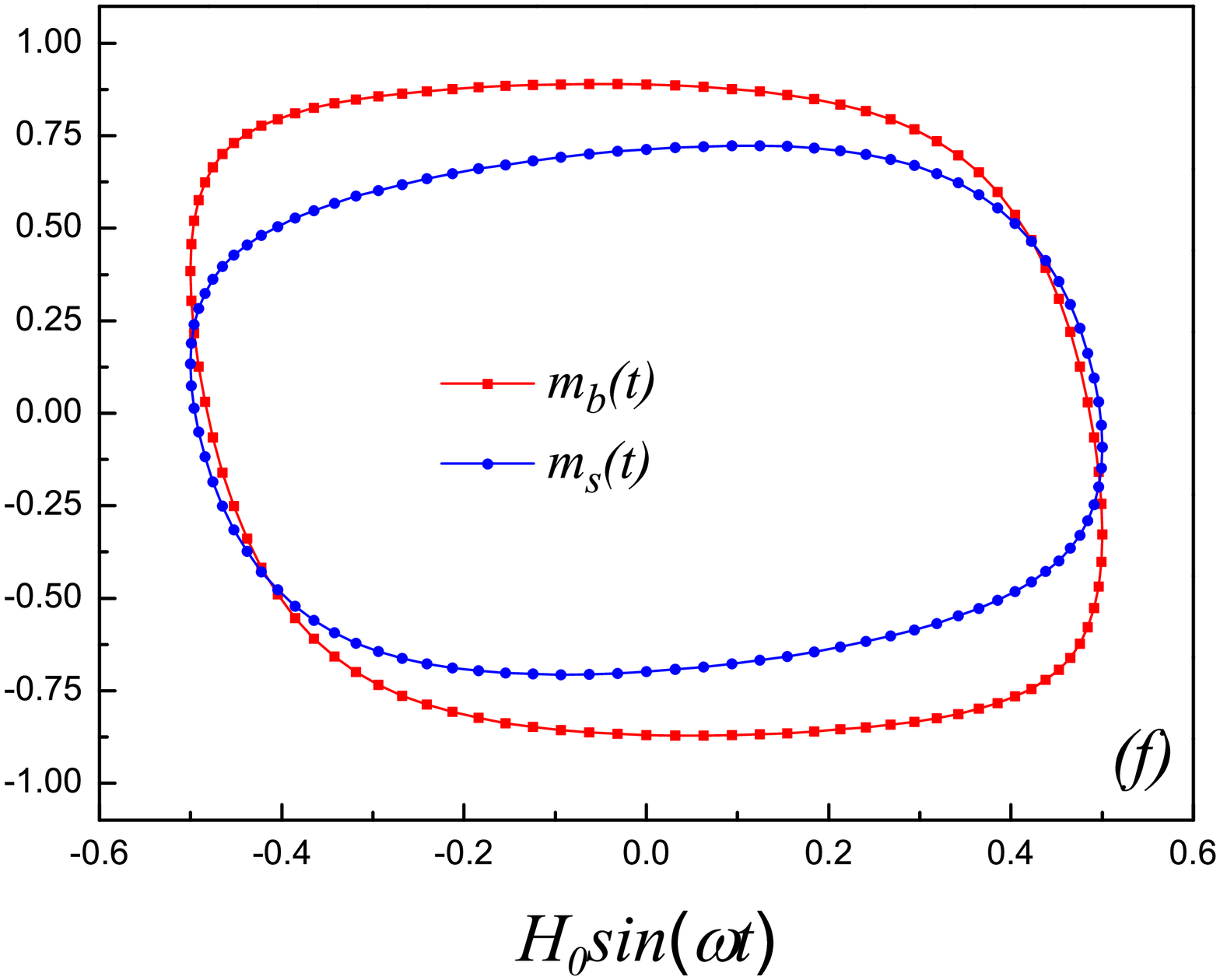}
\includegraphics[width=5.2cm]{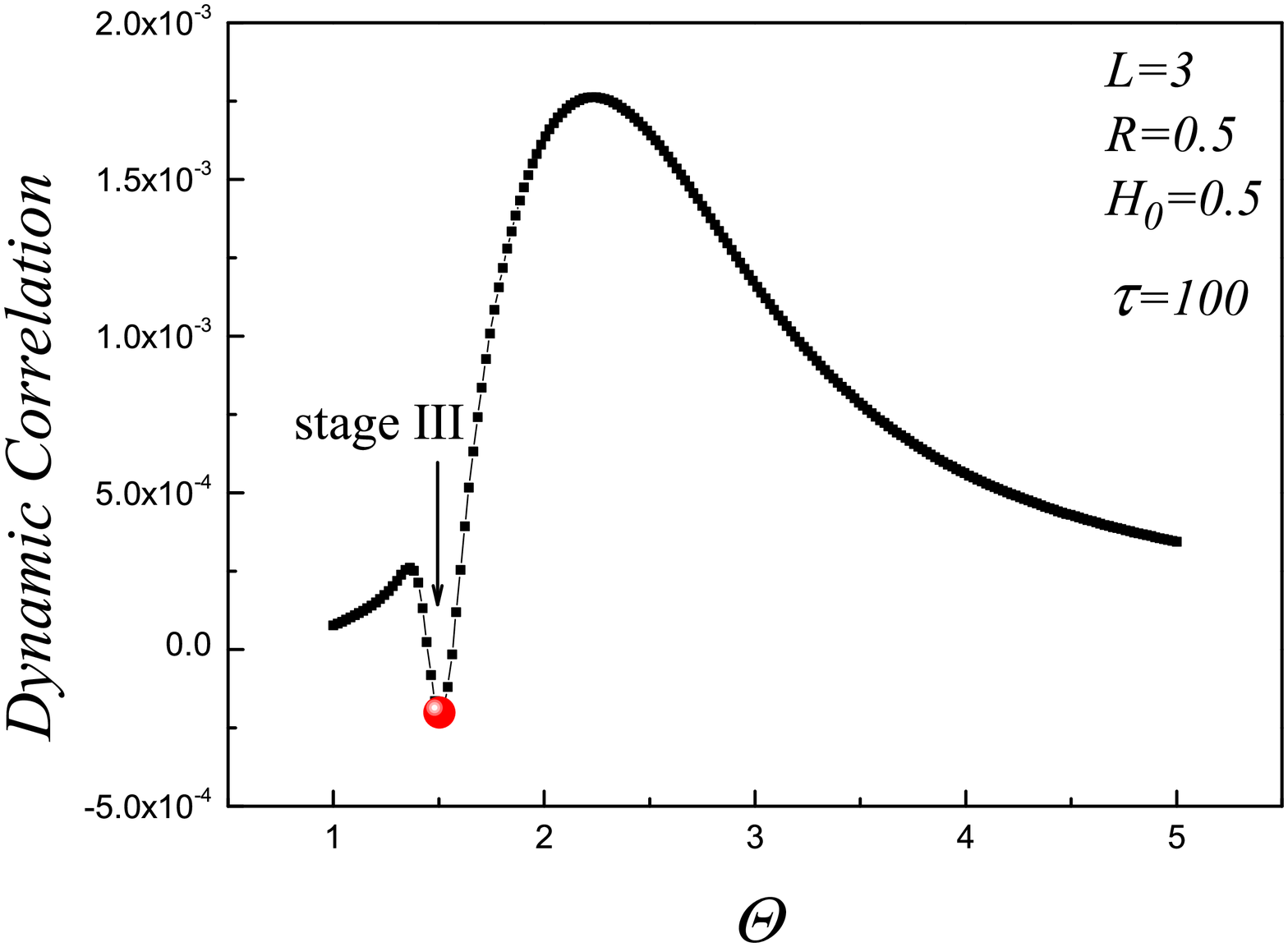}\\
\includegraphics[width=4.5cm]{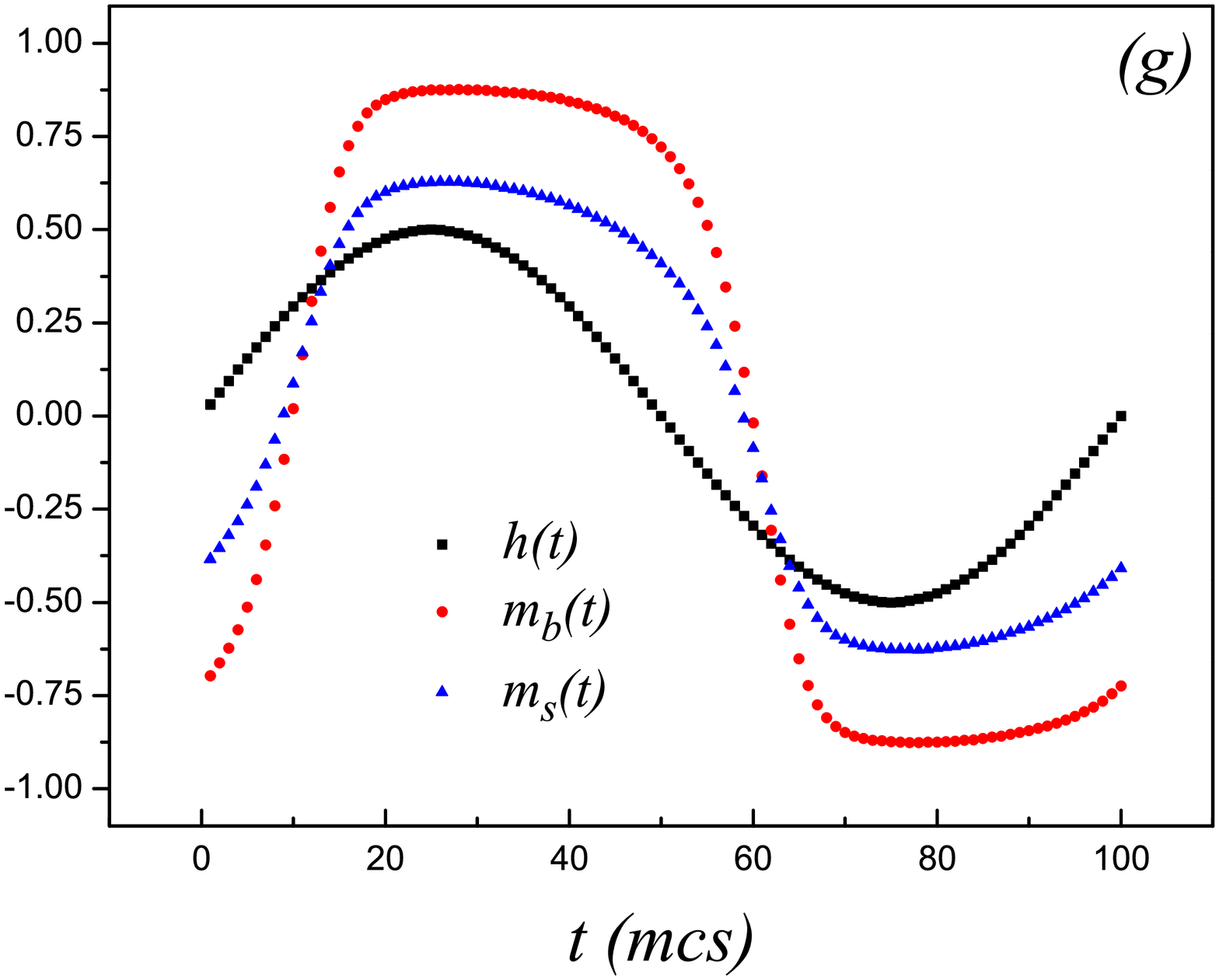}
\includegraphics[width=4.5cm]{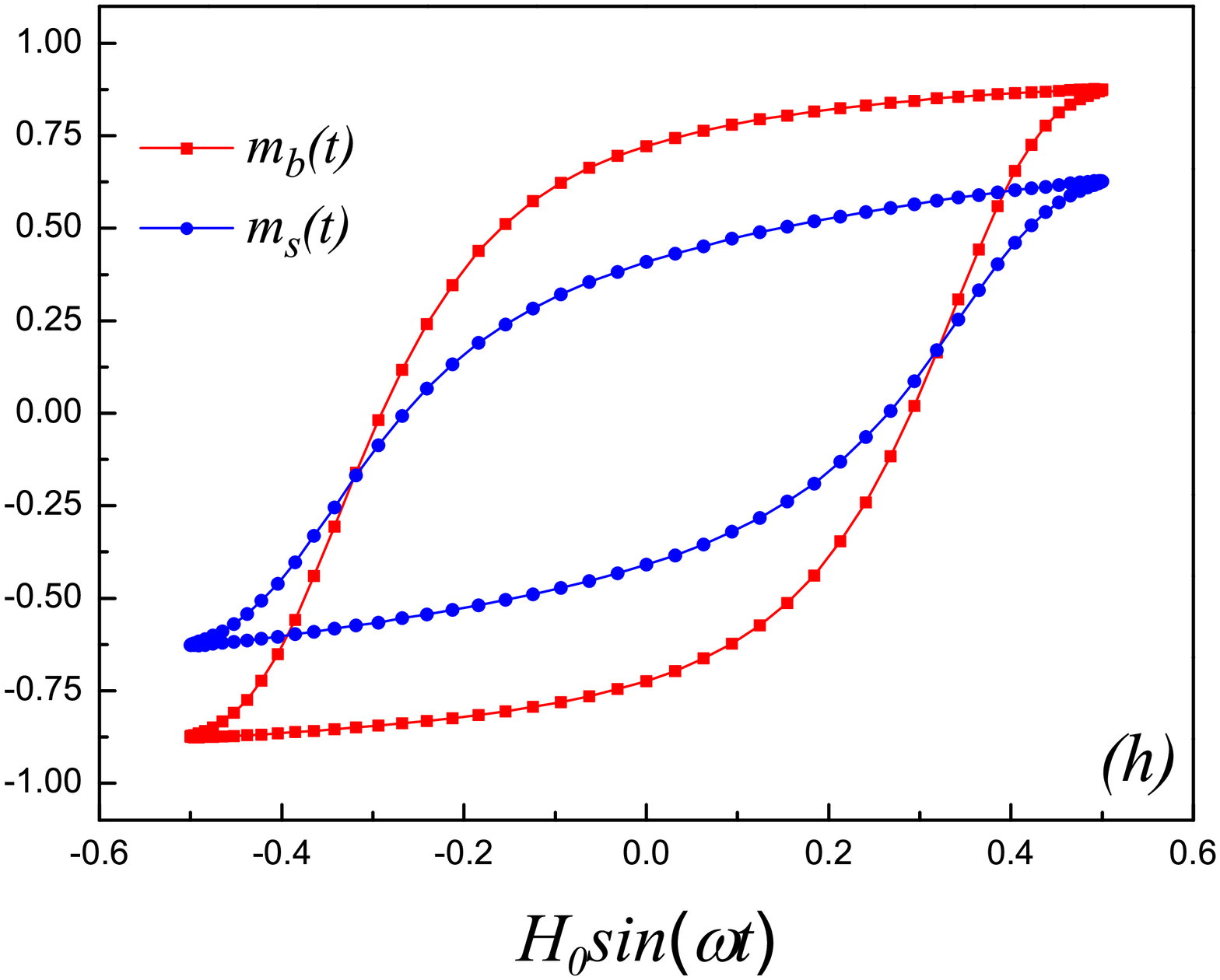}
\includegraphics[width=5.2cm]{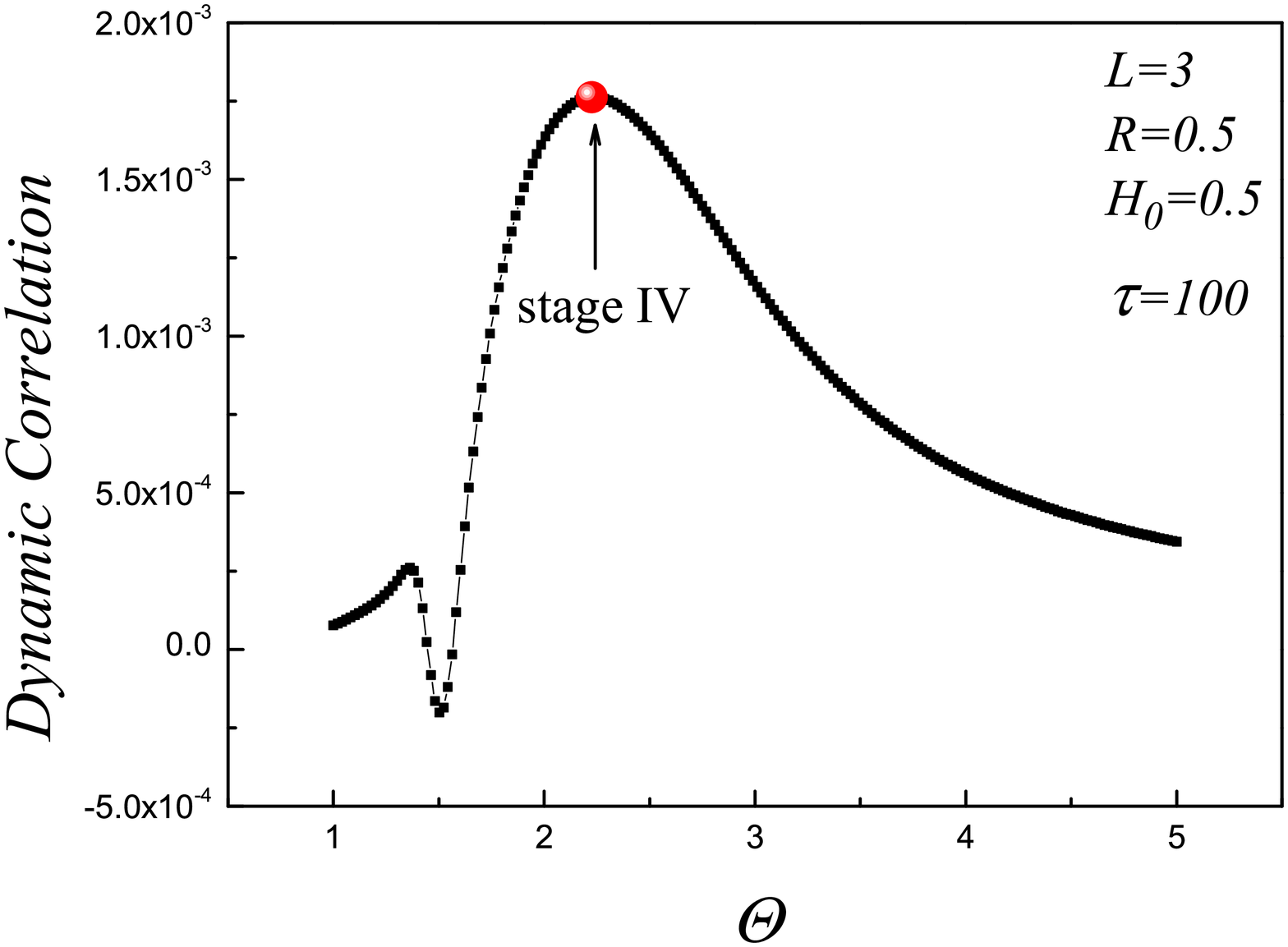}\\
\includegraphics[width=4.5cm]{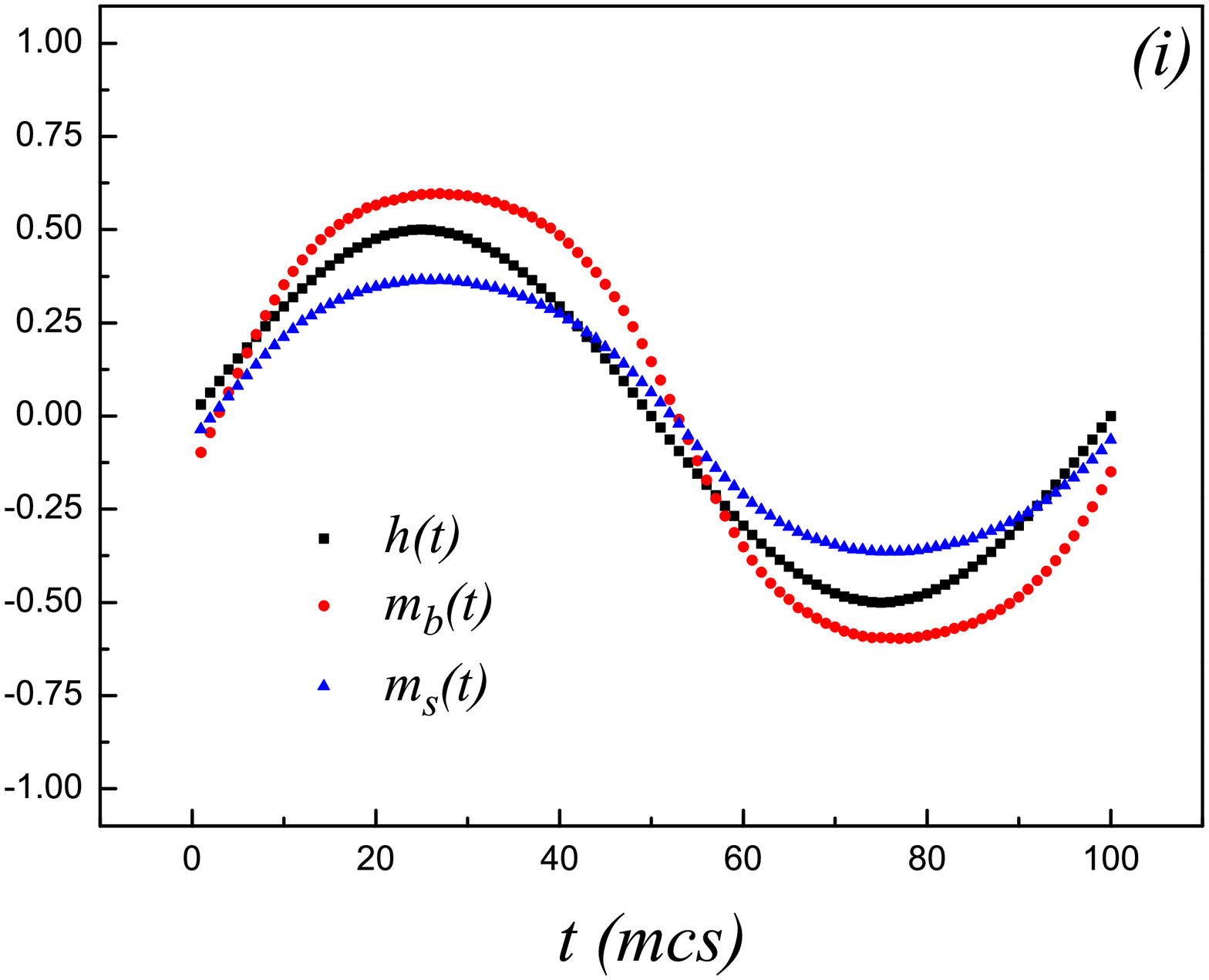}
\includegraphics[width=4.5cm]{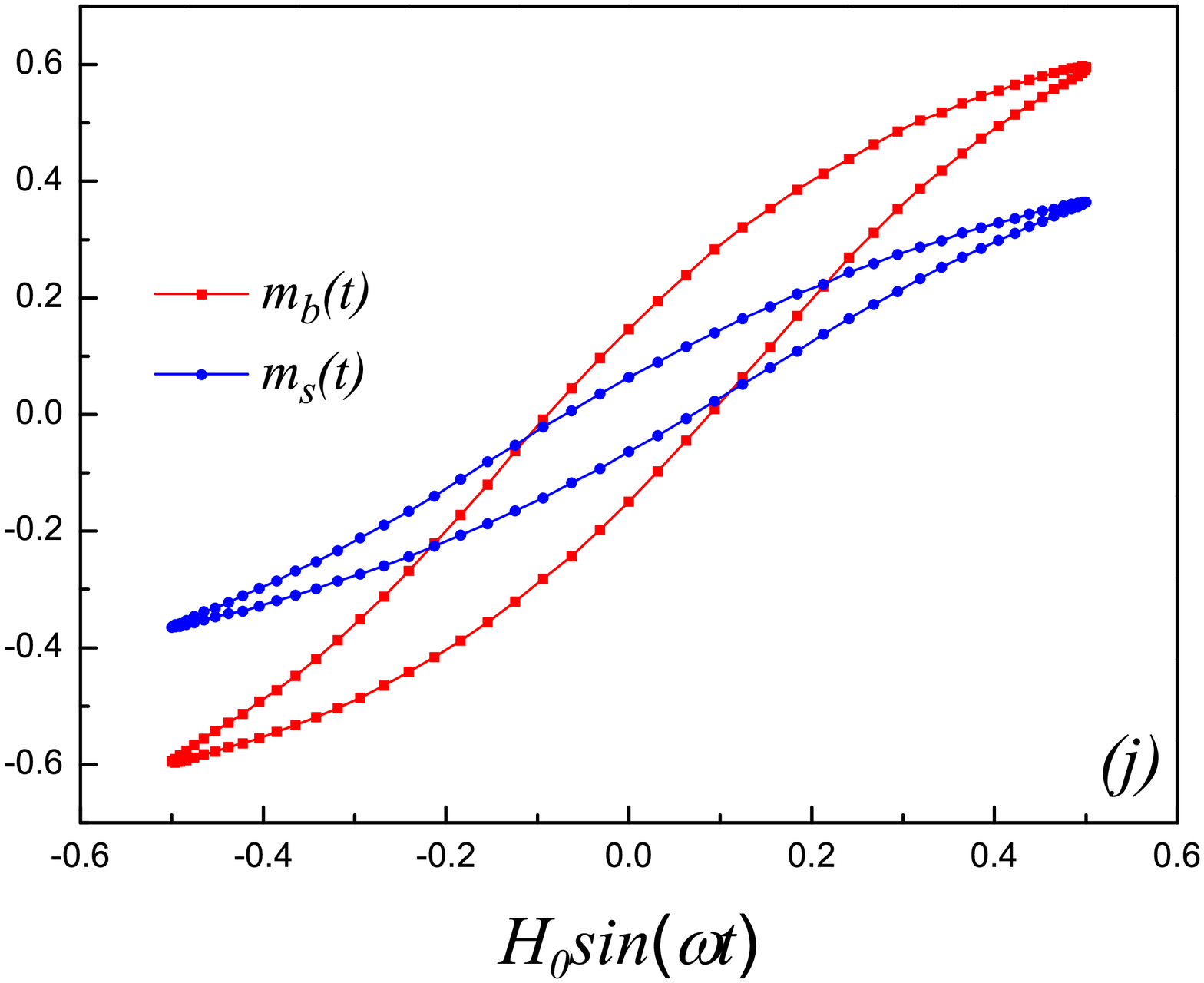}
\includegraphics[width=5.2cm]{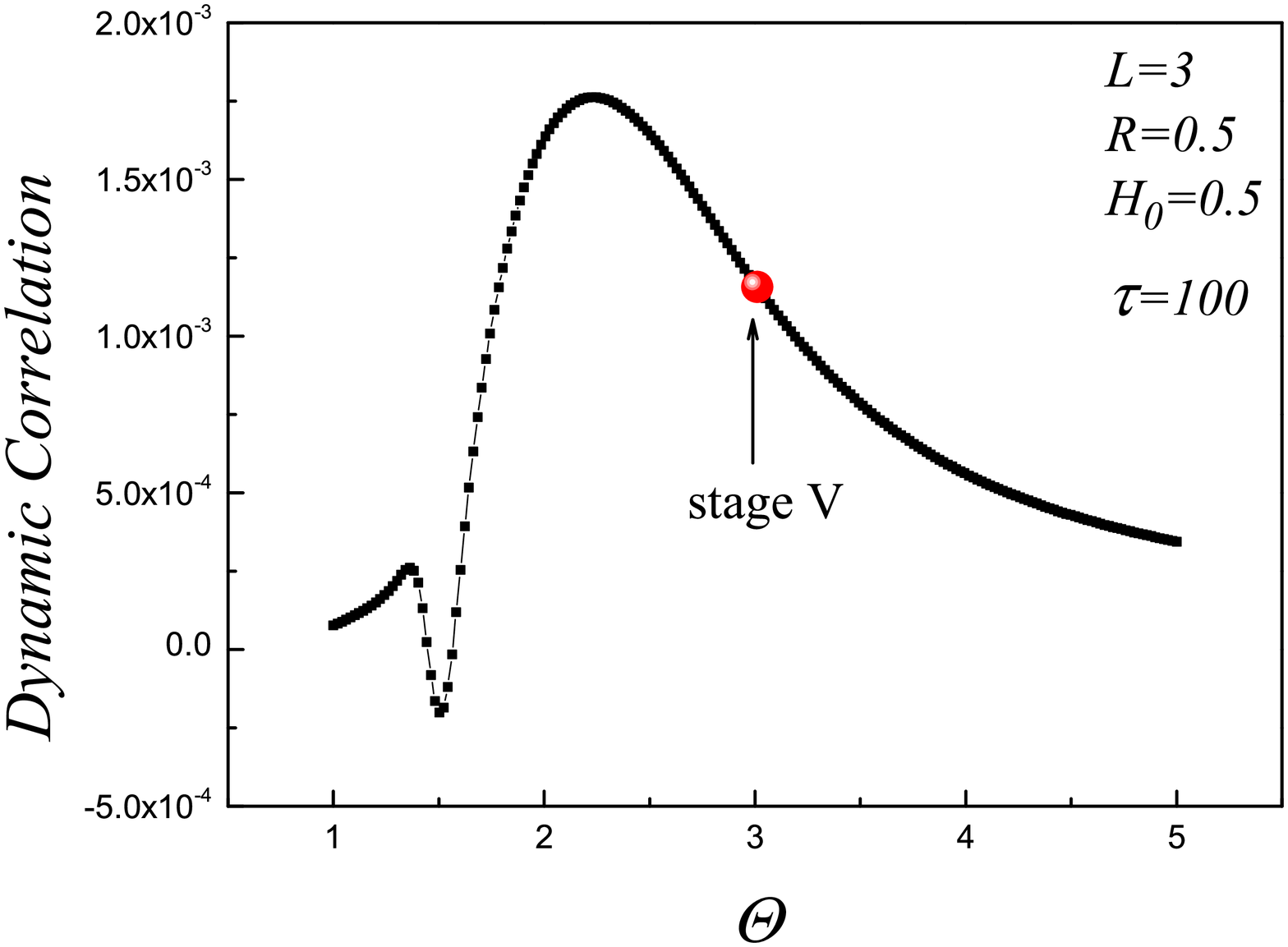}\\
\caption{}\label{fig5}
\end{figure}

\begin{figure}
\center
\includegraphics[width=8cm]{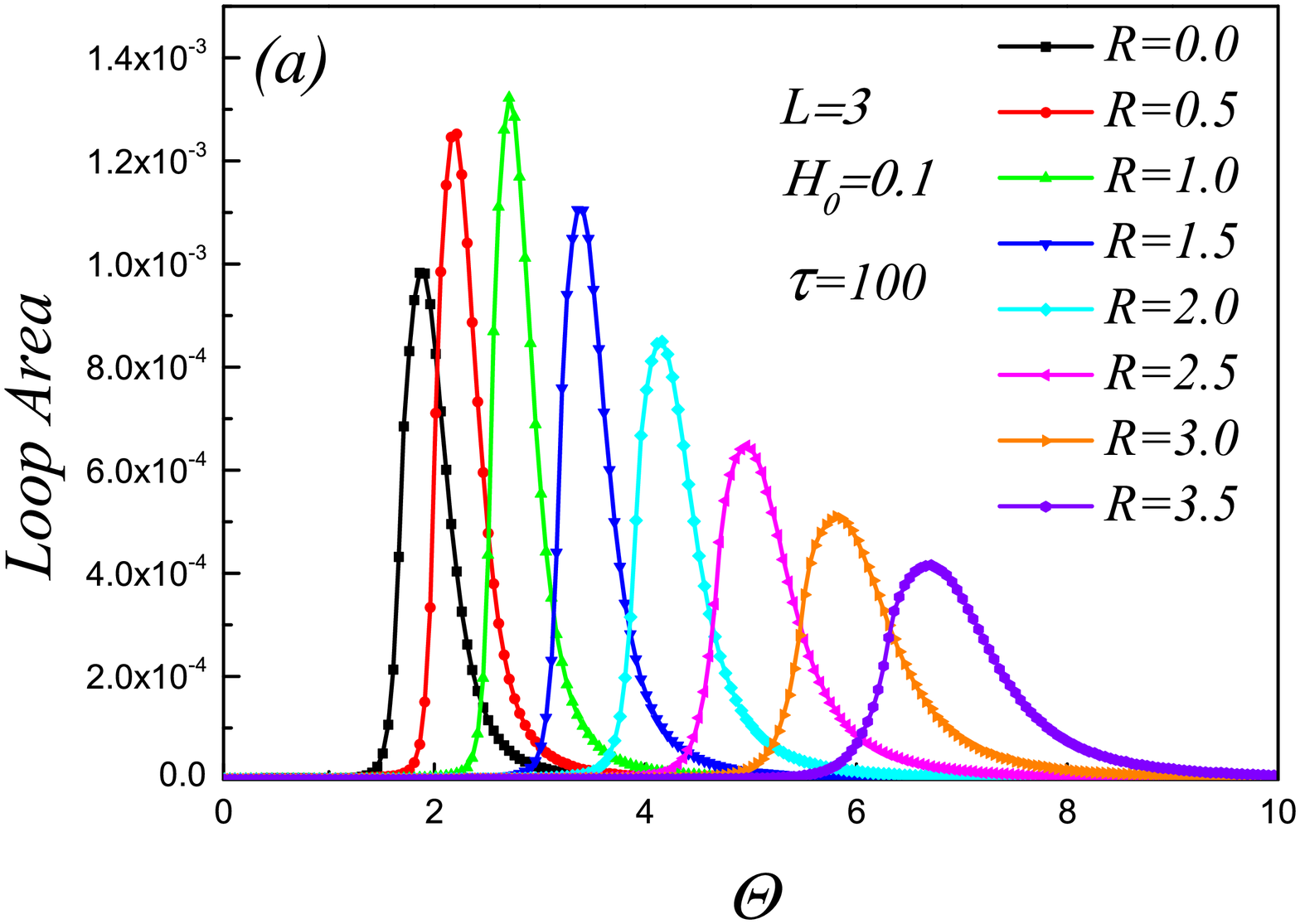}
\includegraphics[width=8cm]{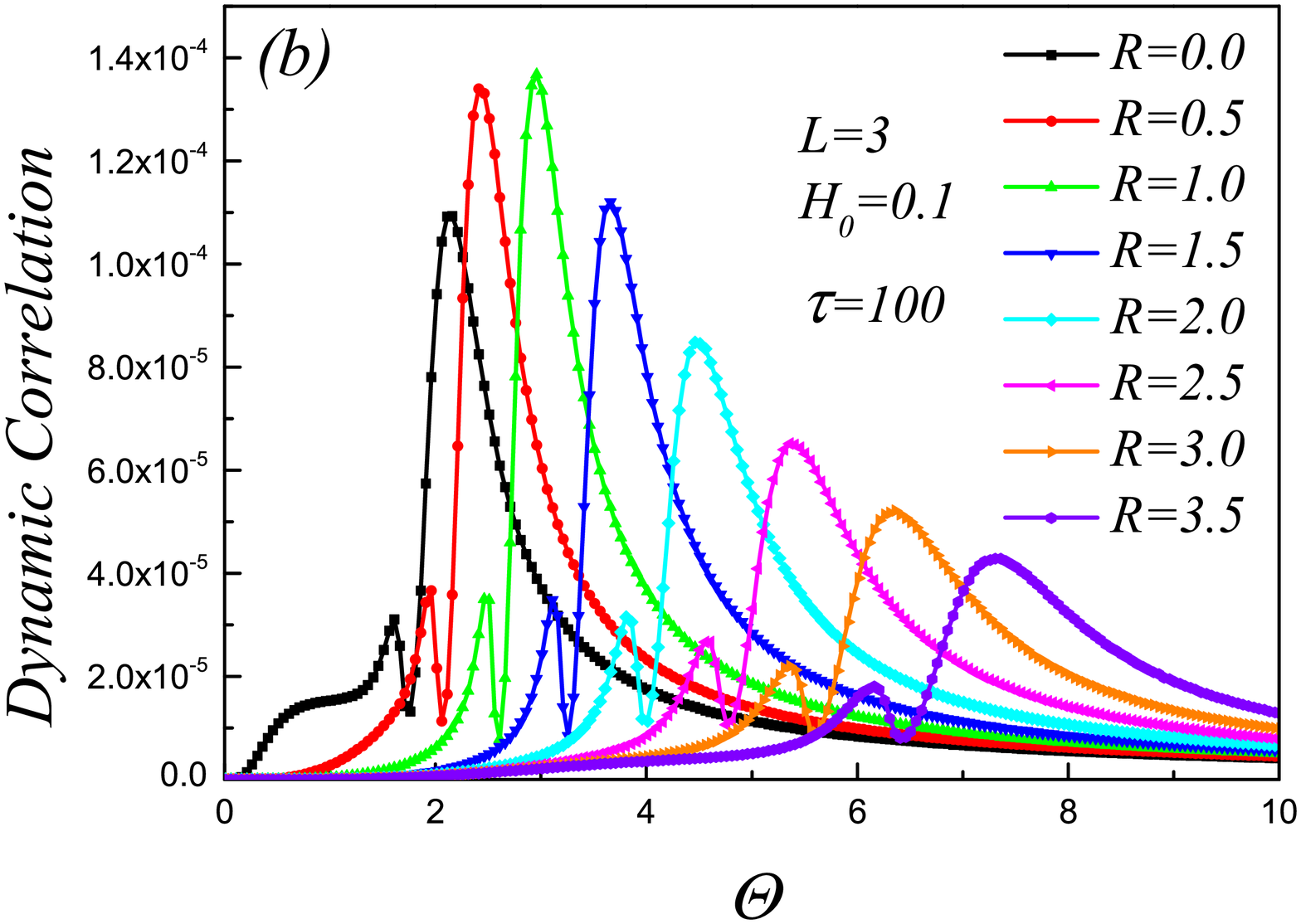}\\
\includegraphics[width=8cm]{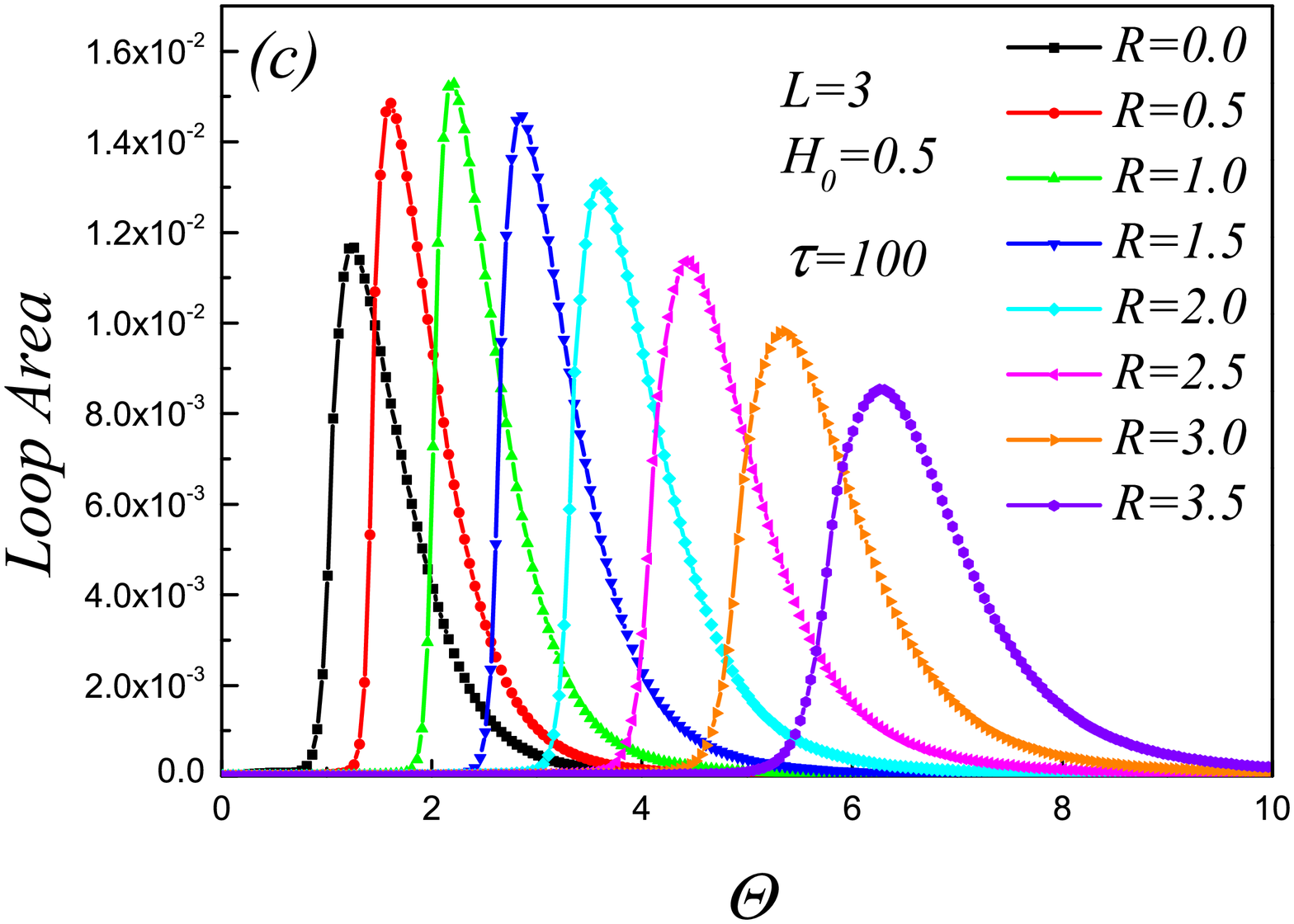}
\includegraphics[width=8cm]{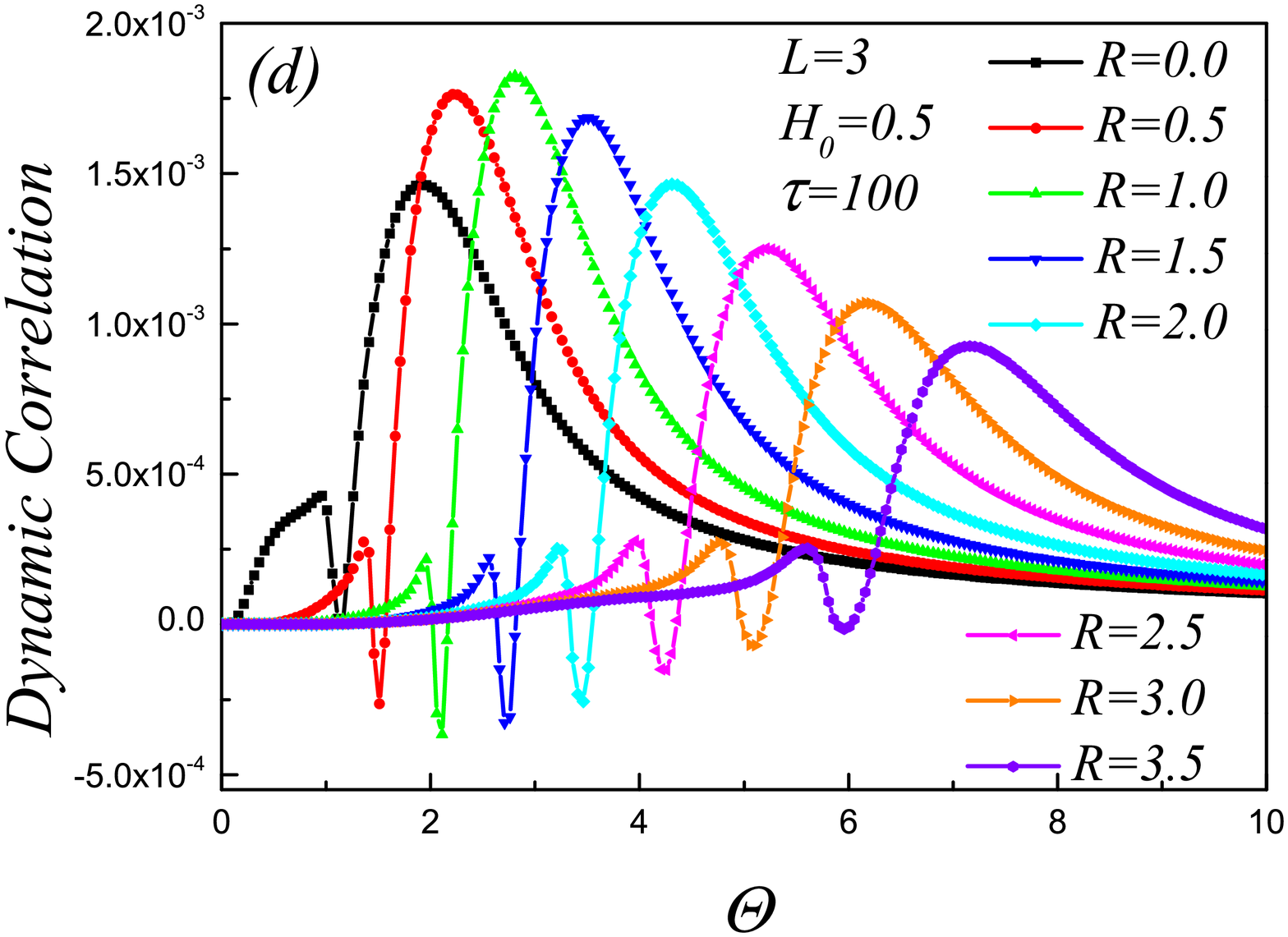}\\
\caption{}\label{fig6}
\end{figure}

\begin{figure}
\center
\includegraphics[width=4cm]{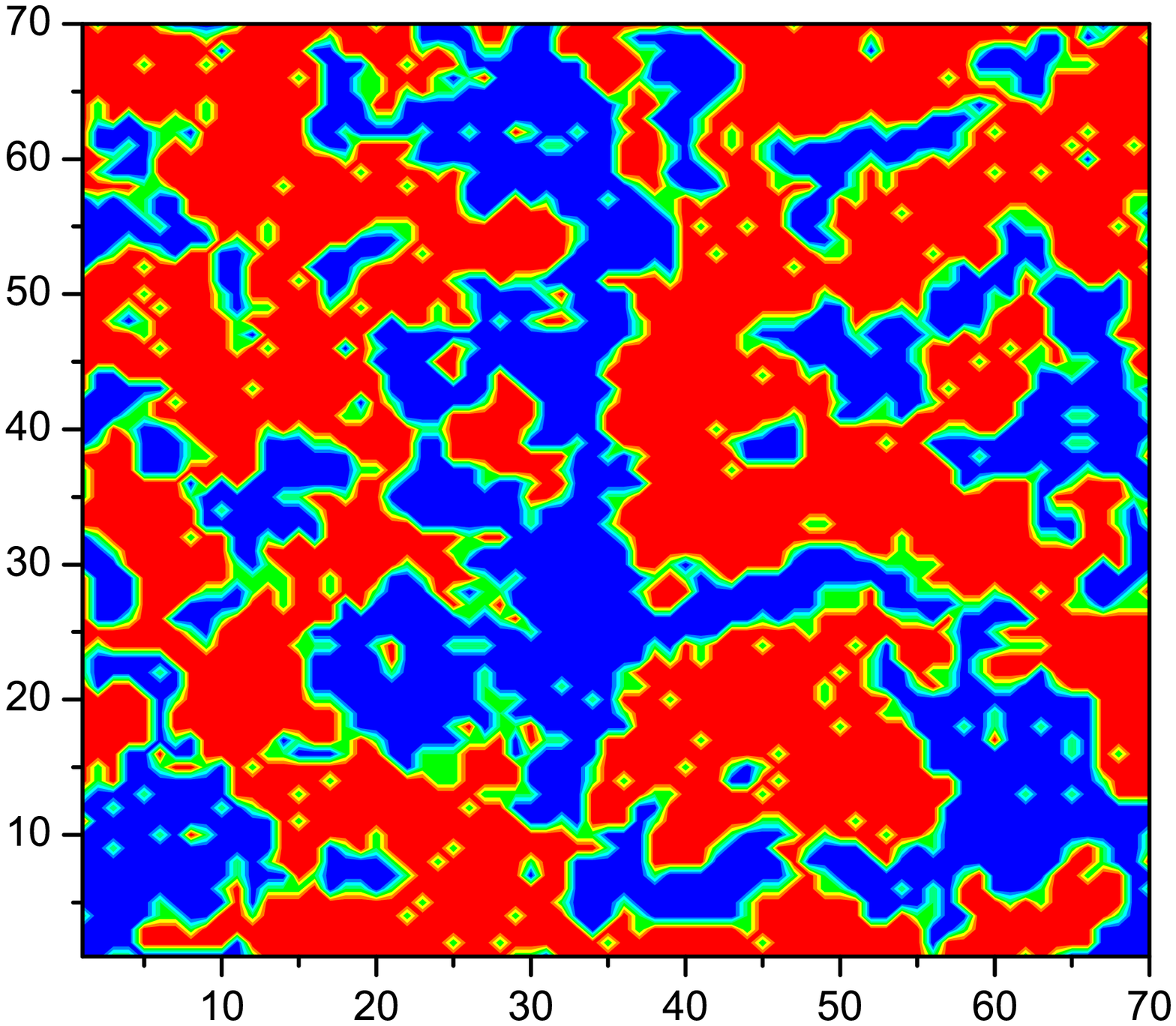}
\includegraphics[width=3.8cm,height=3.5cm]{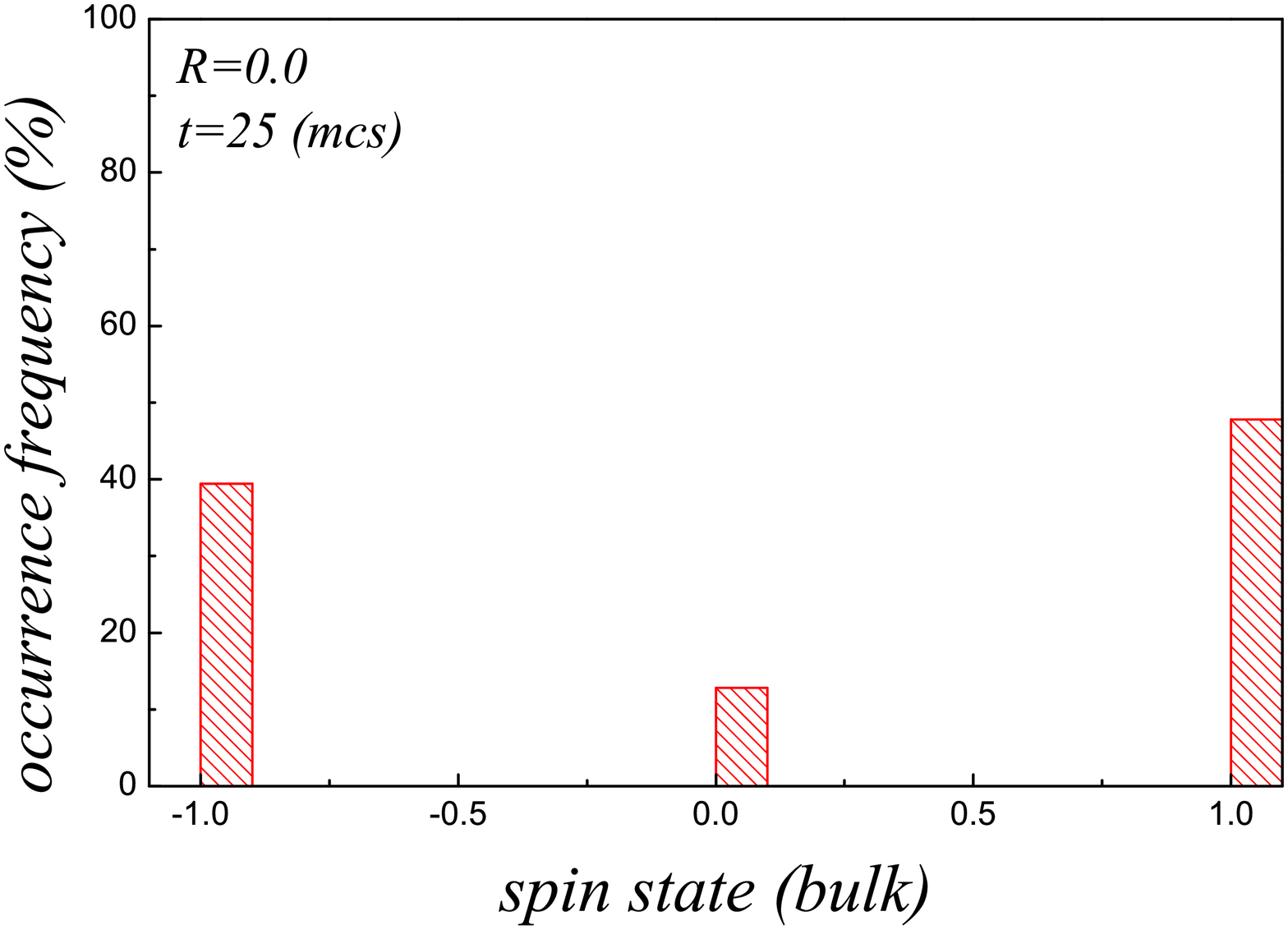}
\includegraphics[width=4cm]{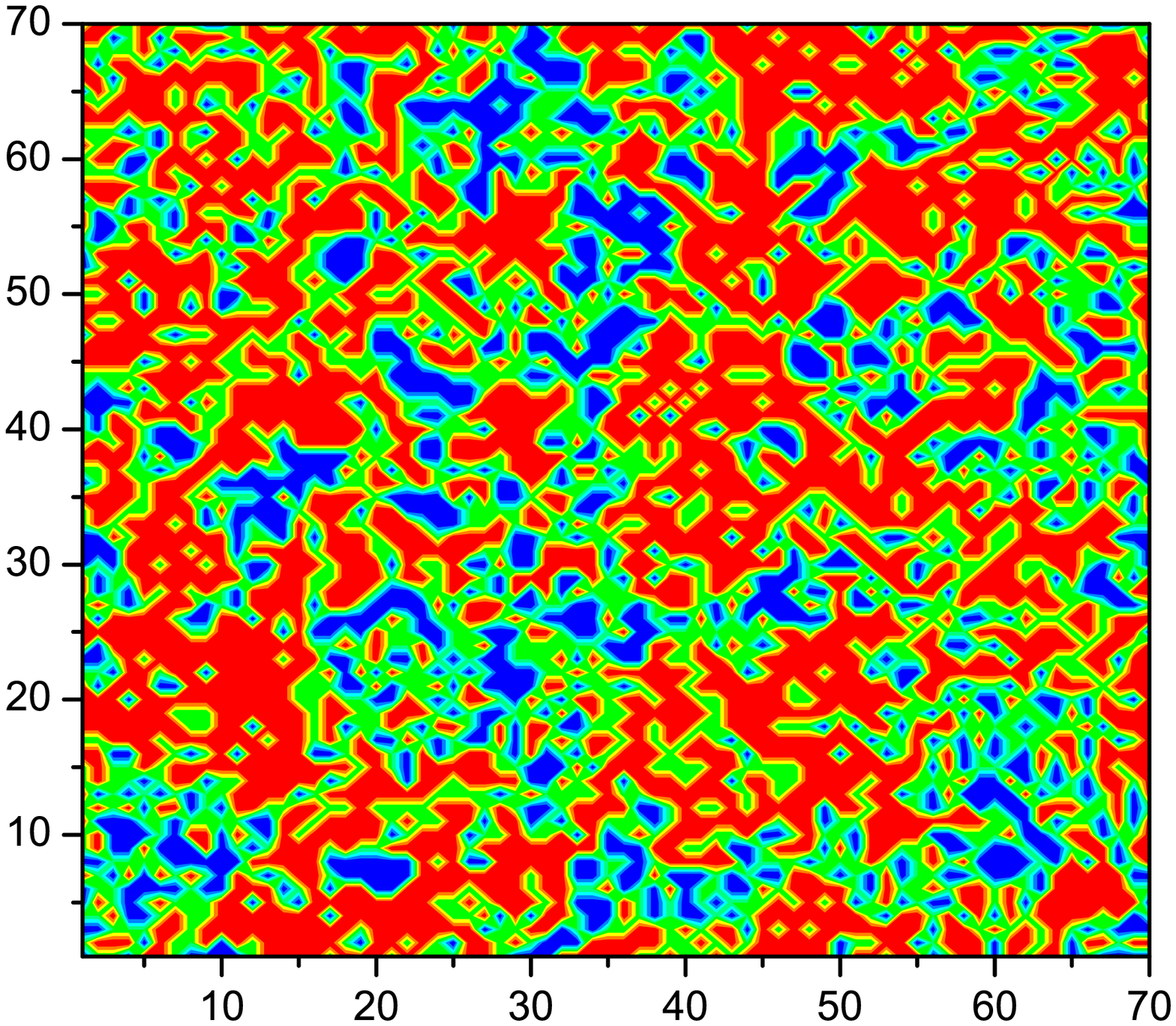}
\includegraphics[width=3.8cm,height=3.5cm]{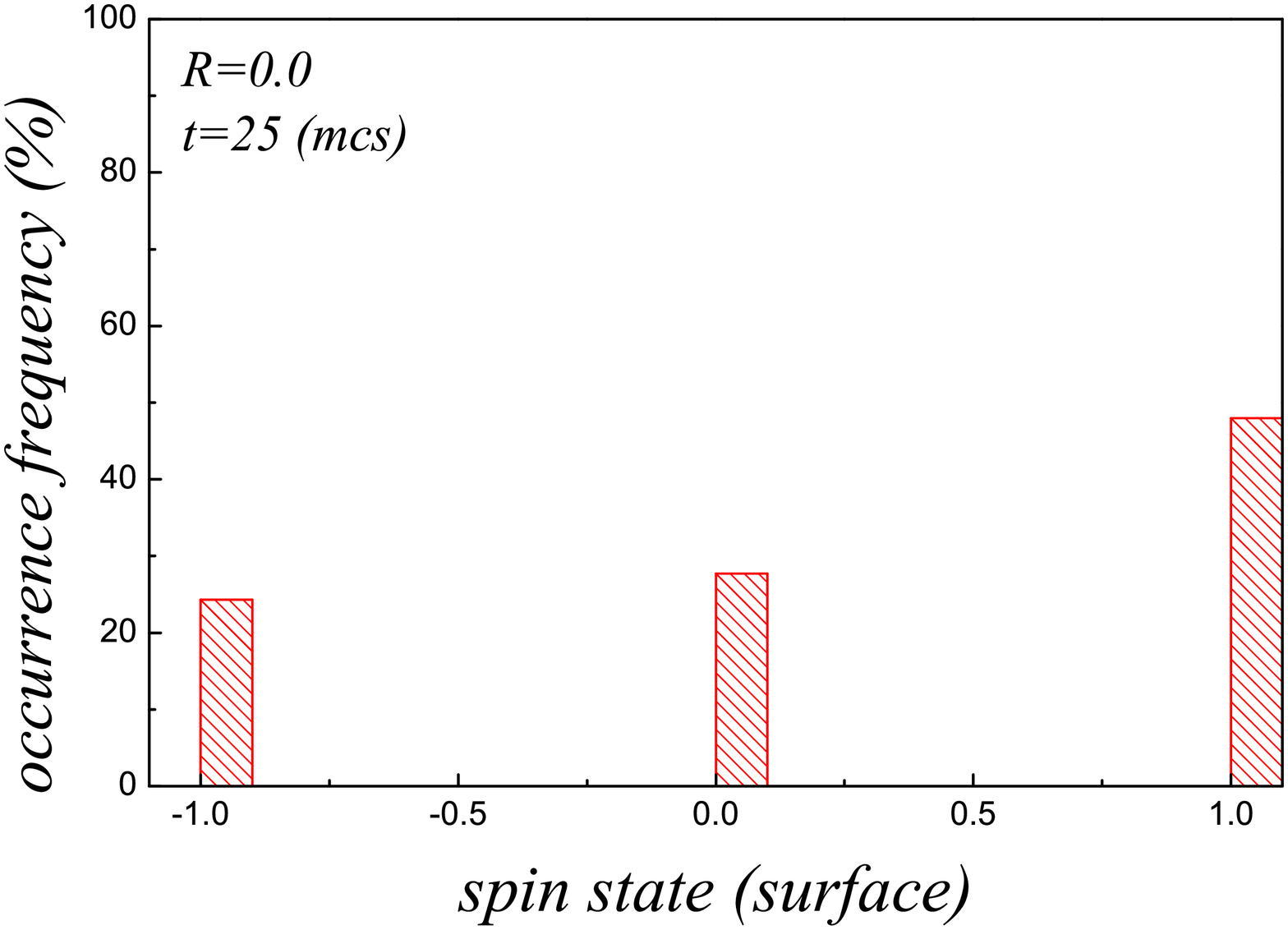}\\
\includegraphics[width=4cm]{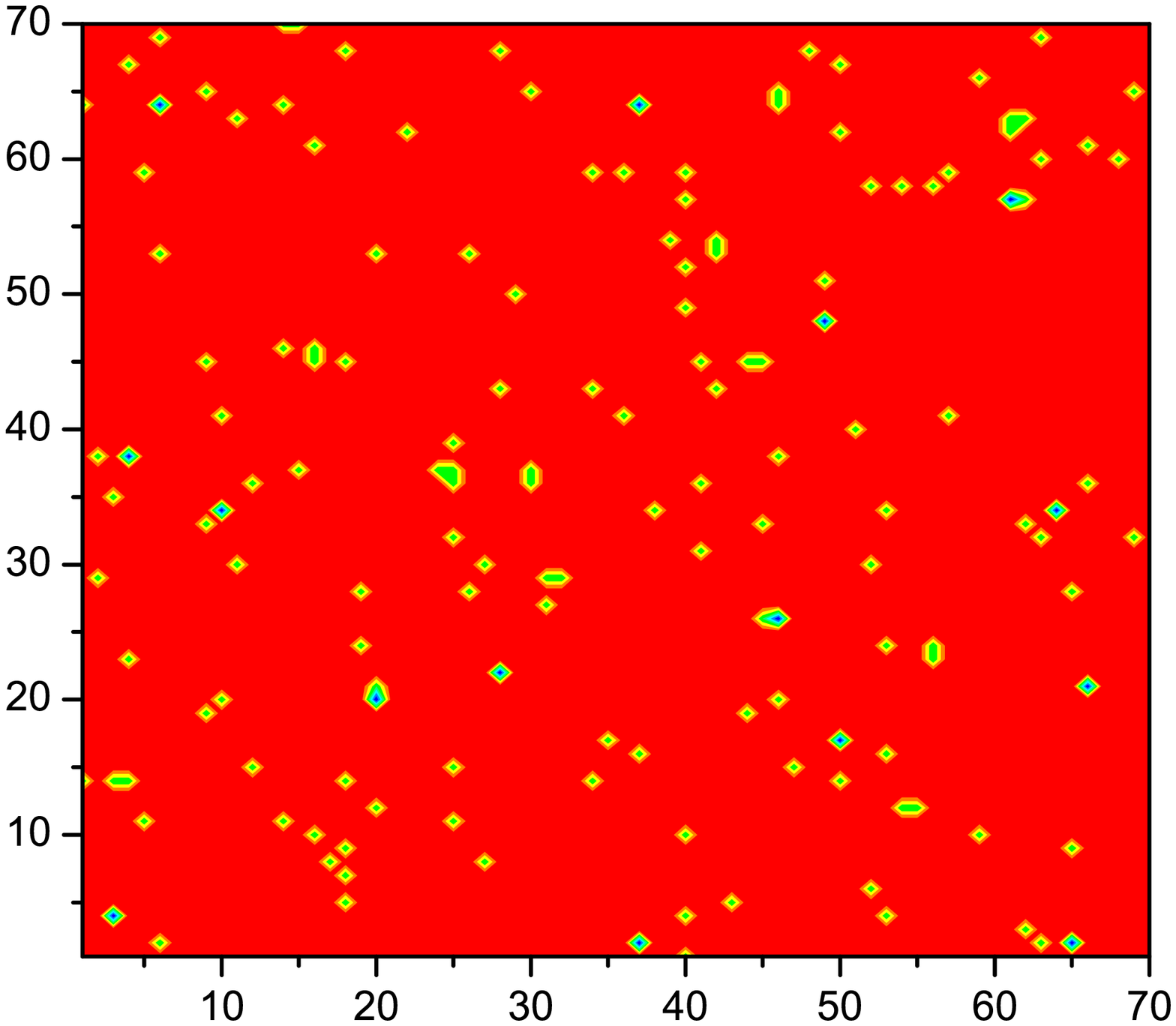}
\includegraphics[width=3.8cm,height=3.5cm]{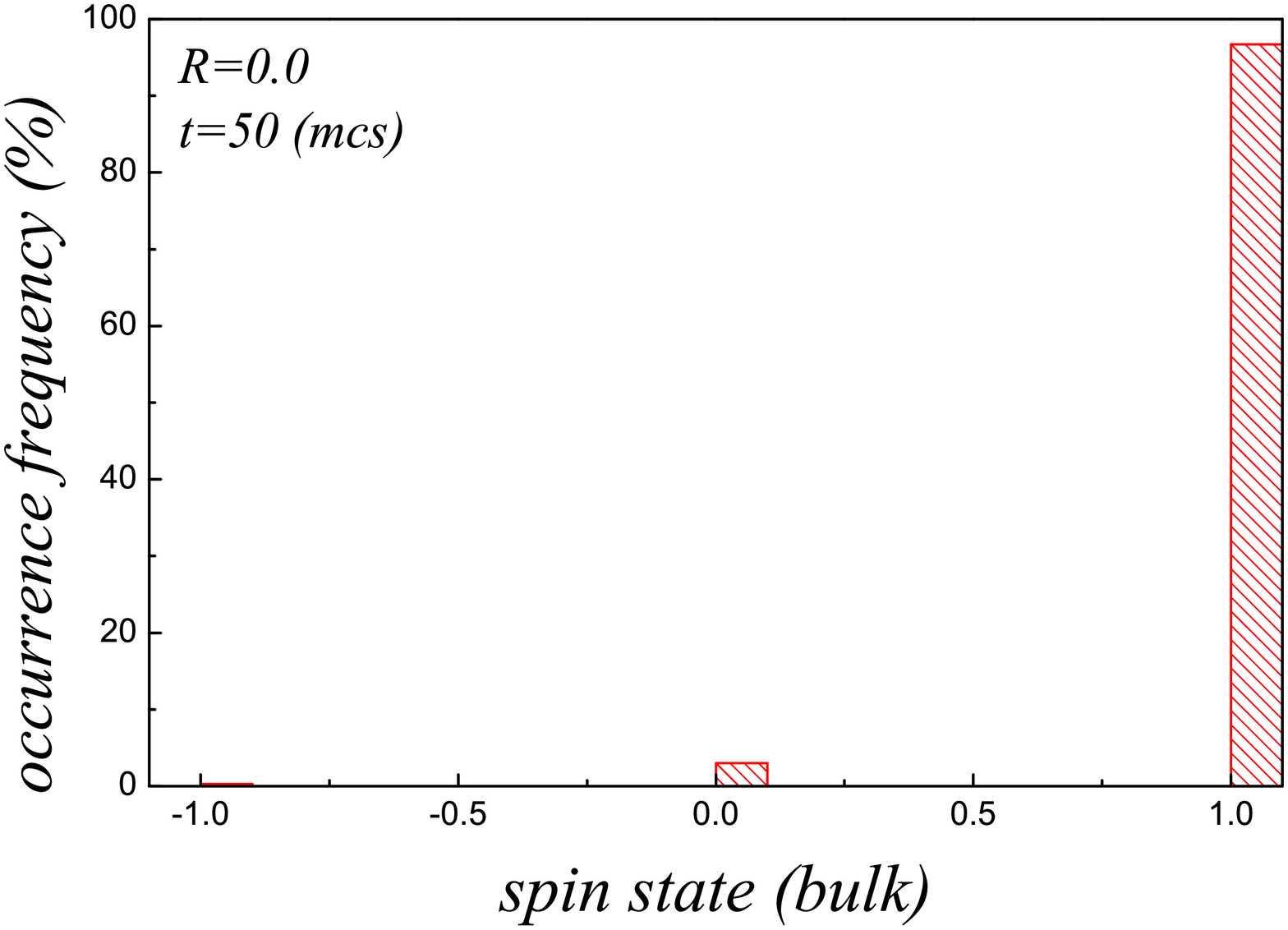}
\includegraphics[width=4cm]{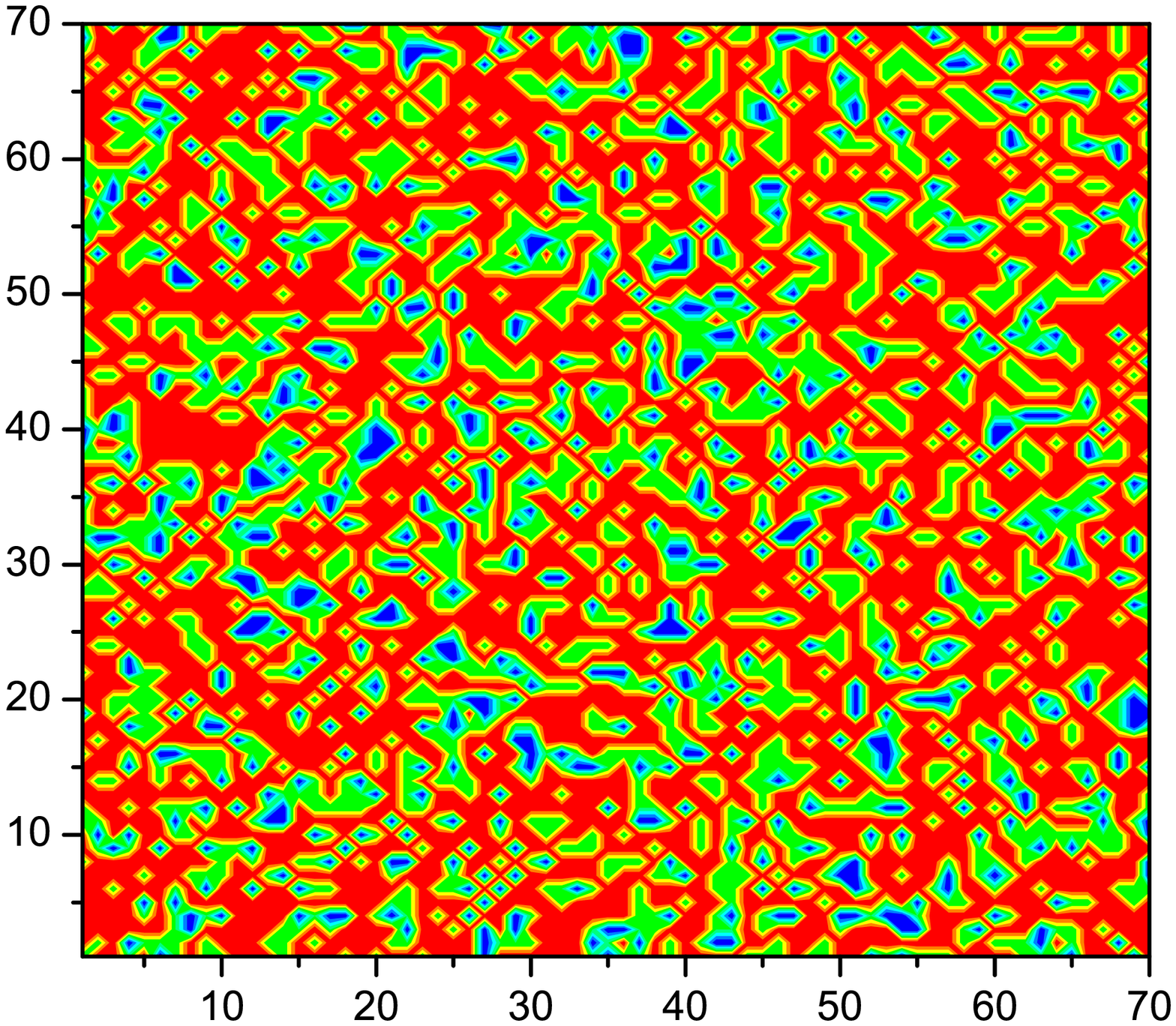}
\includegraphics[width=3.8cm,height=3.5cm]{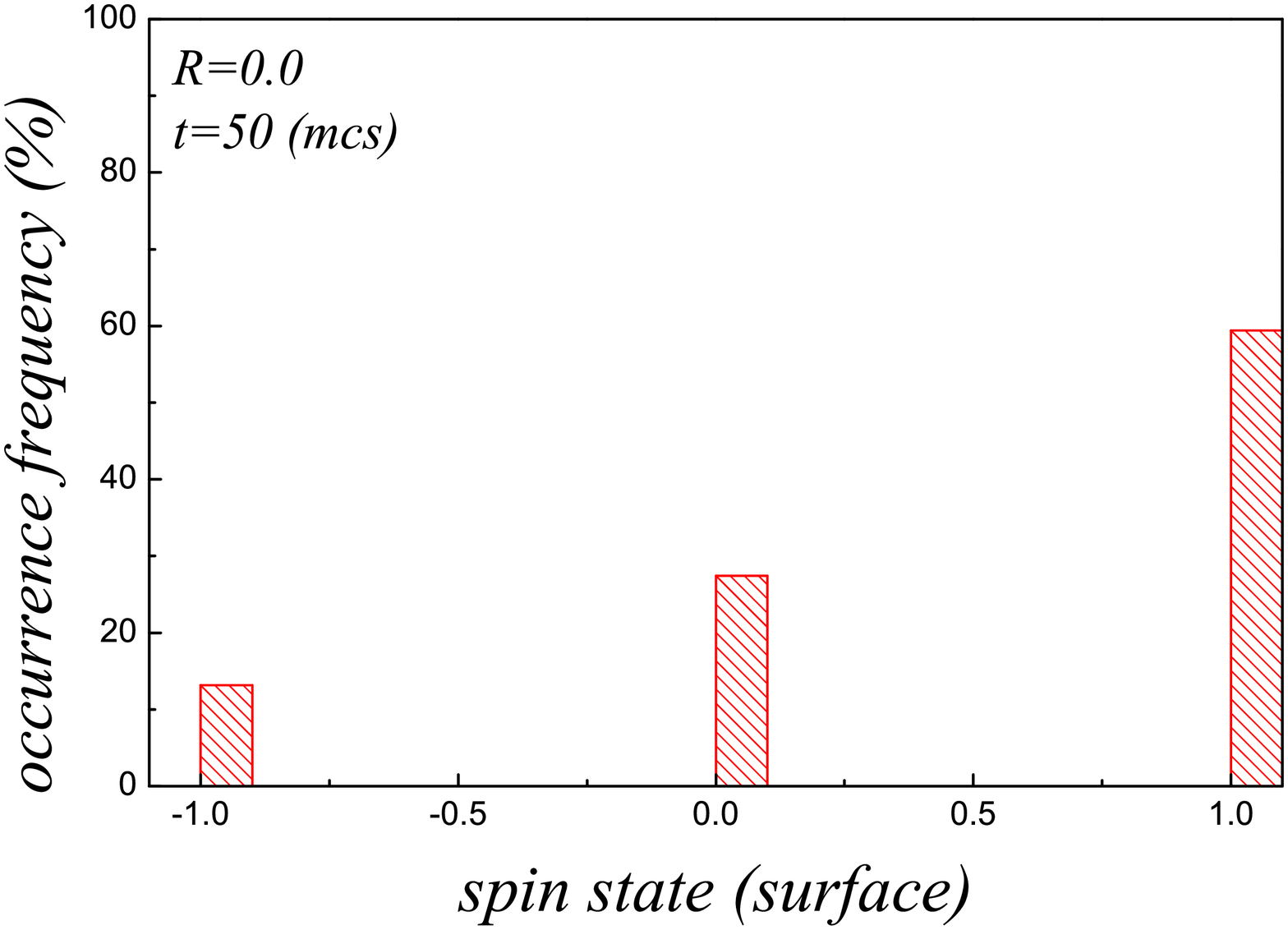}\\
\includegraphics[width=4cm]{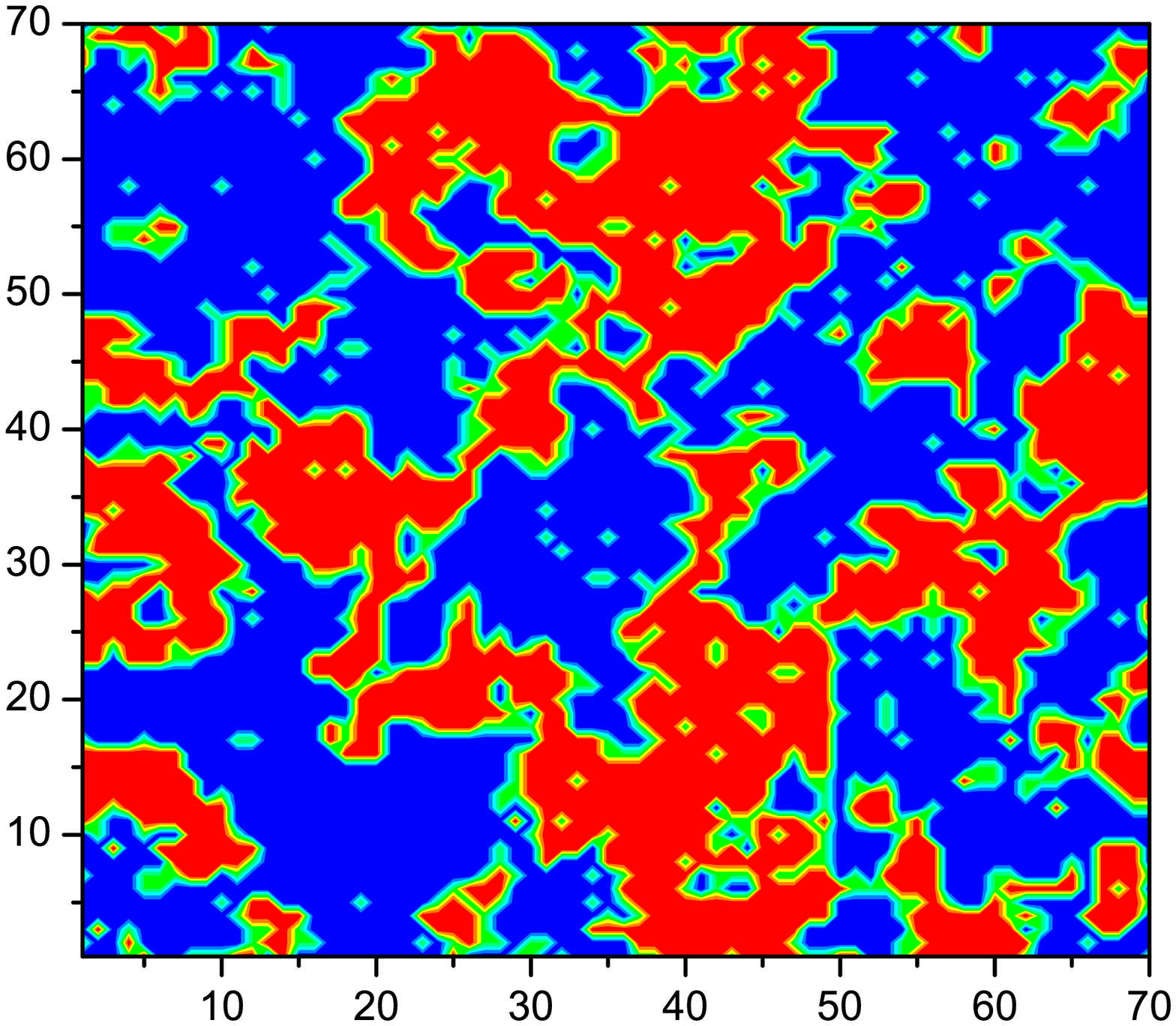}
\includegraphics[width=3.8cm,height=3.5cm]{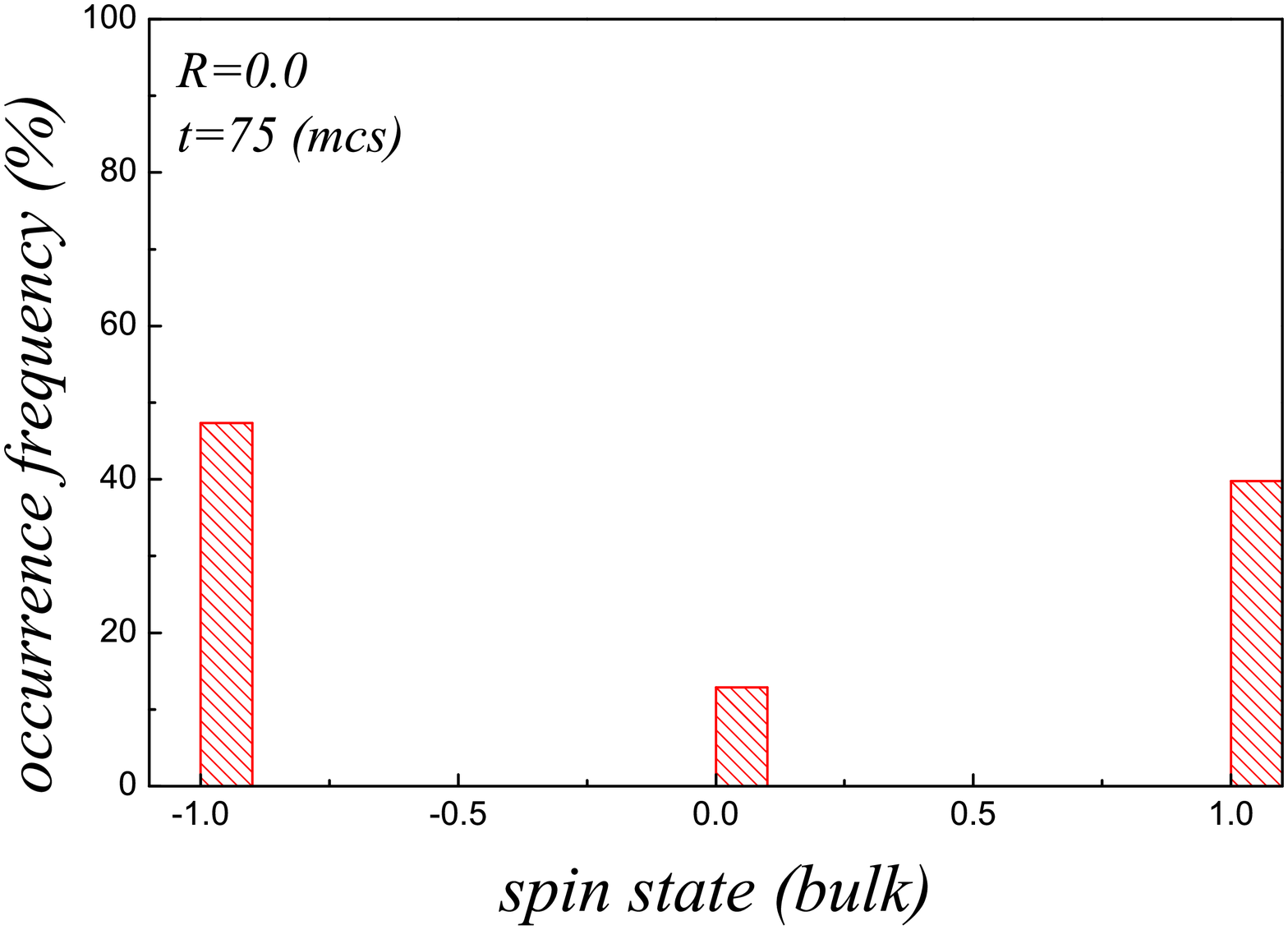}
\includegraphics[width=4cm]{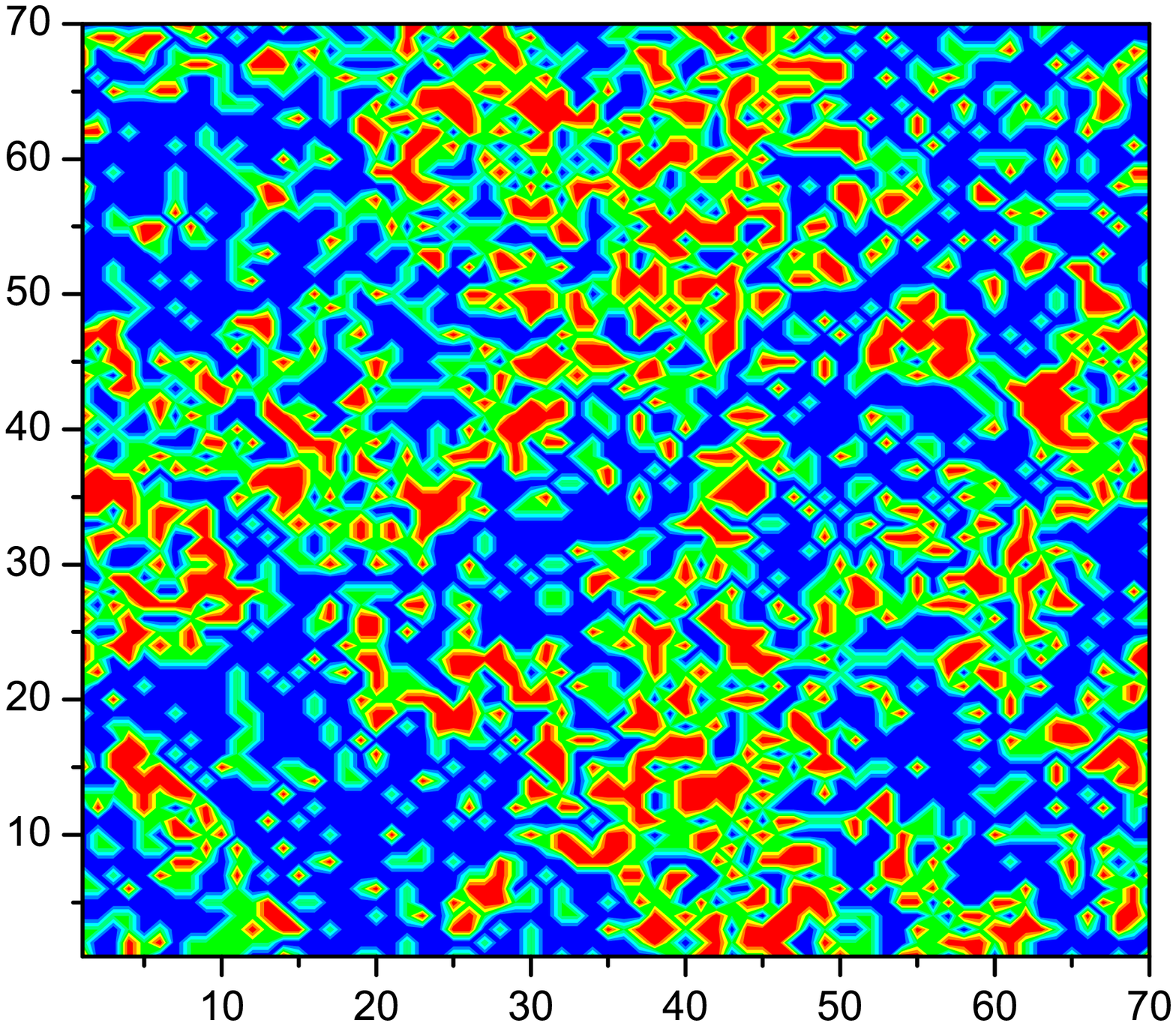}
\includegraphics[width=3.8cm,height=3.5cm]{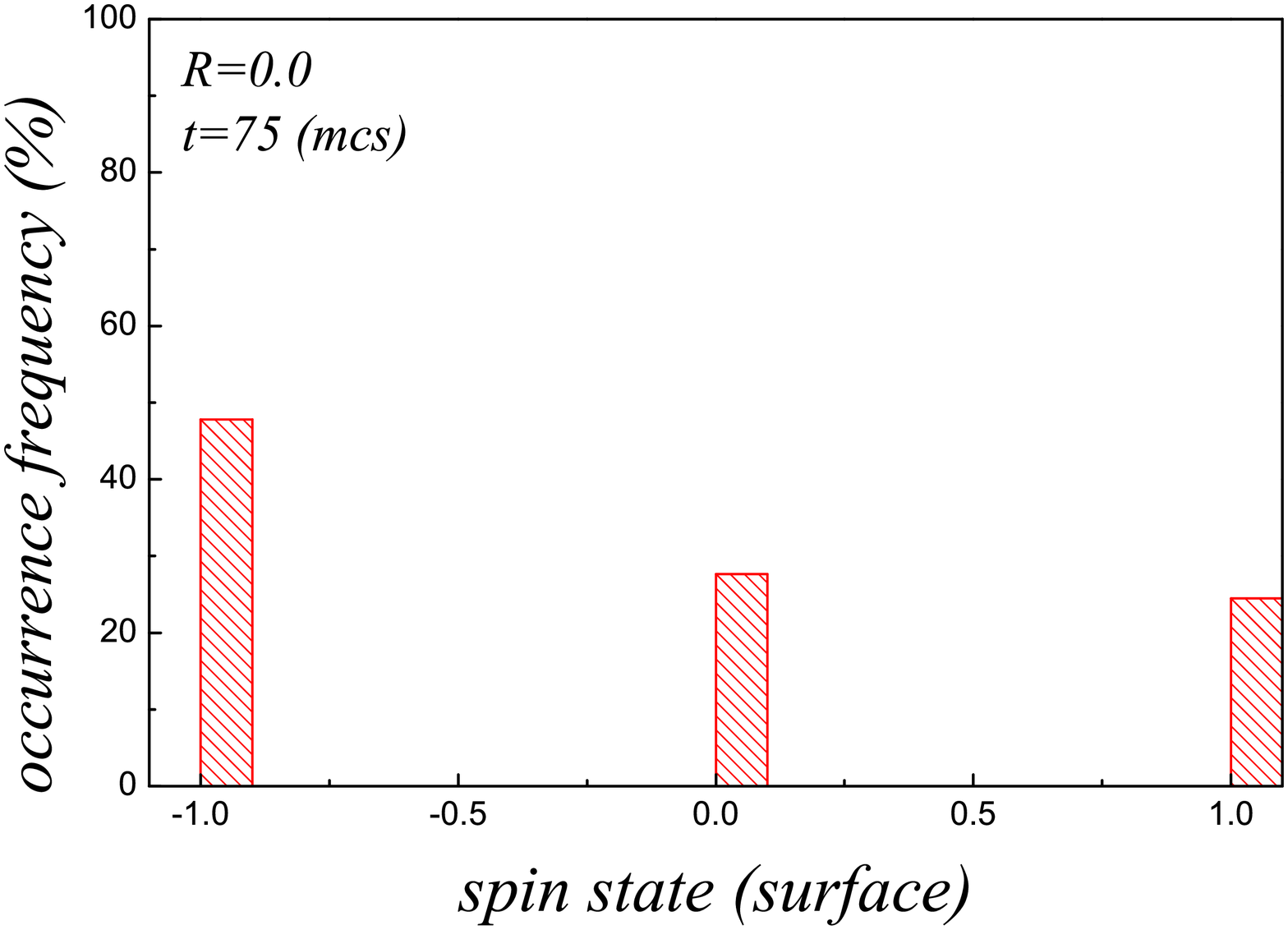}\\
\includegraphics[width=4cm]{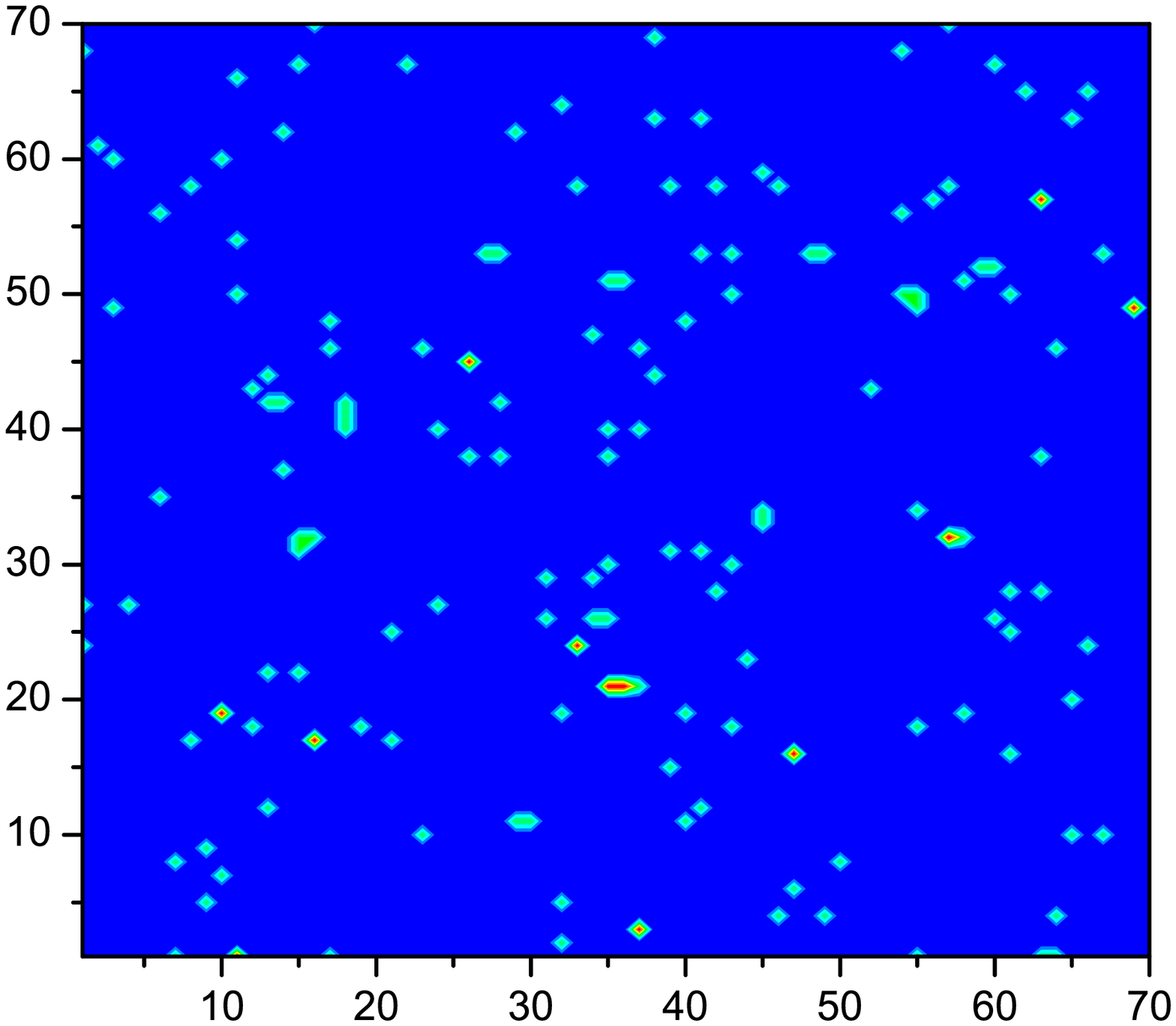}
\includegraphics[width=3.8cm,height=3.5cm]{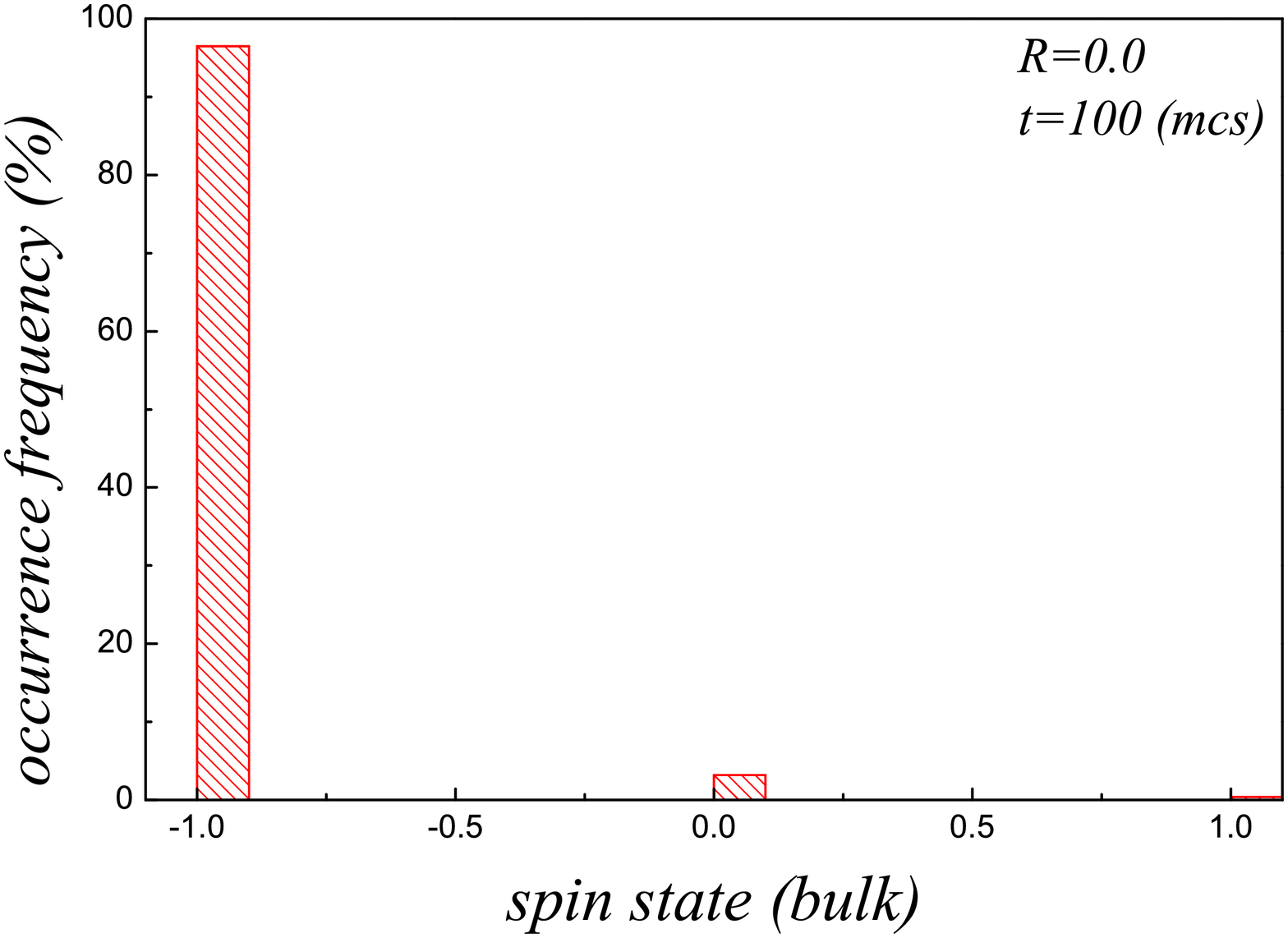}
\includegraphics[width=4cm]{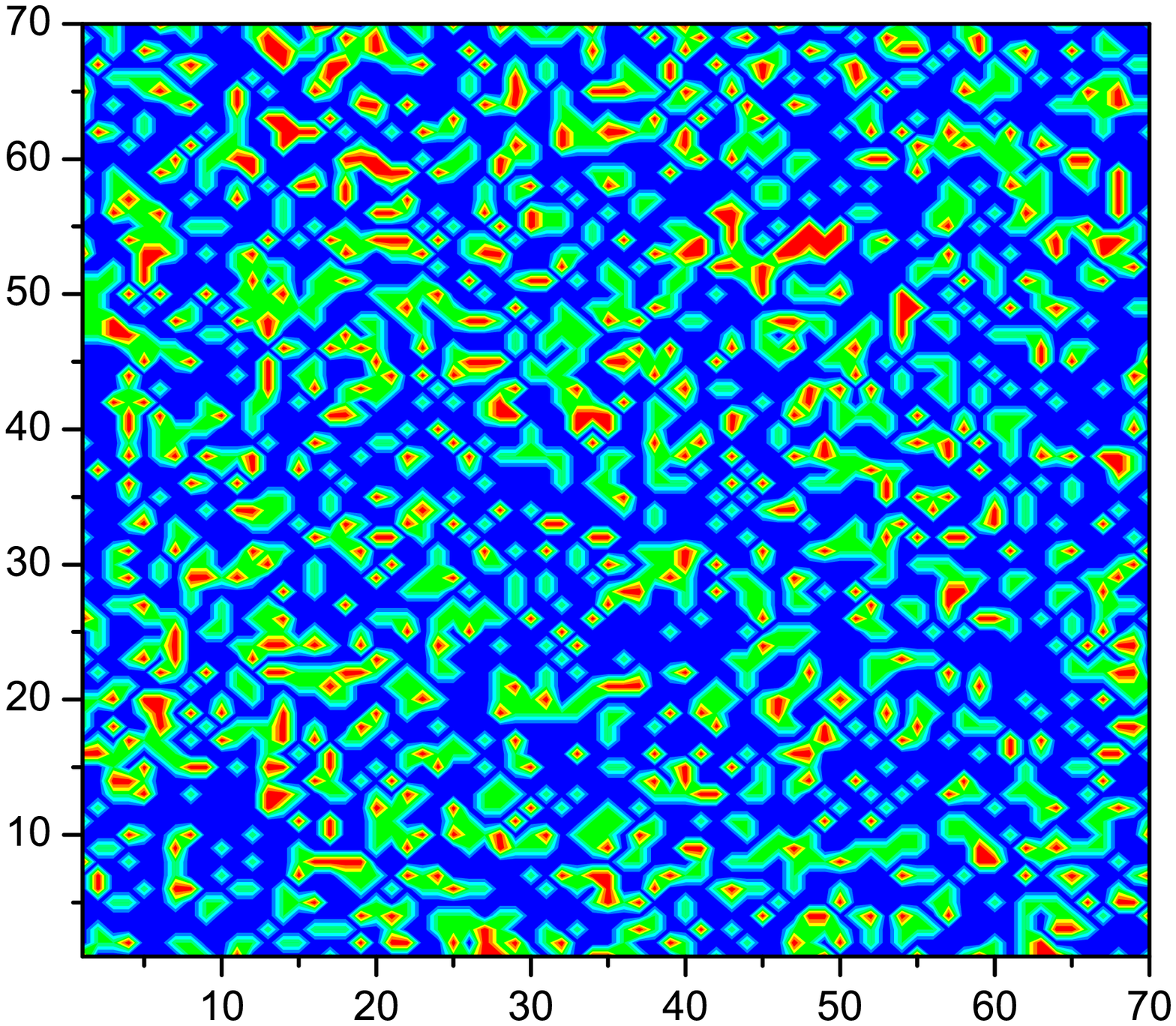}
\includegraphics[width=3.8cm,height=3.5cm]{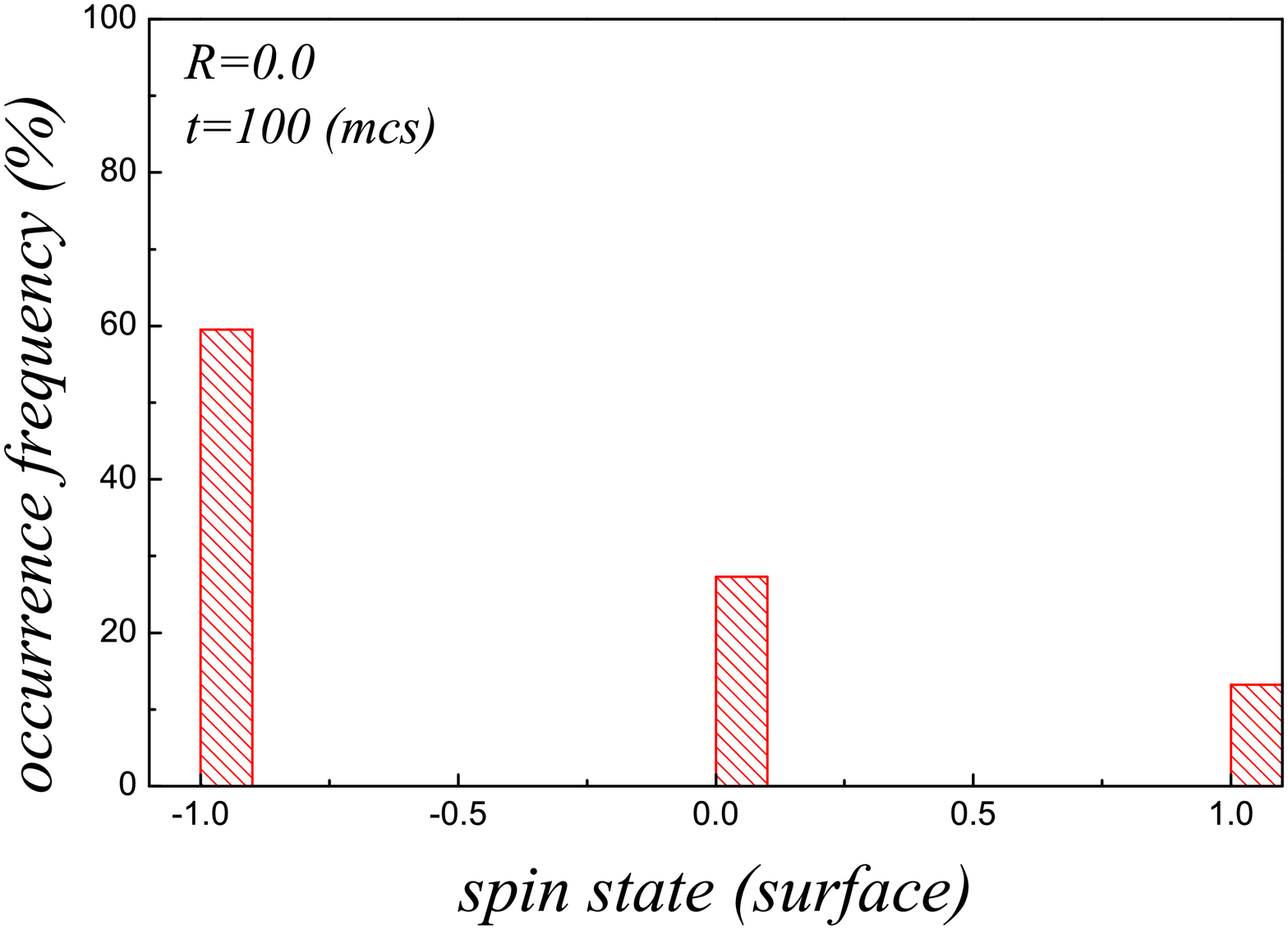}\\
\caption{ }\label{fig7}
\end{figure}

\begin{figure}
\center
\includegraphics[width=4cm]{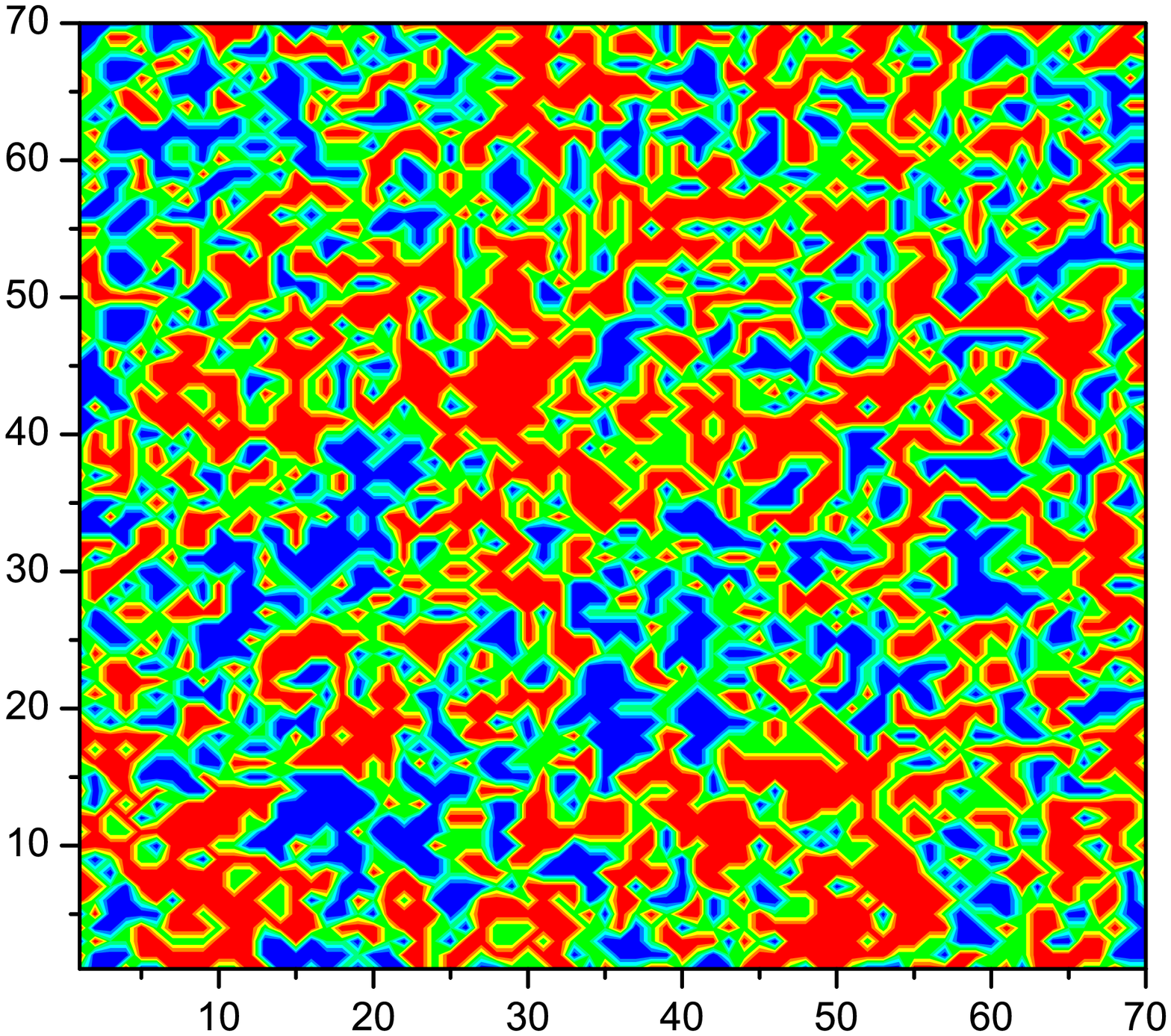}
\includegraphics[width=3.8cm,height=3.5cm]{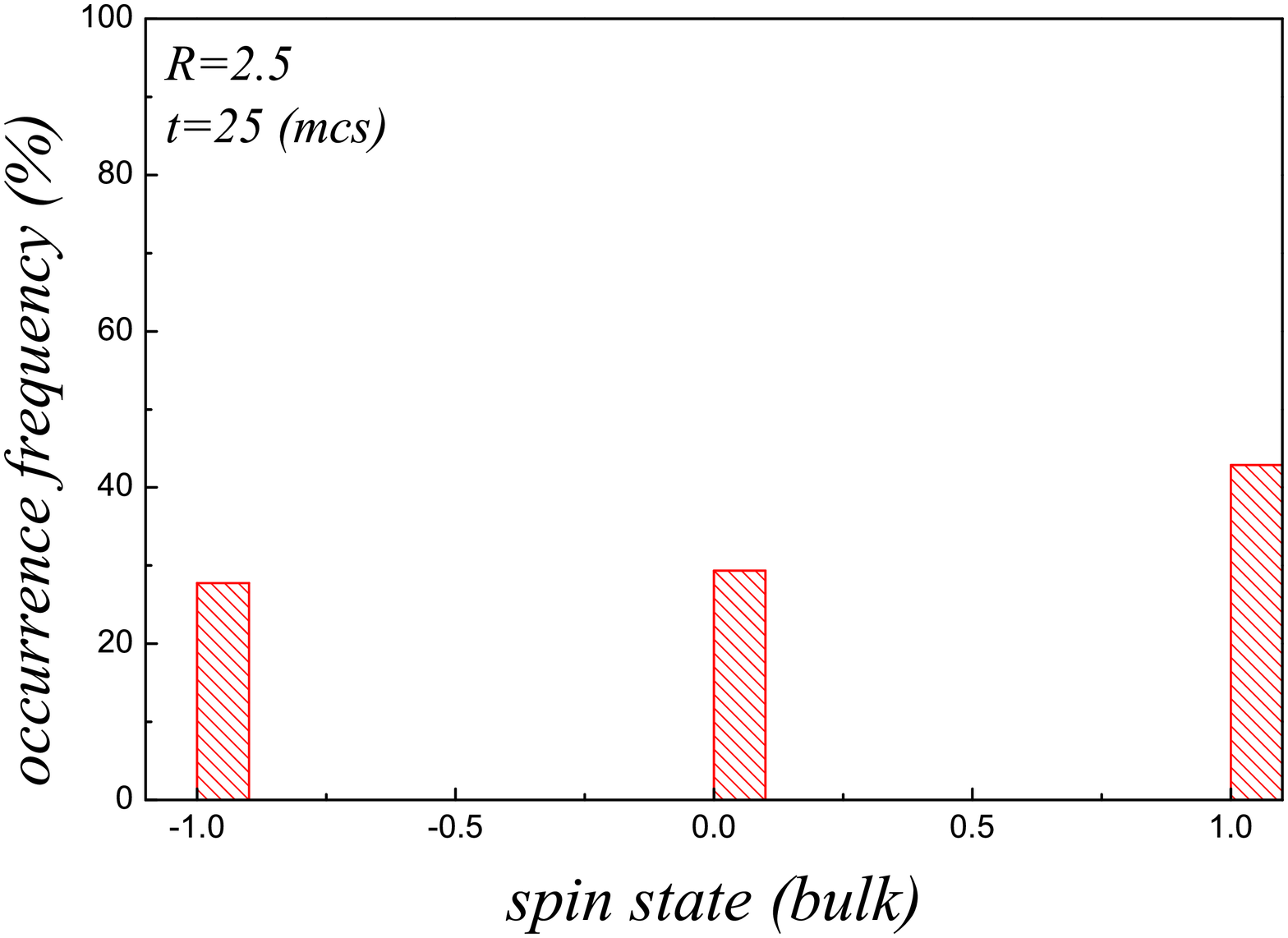}
\includegraphics[width=4cm]{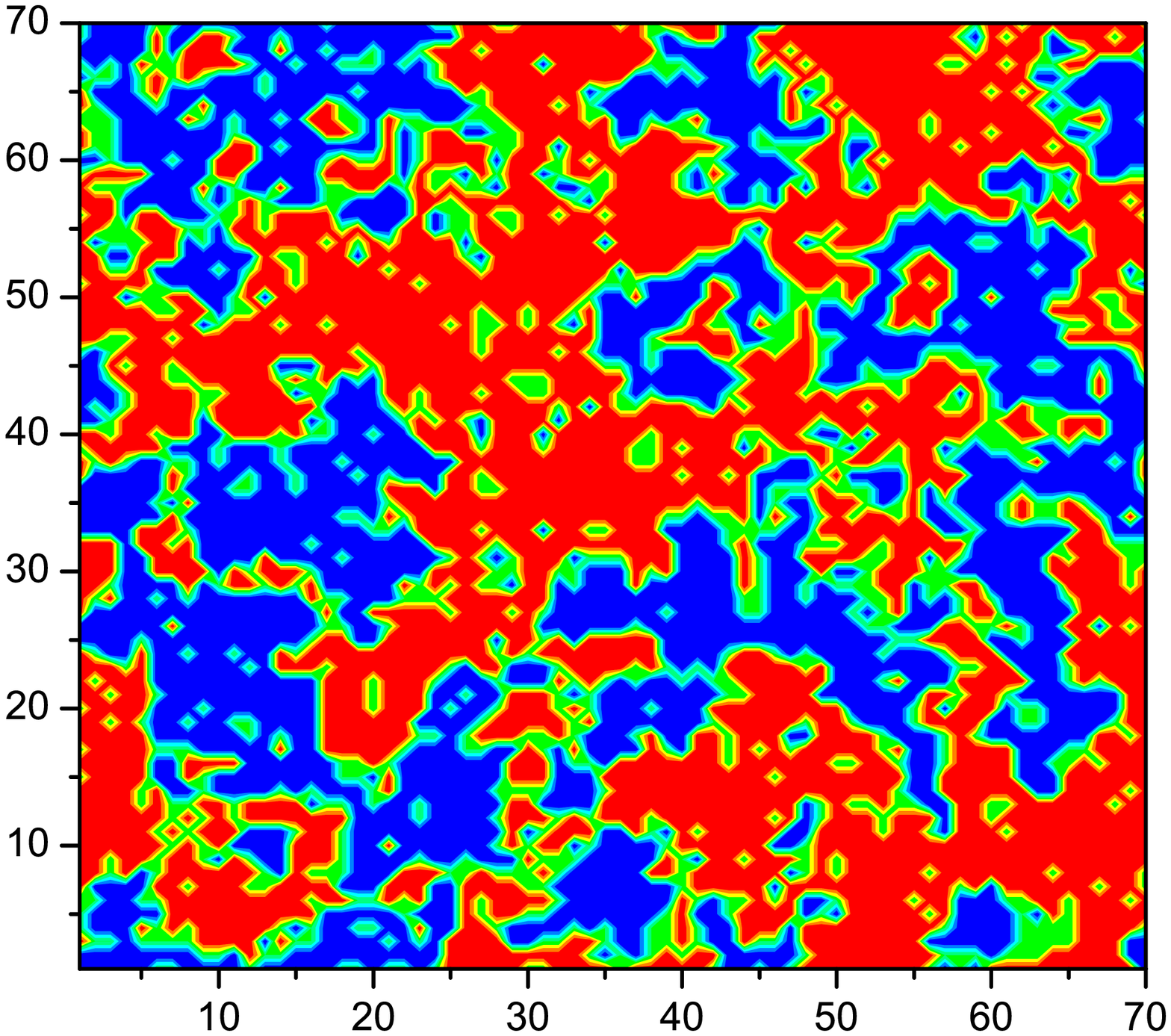}
\includegraphics[width=3.8cm,height=3.5cm]{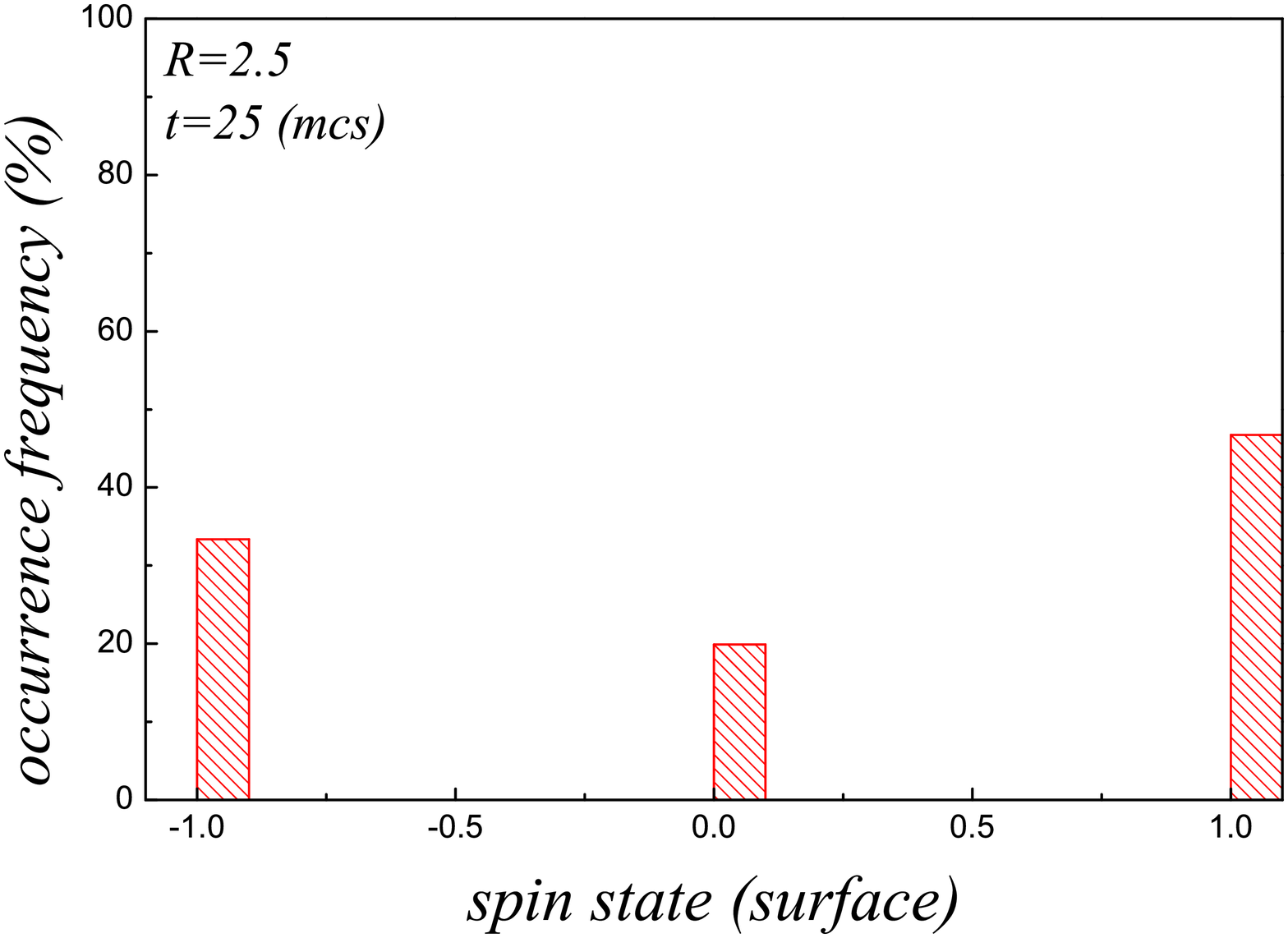}\\
\includegraphics[width=4cm]{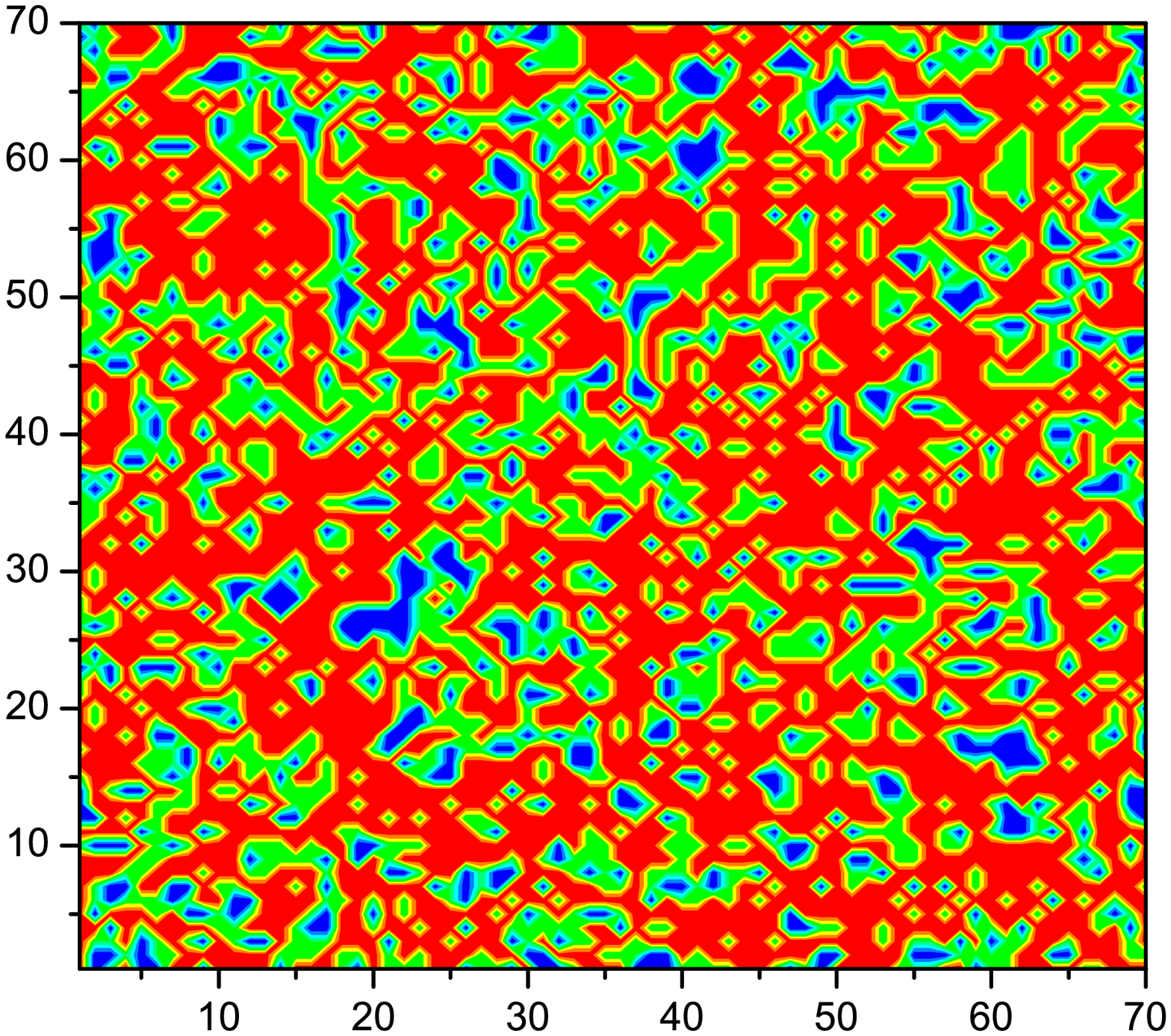}
\includegraphics[width=3.8cm,height=3.5cm]{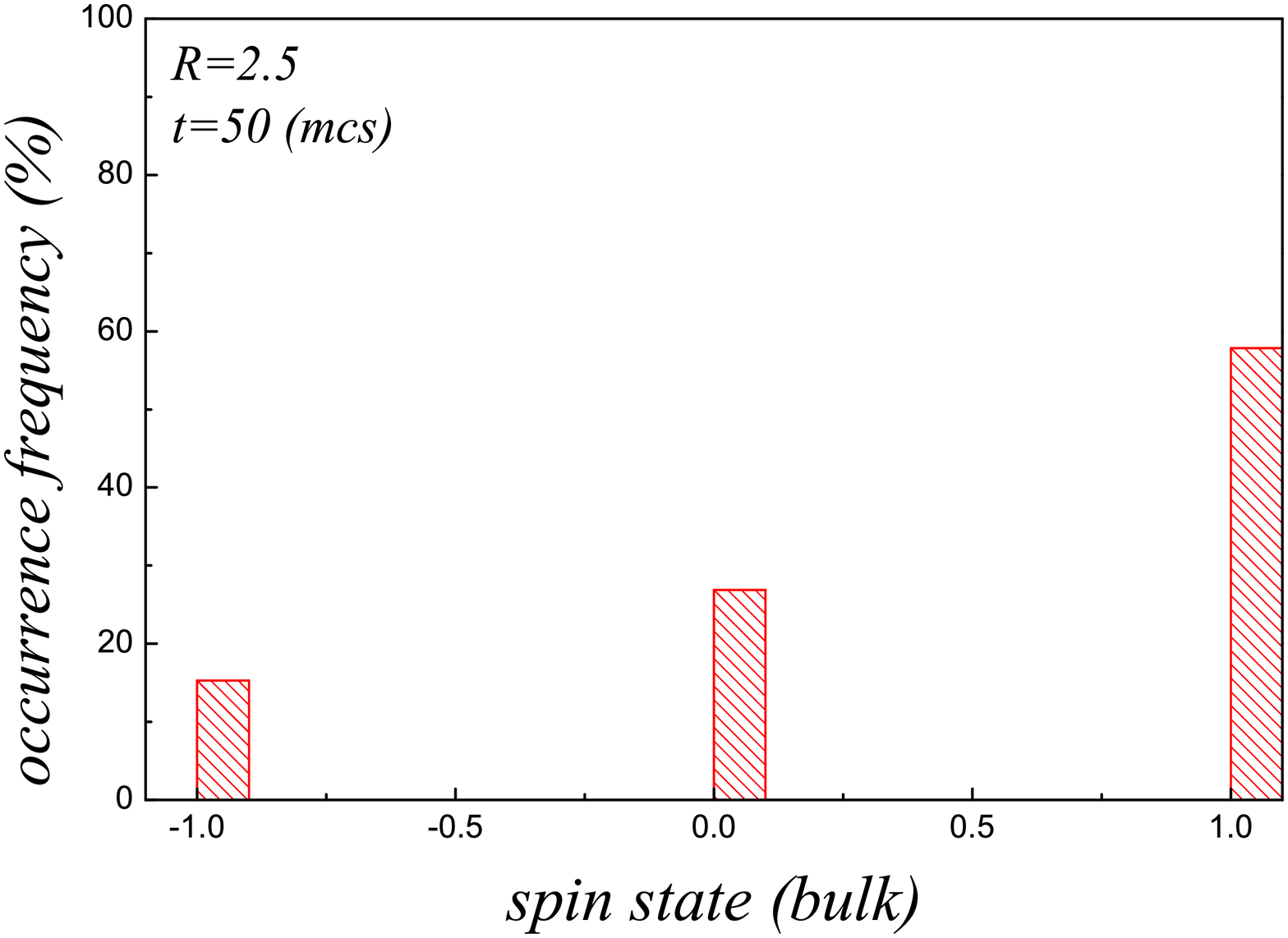}
\includegraphics[width=4cm]{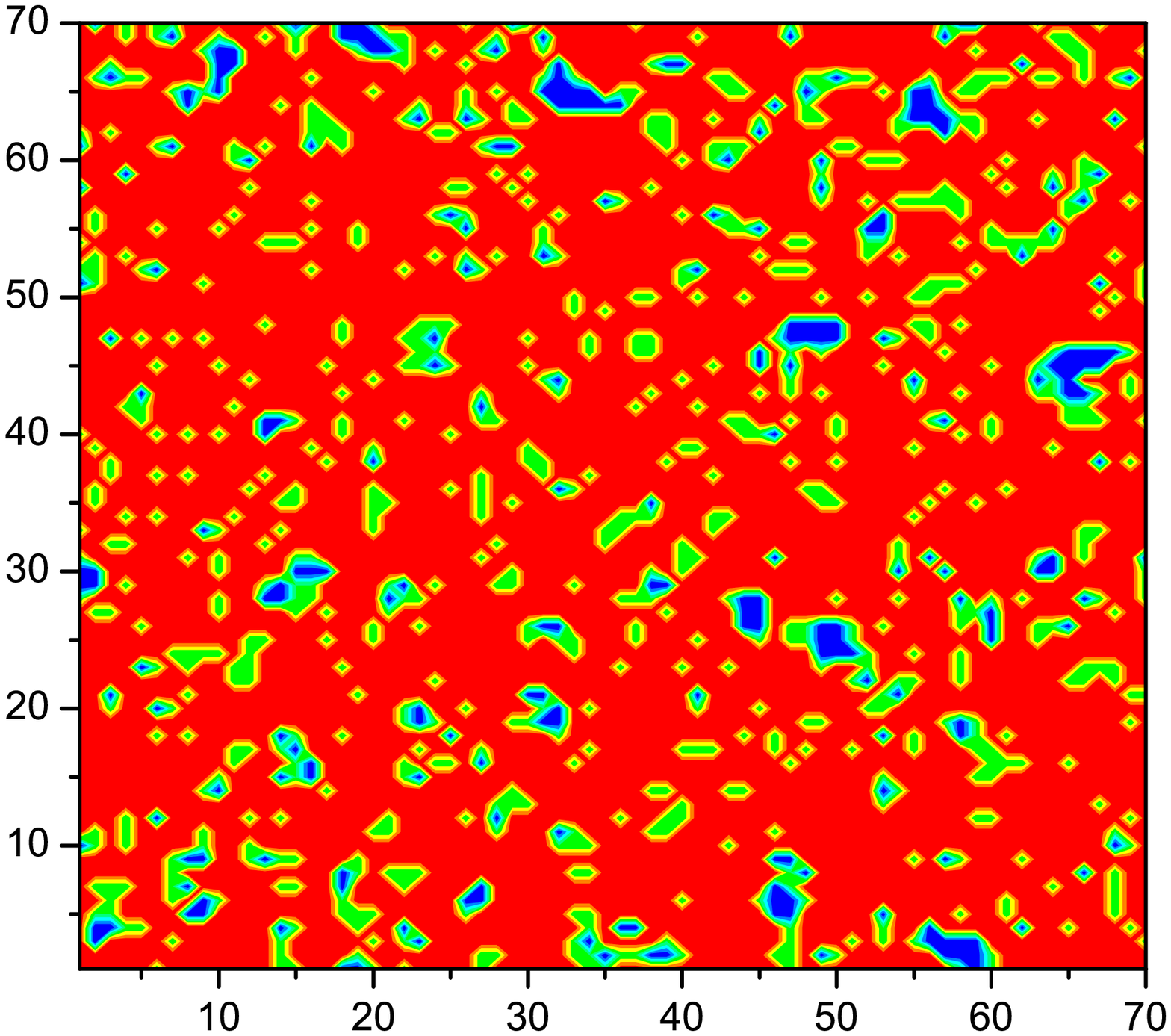}
\includegraphics[width=3.8cm,height=3.5cm]{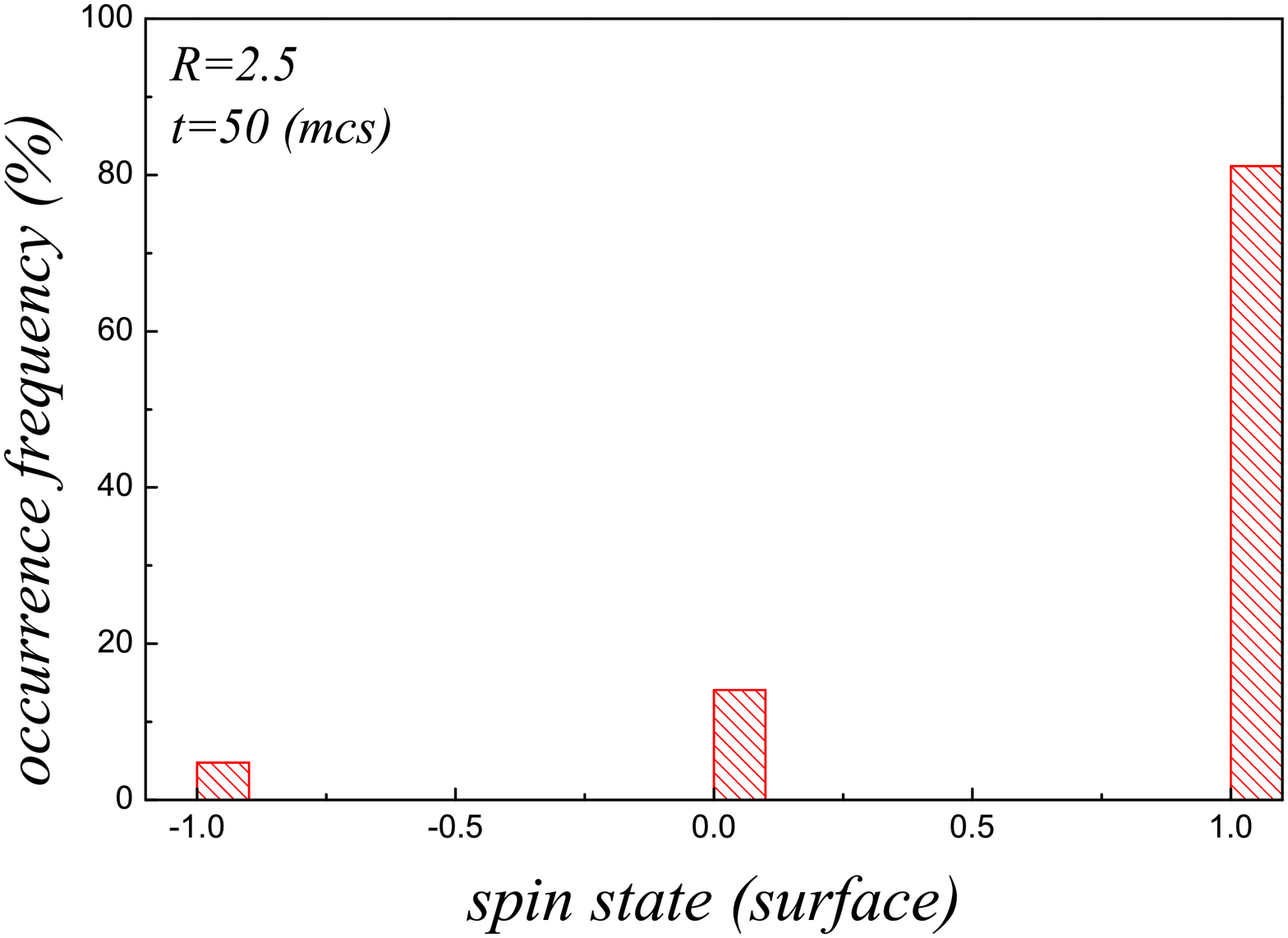}\\
\includegraphics[width=4cm]{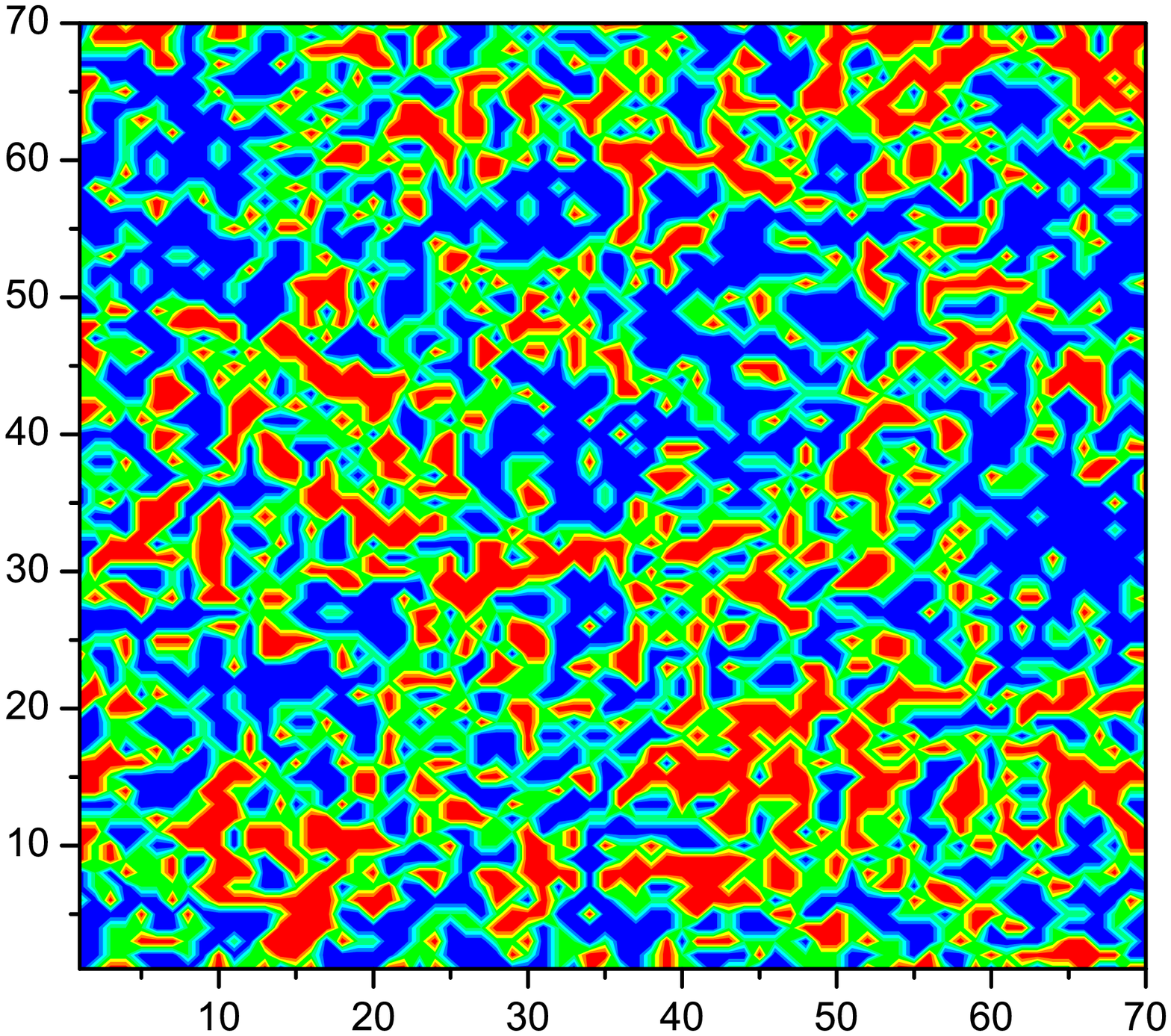}
\includegraphics[width=3.8cm,height=3.5cm]{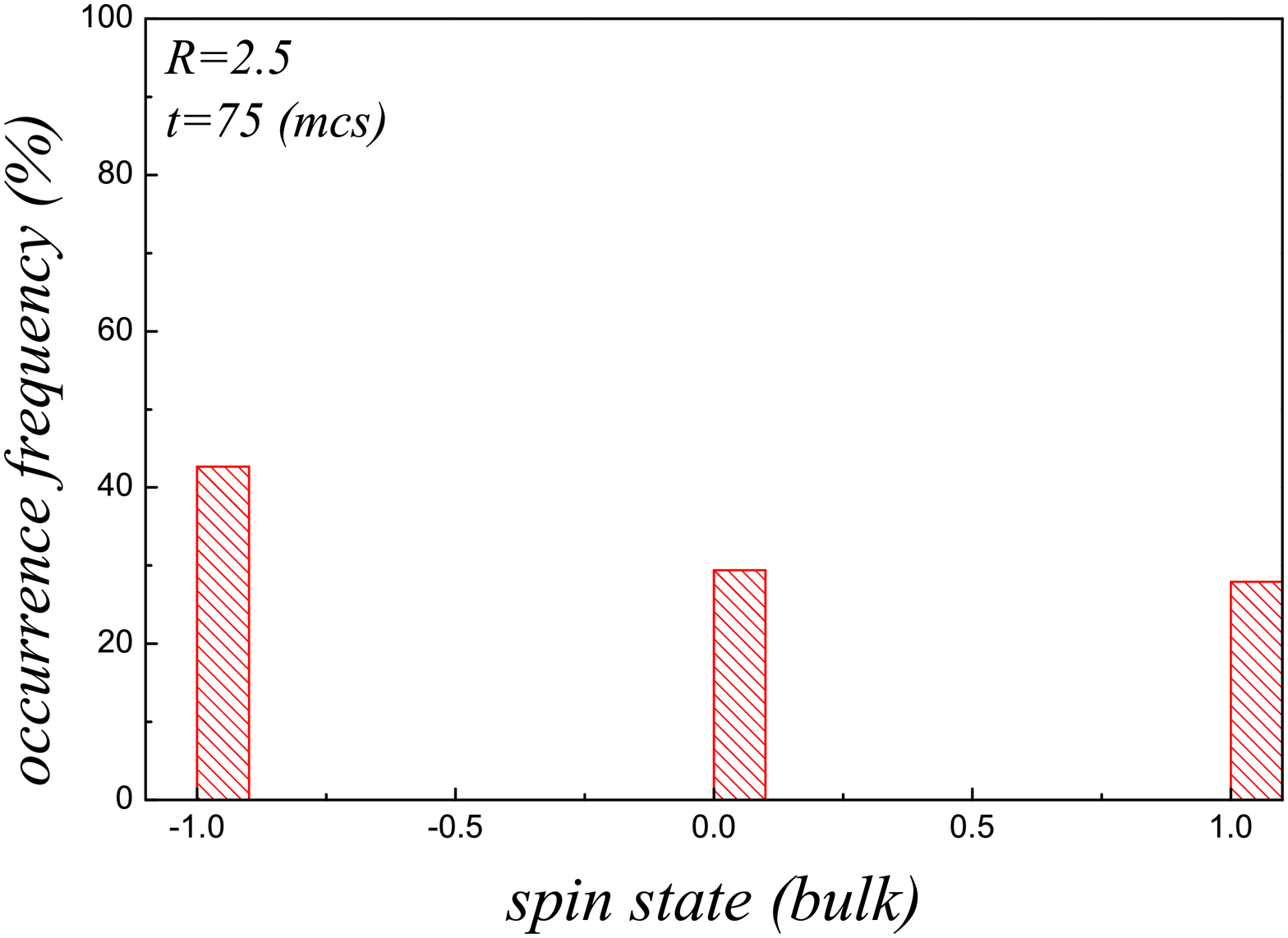}
\includegraphics[width=4cm]{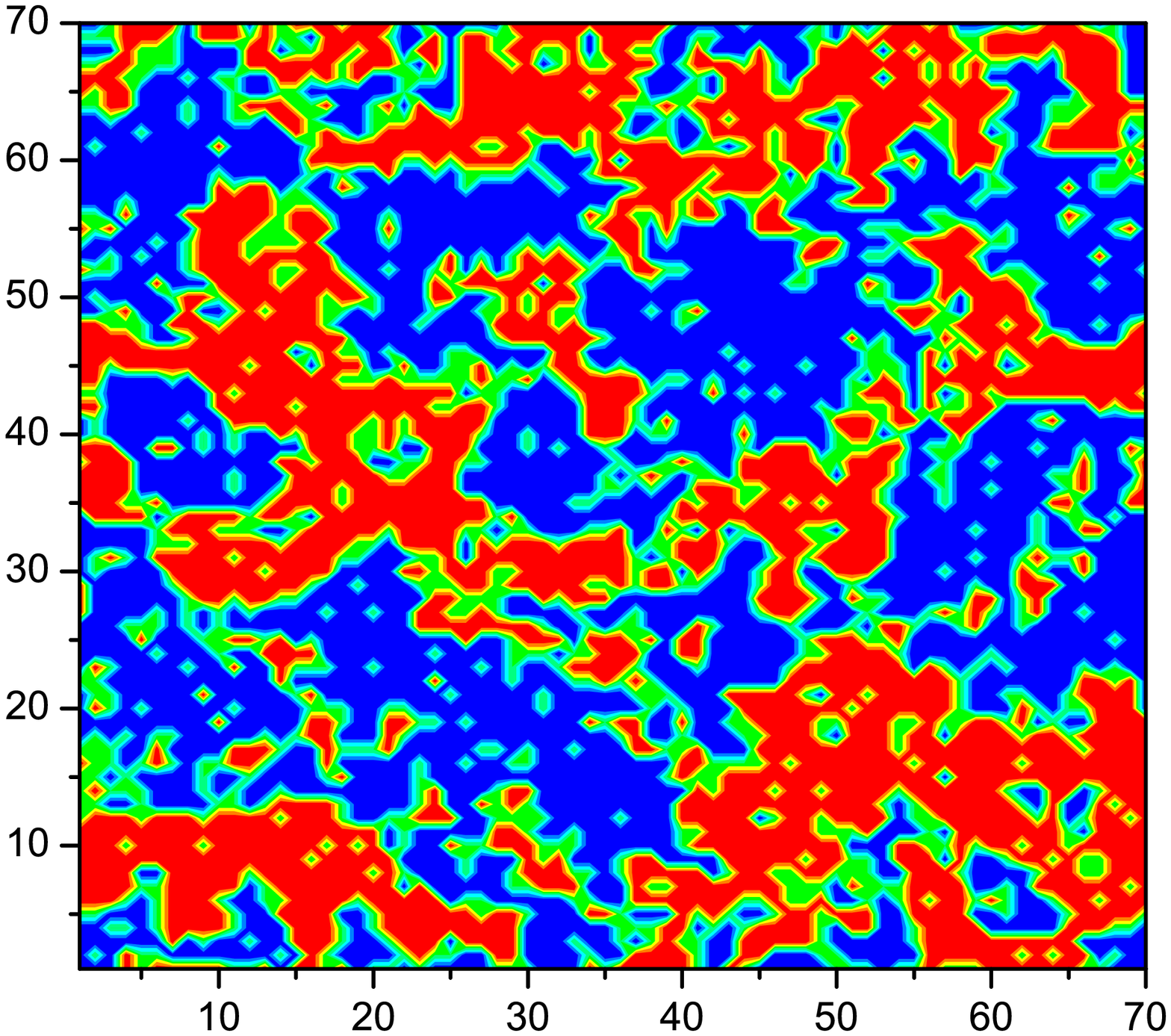}
\includegraphics[width=3.8cm,height=3.5cm]{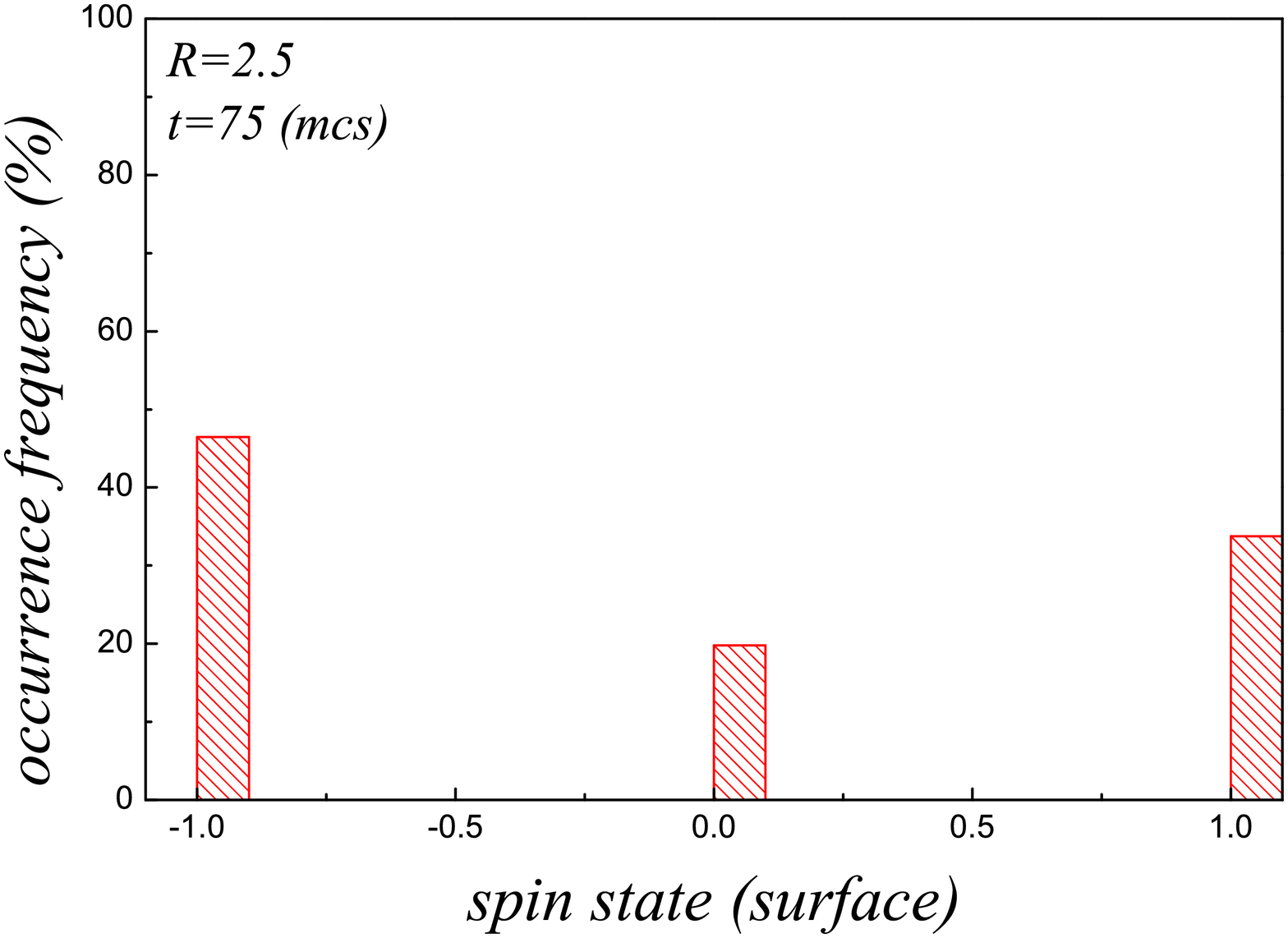}\\
\includegraphics[width=4cm]{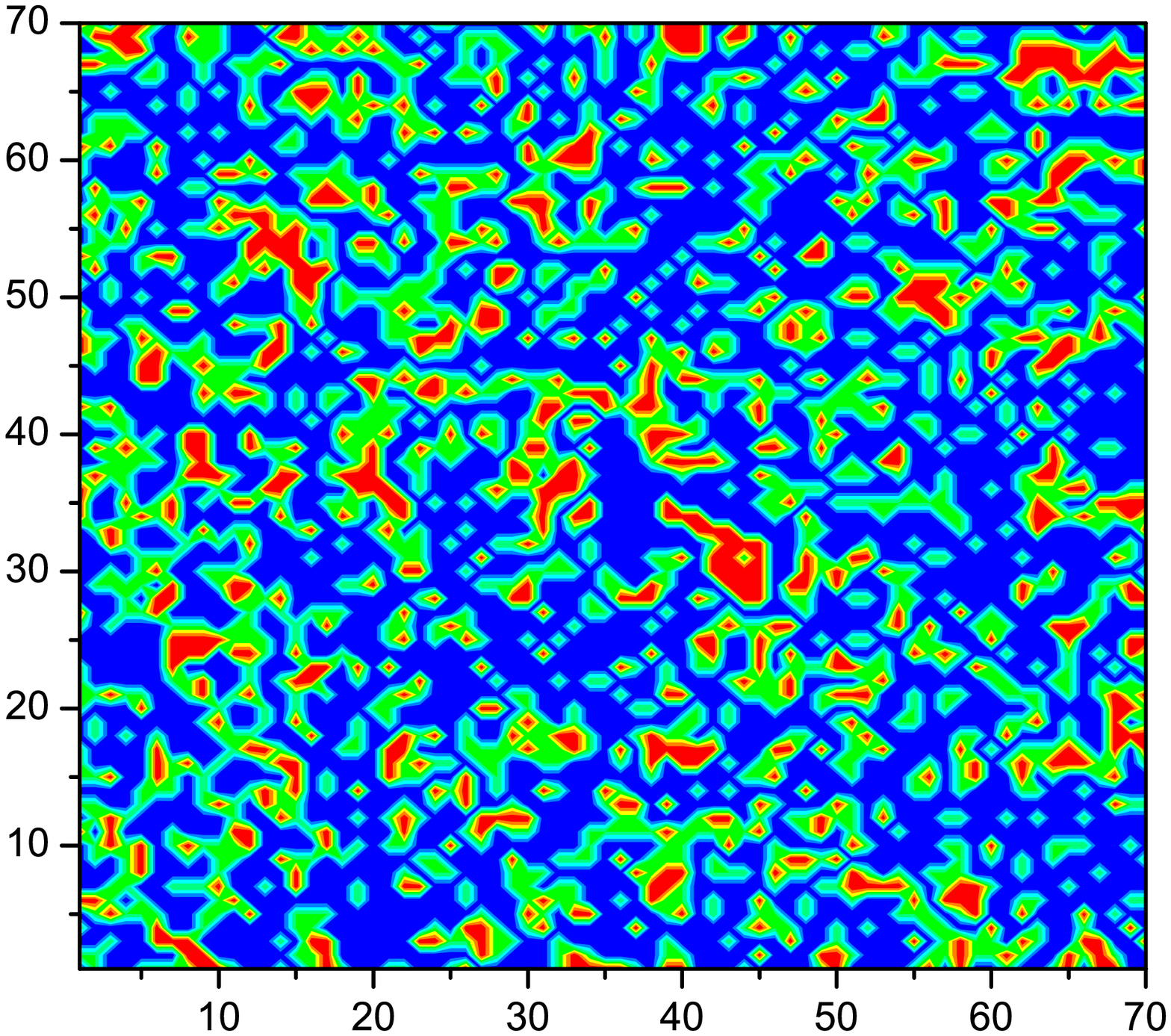}
\includegraphics[width=3.8cm,height=3.5cm]{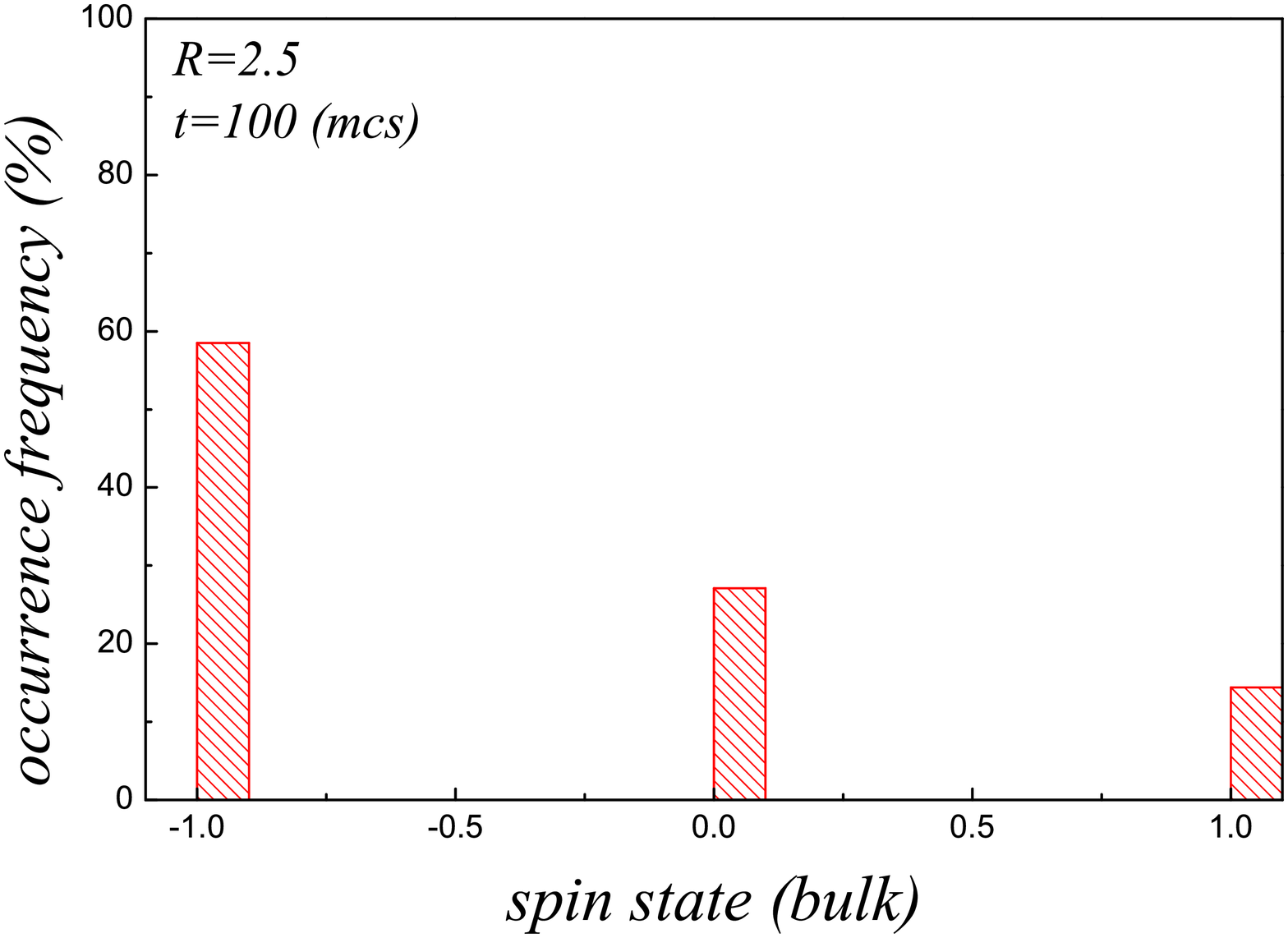}
\includegraphics[width=4cm]{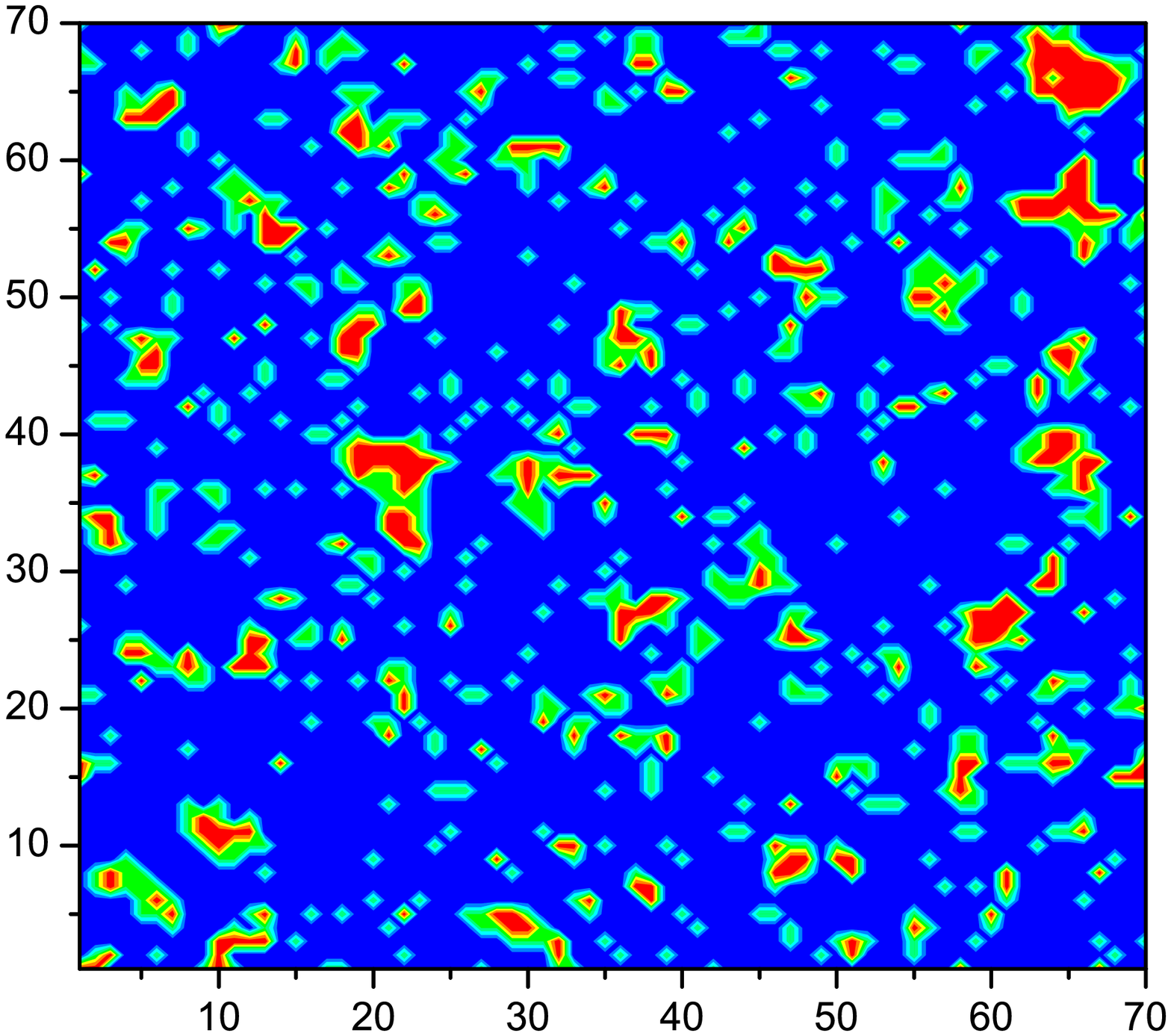}
\includegraphics[width=3.8cm,height=3.5cm]{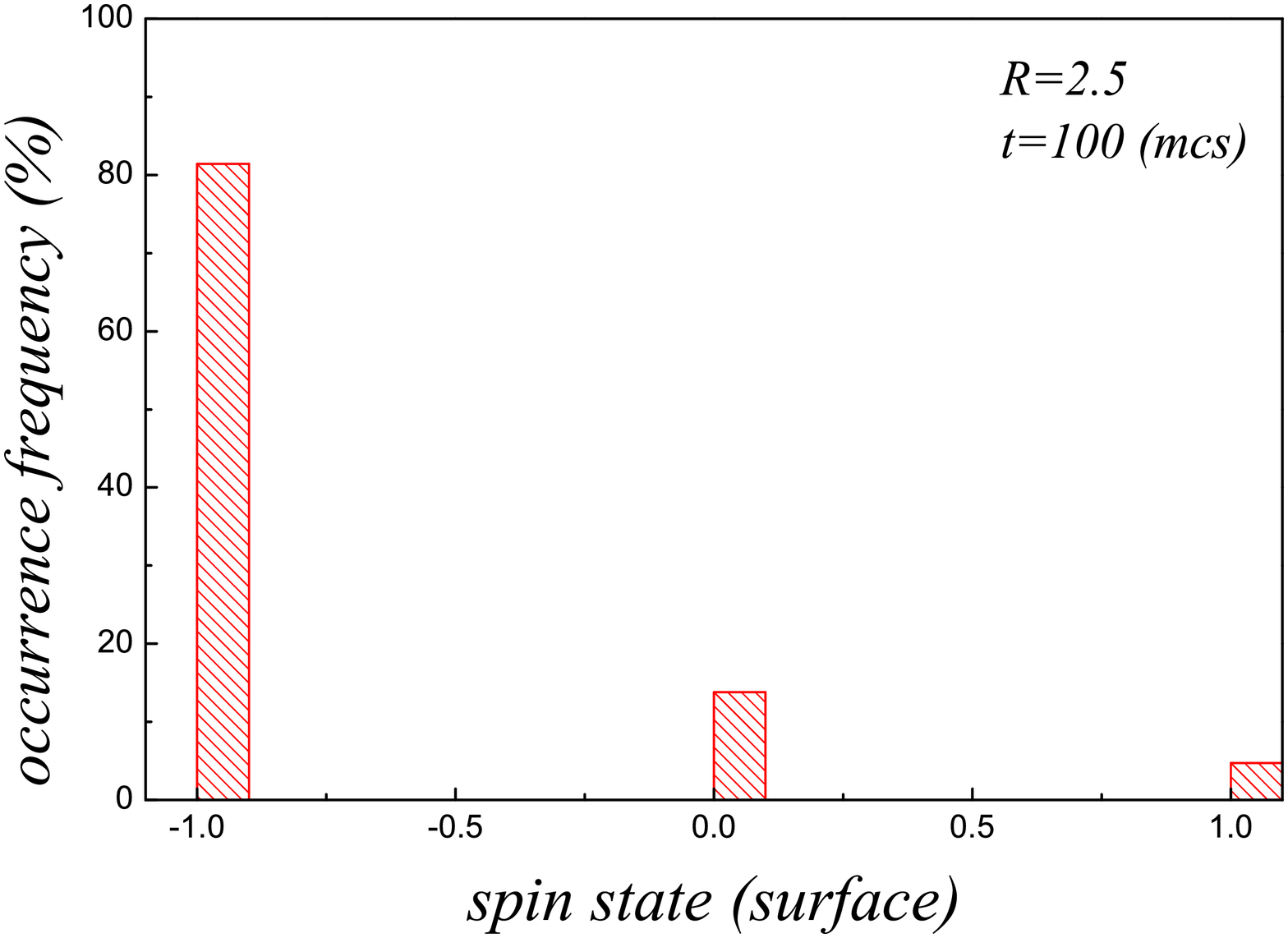}\\
\caption{}\label{fig8}
\end{figure}

\begin{figure}
\center
\includegraphics[width=7cm]{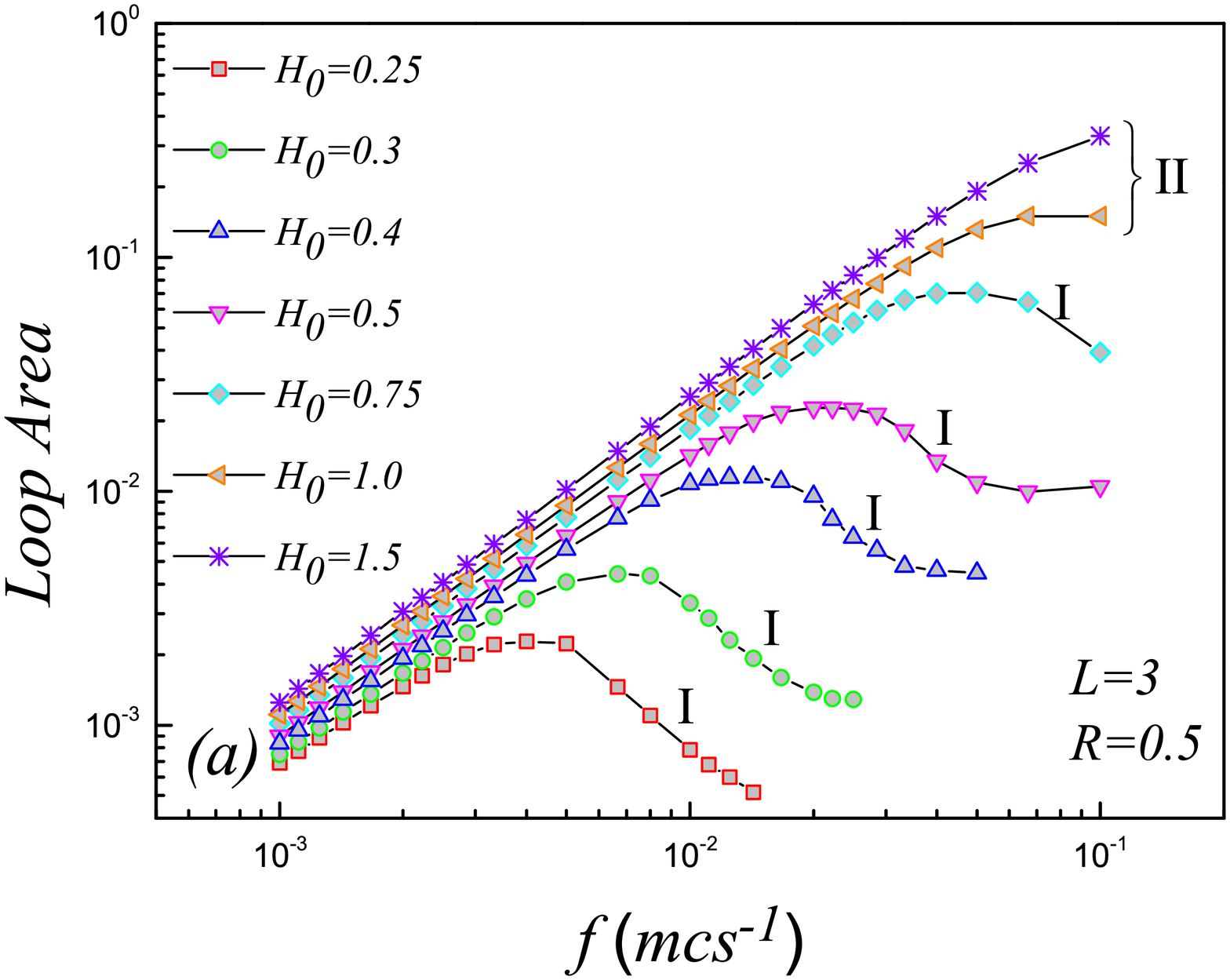}
\includegraphics[width=7cm]{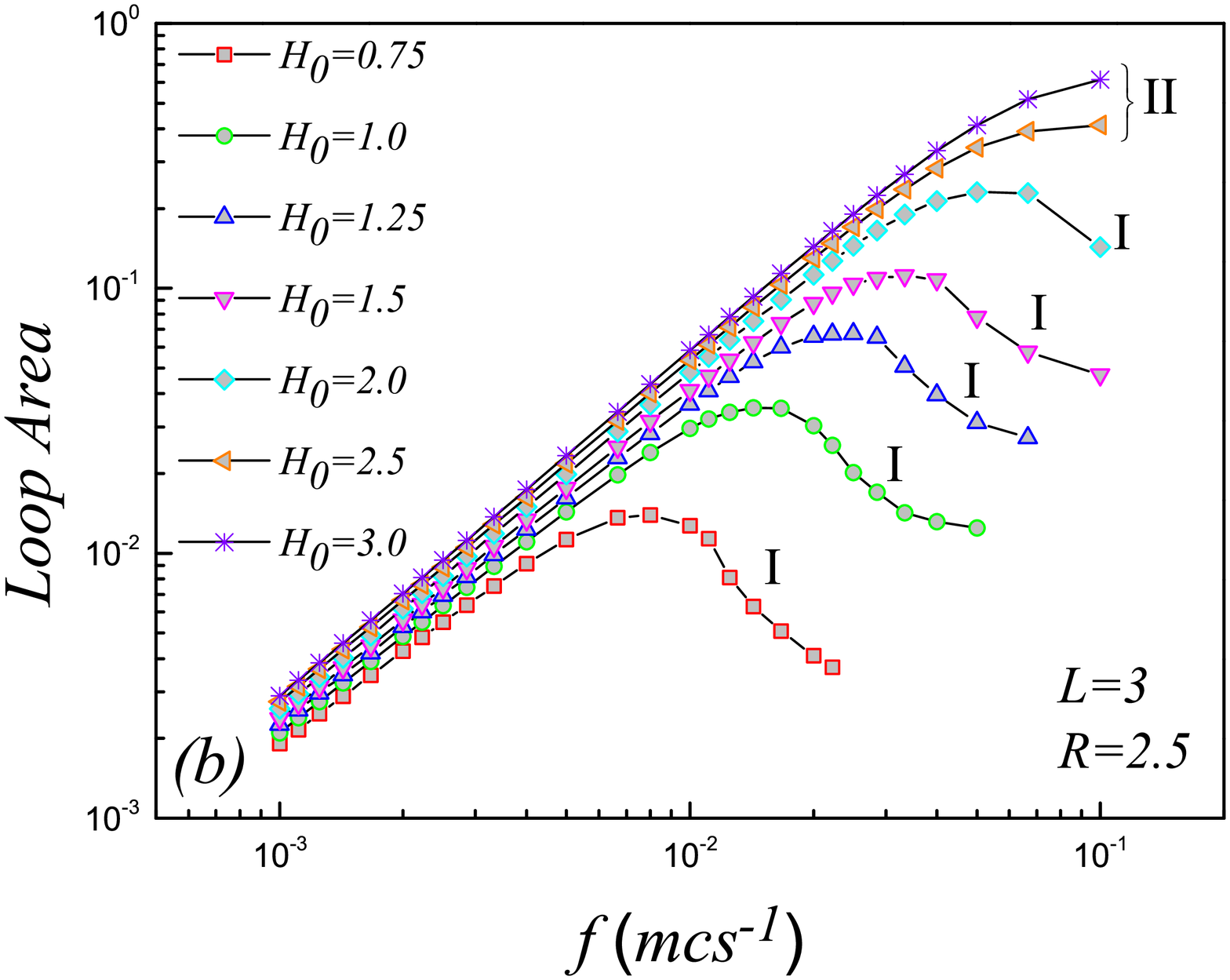}\\
\caption{}\label{fig9}
\end{figure}

\begin{figure}
\center
\includegraphics[width=7cm]{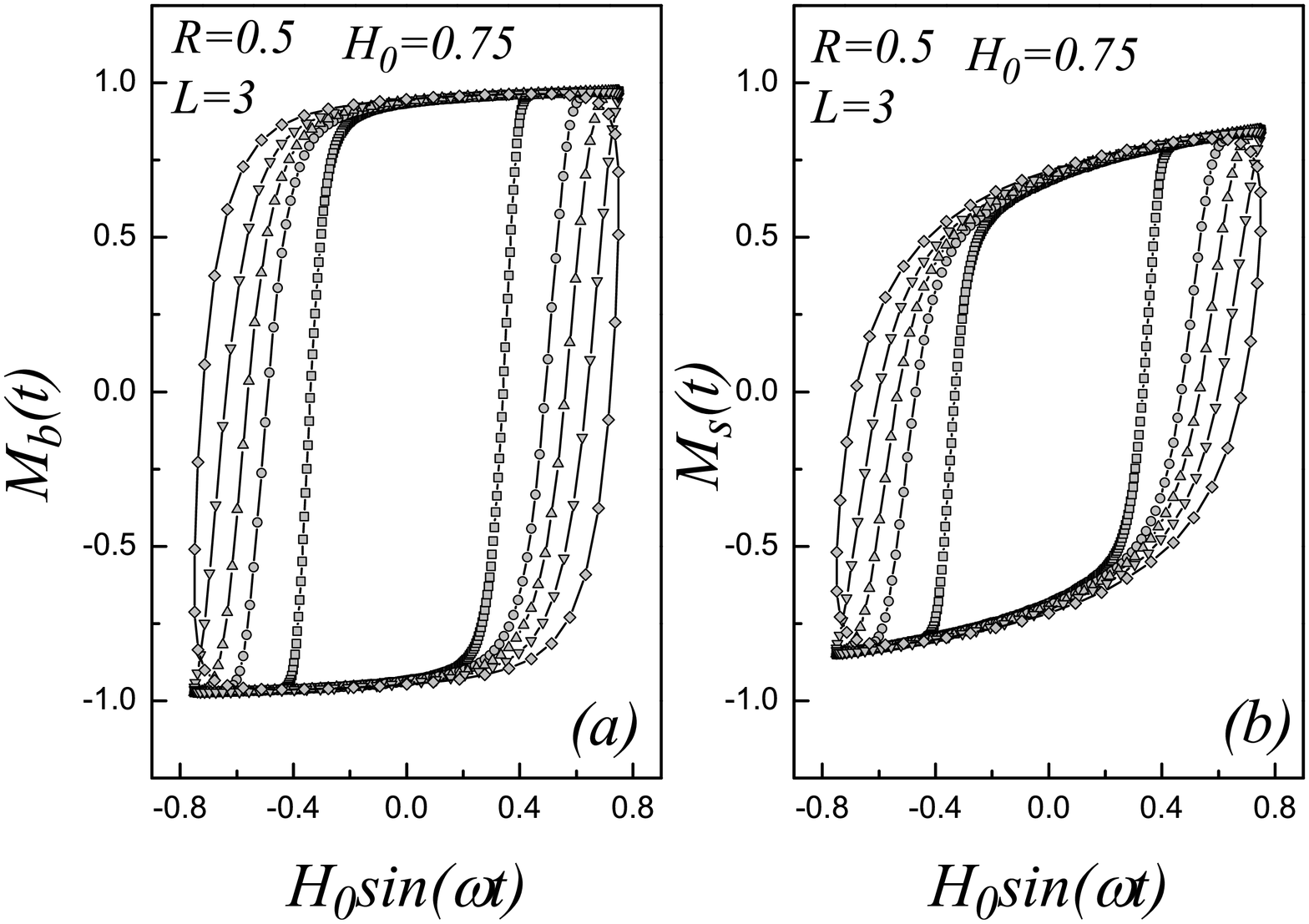}
\includegraphics[width=7cm]{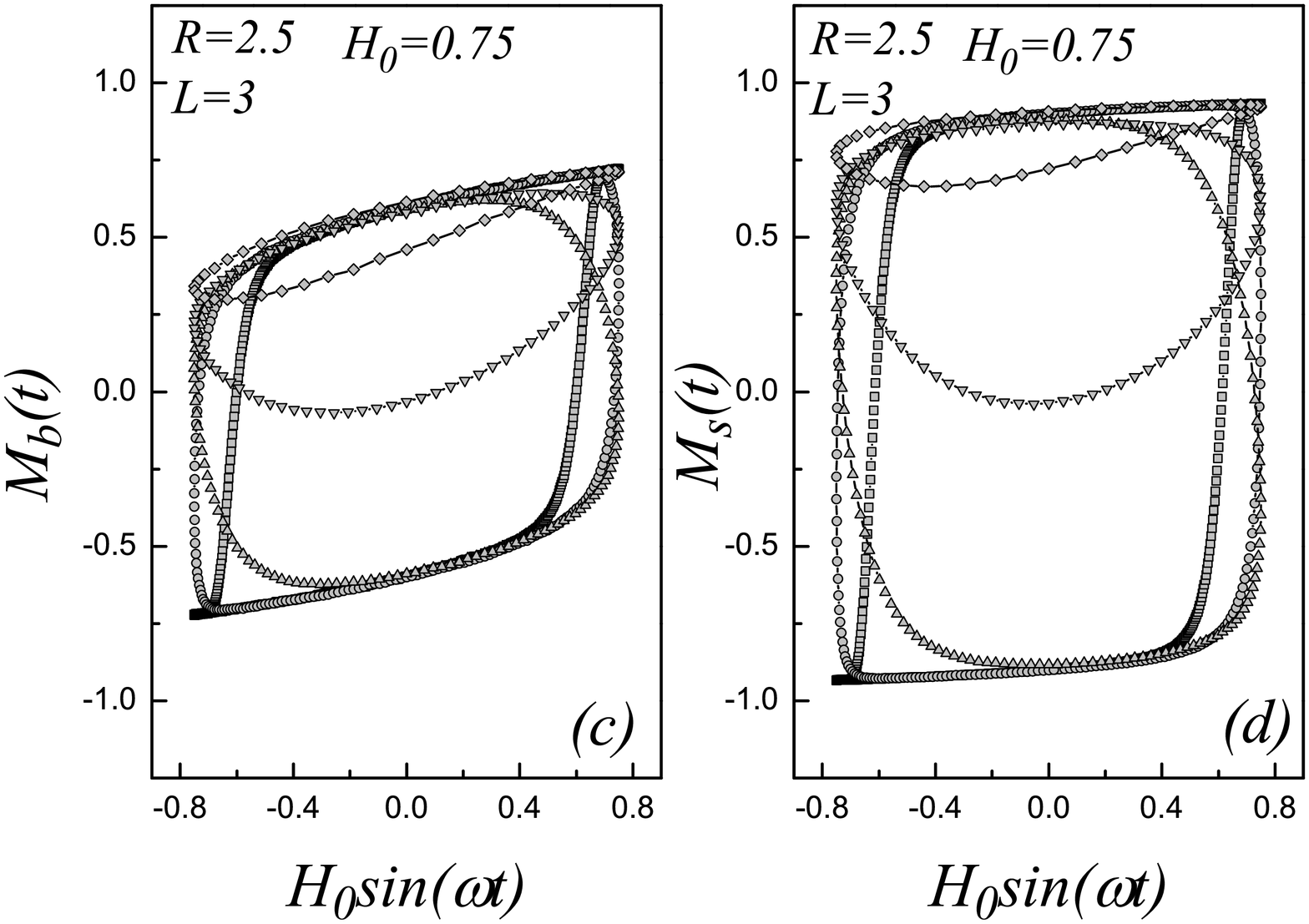}\\
\caption{}\label{fig10}
\end{figure}

\newpage
\section*{Figure Captions}
Fig.1 Phase diagrams of the system in a $(\Theta_{c}-R)$ plane for $\tau=100$ with various film thickness $L=3,4,5$. Three different values of field amplitude have been considered as (a) $H_{0}=0.0$, (b) $H_{0}=0.1$, and (c) $H_{0}=0.5$.

Fig.2 Left: Thermal variation of specific heat and hysteresis loop area curves corresponding to phase diagrams depicted in Fig. \ref{fig1}b with $L=5$. Both ordinary $(R=0.5)$, and extraordinary $(R=2.5)$ transitions are considered. Right: Typical bulk and surface hysteresis loops with $R=0.5$, and  $R=2.5$.

Fig.3 Effect of the film thickness $L$ on the thermal variation of hysteresis loop area curves in the presence of ordinary $(R=0.5)$ and enhanced $(R=2.0)$ surfaces. Oscillation period has been kept fixed as $\tau=100$ while two different field amplitude values have been considered as $H_{0}=0.1$ and $H_{0}=0.5$.

Fig.4 Effect of the film thickness $L$ on the thermal variation of dynamic correlation curves in the presence of ordinary $(R=0.5)$ and enhanced $(R=2.0)$ surfaces. Oscillation period has been kept fixed as $\tau=100$ while two different field amplitude values have been considered as $H_{0}=0.1$ and $H_{0}=0.5$.

Fig.5 The responses of the bulk and surface magnetization curves to the oscillating external magnetic field corresponding to various stages of a typical dynamic correlation versus temperature curve with $L=3$, $R=0.5$, $H_{0}=0.5$, and $\tau=100$.

Fig.6 Effect of the presence of modified surfaces on the thermal variation of hysteresis loop area and dynamic correlation curves for $L=3$ and $\tau=100$ with (a), (b) $H_{0}=0.1$, and (c), (d)  $H_{0}=0.5$.

Fig.7 Micomagnetic domain structures and statistical histograms of spin states at the maximum lossy point for $L=3$, $R=0.0$, $H_{0}=0.5$, and $\tau=100$. The successive rows correspond to the stages $\tau/4$, $\tau/2$, $3\tau/4$, and $\tau$ of the external field oscillation. Illustrated domain patterns consist of $S=1$ (red), $S=-1$ (blue), and $S=0$ (green) states. The first and the third columns in this figure are related to bulk and surface layers of the film, respectively.

Fig.8 Micomagnetic domain structures and statistical histograms of spin states at the maximum lossy point for $L=3$, $R=2.5$, $H_{0}=0.5$, and $\tau=100$. The successive rows correspond to the stages $\tau/4$, $\tau/2$, $3\tau/4$, and $\tau$ of the external field oscillation. Illustrated domain patterns consist of $S=1$ (red), $S=-1$ (blue), and $S=0$ (green) states. The first and the third columns in this figure are related to bulk and surface layers of the film, respectively.

Fig.9 Frequency dispersion of hysteresis loop area curves as a function of the field amplitude $H_{0}$ for (a) ordinary $(R=0.5)$, and (b) extraordinary $(R=2.5)$ surfaces. The film thickness is selected as $L=3$.

Fig.10 Bulk and surface hysteresis loops corresponding to the curves labeled $H_{0}=0.75$ in Fig. \ref{fig9}. Selected oscillation frequency values are $1000^{-1} (\blacksquare)$, $250^{-1} (\bullet)$, $150^{-1} (\blacktriangle)$, $90^{-1} (\blacktriangledown)$, $50^{-1} (\blacklozenge)$.
\end{document}